\documentclass[pra, preprint, superscriptaddress,preprintnumbers,showpacs,amsmath,amssymb]{revtex4-1}
\usepackage{morefloats}
\usepackage{psfrag}
\usepackage{graphicx}
\usepackage{subfigure}
\def\be{\begin{equation}}       \def\ee{\end{equation}}
\def\bea{\begin{eqnarray}}      \def\eea{\end{eqnarray}}

\usepackage{natbib}   

\begin{document}
\title{Entanglement robustness in Heisenberg spin chains coupled to dissipative environment at finite temperature}
\author{Gehad Sadiek \footnote{Corresponding author: gsadiek@sharjah.ac.ae}}
\affiliation{Department of Applied Physics, University of Sharjah, Sharjah 27272, UAE}
\affiliation{Department of Physics, Ain Shams University, Cairo 11566, Egypt}
\author{Samaher Almalki}
\affiliation{Department of Physics, King Saud University, Riyadh 11451, Saudi Arabia}
\date{\today}
\begin{abstract}
We consider a finite one-dimensional Heisenberg XYZ spin chain under the influence of dissipative Lindblad environment obeying the Born-Markovian constrain in presence of an external magnetic field. We apply both closed and open boundary conditions at zero and finite temperature. We present an exact numerical solution for the Lindblad master equation of the system in the Liouville space. we find that, in the free spin chain (in absence of any environment), the entanglement at all ranges evolve in time in a non-uniform oscillatory form that changes significantly depending on the initial state, system size and the spatial anisotropy. The oscillatory behavior is suppressed once the system is coupled to the environment. Furthermore, the asymptotic behavior of the entanglement, nearest neighbor and beyond, in the system under the influence of the environment at zero temperature is very sensitive to the x-y spatial anisotropy, which causes them to reach either a zero or a finite sustainable steady state value regardless of the initial state of the system. The anisotropy in the $z-$direction may enhance the entanglement depending on the interplay with the magnetic field applied in the same direction. As the temperature is raised, the steady state of the short range entanglements is found to be robust within very small non-zero temperature range, which depends critically on the spatial anisotropy of the system. The entanglement at each range depends differently on the spatial anisotropy. Moreover, the end to end entanglement transfer time and speed through the open boundary chain vary significantly based on the degree of anisotropy and the temperature of the environment.
\end{abstract}
\maketitle
\section{Introduction}

Quantum entanglement plays a vital role in the static and dynamic behavior of many body systems \cite{Peres1993}. It is considered as the physical resource responsible for manipulating the linear superposition of the quantum states in quantum systems. Entanglement, and its derivatives, show scaling behavior as the physical system experiences a quantum phase transition \cite{Sachdev2001}. Particularly, it is considered as a crucial resource in quantum information processing fields such as quantum teleportation, cryptography, and quantum computation where it provides the physical basis for implementing the different needed algorithms \cite{Nielsen2000}. Therefore, creating, quantifying, transferring and protecting entanglement in quantum states of multiparticle systems is in the focus of interest of both theoretical and experimental research. However, quantum entanglement is very fragile due to the induced decoherence caused by the inevitable coupling of the quantum system to its surrounding environment \cite{Zurek1991, Bacon2000}. The main effect of decoherence is to randomize the relative coherent phases of the possible states of the quantum system diminishing its quantum aspects. It is considered as one of the main obstacles toward realizing an effective quantum computing system. Offering a potentially ideal protection against environmentally induced decoherence is found to be a very difficult task. The decoherence in the system causes sweeping out of entanglement between the different parties of the system. Therefore, monitoring the entanglement dynamics in the considered system helps us understand the behavior of the decoherence as well. 

The Heisenberg interacting spin systems have been in focus of interest for their own sake as they describe the novel Physics of localized spins in magnetic systems as well as for their successful role in representing many of the physical systems that are very promising candidates for quantum information processing such as the solid state systems\cite{Loss1998, Buckard1999, Imamoglu1999}, NMR \cite{Kane1998, Ernst1988}, optical lattices \cite{Sorensen2001, Liu2002}, electronic spins \cite{Vrijen2000}, superconducting arrays \cite{Heule2011}. 
Entanglement properties and dynamics in Heisenberg spin chains in absence of dissipative environments have been studied intensively \cite{Sadiek2010,Barouch1970,Sen(De)2004,HuangZ2006,Lieb1961,XuQ2010,XuQ2011,Sadiek2013}. There have been several interesting works that focused on the dynamics of a system of interacting qubits, represented by Heisenberg spin model, coupled to dissipative environments. Particularly, the problem of two qubits coupled to dissipative environments has been intensively studied. Analytic and numerical solutions were provided for a two-qubit XY system in an external magnetic field coupled to a population relaxation environment as well as a thermal environment \cite{Wang2006}. It was shown that the system reaches a steady state value though it is coupled to a population relaxation environment, which causes decoherence, provided that the spatial anisotropy of the system is maintained. The steady state value may vanish as the temperature of the thermal environment is raised. The anisotropic two-qubit XYZ Heisenberg model in an inhomogeneous magnetic field coupled to a population relaxation environment at zero temperature was investigated too, both analytically and numerically \cite{ABLIZ2006}. It was demonstrated that the two-qubit system reaches a steady state starting from an initial separable state as long as the anisotropy of the spin coupling in the $x$ and $y$ direction is non-zero regardless of the value of the coupling in the $z$-direction. The spin relaxation in a two-qubit Ising system under a single spin flip inducing environment was investigated and the relaxation rates were calculated \cite{Dubi2009}.

The one-dimensional multiqubit chains, $N > 2$, coupled to dissipative environments were investigated as well at different degrees of anisotropy, magnetic field strength and temperatures \cite{Hein2005, Tsomokos2007_NJP9_79, Buric2008, Hu2009, Hu2009a, Buric2009, Pumulo2011, Zhang2013}. Of most relevance, the time evolution of the concurrence of the nearest neighbor spins in a one-dimensional $XX$ spin chain in absence of any external magnetic fields coupled to a thermal and dephasing environments were studied \cite{Hu2009}. It was shown that in all cases the entanglement vanishes within a finite time that depends on the system-environment coupling parameter and temperature. The dynamics of entanglement in the Ising and isotropic ($XXX$) one-dimensional spin chains has been investigated \cite{Buric2008} using numerical stochastic approach by applying the quantum state diffusion theory \cite{Lakshminarayan2001}, which reduces the needed huge storage space from $2^{2N}$ to $2^N$ for $N$ interacting spins. They focused on the influence of noise during short periods of time. The effect of the initial state of the system on the time evolution behavior under coupling with the environment was considered and was shown that most of the time the main effect of the noise is to reduce the amplitude of the large oscillation of the entanglement. An Ising one-dimensional spin system in an external magnetic field with two non-vanishing components in the $x$ and $z$ direction and coupled to a Markovian environment was investigated using stochastic calculations too \cite{Buric2009}. One particular work of special interest considered a one-dimensional chain of superconducting Josephson qubits with experimentally realistic conditions \cite{Tsomokos2007_NJP9_79}. The effect of the environmental noise on the entanglement in the chain was tested. The influence of the noise was introduced as a set of bosonic baths such that each one of them is coupled to a single qubit. It was shown that this noise environment causes significant change to the entanglement dynamics of the Josephson qubits. In the limiting case when the internal degrees of freedom of the bath's were traced out the system behaves as an Ising spin chain coupled to a Born-Markovian environment with an asymptotic steady state entanglement. Other recent works have investigated the entanglement dynamics in spin systems under different environmental and external effects and focused on the entanglement and information transfer through the system \cite{Petrosyan2010_PRA81_042307, Ronke2011_PRA83_012325, Alkurtass2013_PRA88_062325, Wu2014_PRA89_062105}.

In this paper, we investigate the time evolution and transfer of quantum entanglement in a finite one-dimensional Heisenberg $XYZ$ spin-1/2 chain with nearest-neighbor spin interaction under the influence of dissipative Lindblad environment in presence of an external magnetic field at zero and finite temperature. We consider both cases of closed and open boundary spin chains with maximum number of 7 spins. We provide an exact numerical solution of the Lindblad master equation of the system. In the closed boundary case, we show how the nearest neighbor (nn) and beyond nearest neighbor entanglement (nnn, nnnn, ...) as well as the one-tangle $\tau_1$ and the overall bipartite entanglement $\tau_2$ in the free (isolated) system evolve in time in a non-uniform oscillatory form that changes significantly depending on the initial state of the system, the number of spins and the degree of spatial anisotropy but disappears in presence of the environment. Also, we investigate the asymptotic steady state of the entanglement at the different ranges in the system under the influence of the environment at zero temperature and show how it varies strongly and differently based on the degrees of anisotropy of the spin coupling strength, leading to either a vanishing or a constant steady state value. We emphasis the important role played by the interplay between the spin coupling in the $z-$direction and the external magnetic field applied in the $z-$direction. We explore the robustness of the quantum effects and the steady state of the entanglement at finite temperature and its critical dependence on the degree of anisotropy. We study the end to end entanglement transfer through the open boundary chain starting from an initial state with a maximum entanglement at one terminal of the chain and disentanglement over the rest of it. We discuss how the entanglement transfer time, speed and residue through the chain vary depending on the degrees of anisotropy, the temperature and the separation from the maximally entangled end.
This paper is organized as follows. In the next section, we present our model and calculations. In sec. III, we study the time evolution of the entanglement in Heisenberg spin chains with closed boundary condition in absence and presence of the Lindblad environment at zero and finite temperature. In sec. IV, we investigate the entanglement transfer in a Heisenberg chain with open boundary condition under the influence of thermal and dissipative environments. We conclude in sec. V.
\section{The Model}
We consider a one-dimensional system of N spin-1/2 particles with nearest neighbor coupling $J$ in an external magnetic field in the $z$-direction $B$. The system is described by the Heisenberg Hamiltonian 
\begin{equation}
H=\frac{(1+\gamma)}{2} J \sum_{i=1}^{N} S_{i}^{x} S_{i+1}^{x} 	+  \frac{(1-\gamma)}{2} J \sum_{i=1}^{N} S_{i}^{y} S_{i+1}^{y} 	+   \delta J \sum_{i=1}^{N} S_{i}^{z} S_{i+1}^{z}	 +  \sum_{i=1}^{N} B^{z} S_{i}^{z}\;, 
\label{eqn:H}
\end{equation}
where: $S^{\alpha}_{i}$ = $\frac{1}{2}$ $\sigma^{\alpha}_{i}$ ($\alpha$ = $\textit{x}$, $\textit{y}$ or $\textit{z}$) and $\sigma^{\alpha}_{i}$ are the local spin-$\frac{1}{2}$ operators and Pauli operators, respectively (for convenience we set $\hbar = k = 1$). When we apply the periodic boundary condition we set $S_{N+1}=S_{1}$. $\gamma$ and $\delta$ are the anisotropy parameters which determines the relative strength of the spin coupling in the $x$- , $y$- and $z$-directions. We study different classes of the Heisenberg spin system by changing the values of the parameters $\gamma$ and $\delta$ such as the Ising ($\gamma=1$ and $\delta=0$), $XX$ ($\gamma=0$ and $\delta=0$), $XXX$ ($\gamma=0$ and $\delta=0.5$), $XYZ$ ($\gamma=0.5$ and $\delta=0.5$), etc. The system is subject to an external homogeneous static magnetic field $\textbf{\textit{B}} = B^{z} \; \hat{z} = \omega \; \hat{z}$ in the $z$-direction, where $\omega$ represents the magnitude of effective applied external magnetic field as well as the energy gap of each spin.

The dynamics of an isolated quantum system is described by the time evolution of its density matrix $\rho(t)$ according to the Liouville equation $\dot{\rho}\left(t\right) = -i \left[\textit{H},\rho\right]$. But for an open quantum system that is interacting with its environment, the Liouville equation has to be modified to account for the dissipative effects of the environment on the system. If the system and the environment satisfy the conditions of weak coupling as well as short relaxation time within the environment excitation modes, the Born-Markovian approximation can be applied and the time evolution of the system is best described by the Lindblad Master equation \cite{Lindblad1976, Breuer2002}, which preserves the hermiticity and  unit trace of the density matrix and guarantees positive continuous evolution of the system under the effect of the environment, defined by  
\begin{equation}
\dot{\rho}\left(t\right) = -i \left[\textit{H},\rho\right] + \mathcal{D}_{\rho}\;,
\label{eqn:Lindblad}
\end{equation}
where $\mathcal{D}_{\rho}$ is the extra term that describes the dissipative dynamics and is represented in the Lindblad form as
\begin{equation}
\mathcal{D}_{\rho} = -\frac{1}{2} \sum_{k=1} \left\{ [L_{k} \rho, L_{k}^{\dagger}] + [L_{k}, \rho L_{k}^{\dagger}]\right\} \;,
\label{eqn:dissipative}
\end{equation}
where the Lindblad operator $L_{k}$ represents all the effects of the considered environment on the system site $k$, where the environment is assumed to couple to each site independently of the other sites and therefore is represented by
\begin{equation}
 L_{k} = {\bf 1}_1  \otimes {\bf 1}_2 \otimes \cdots \otimes L_k \otimes \cdots \otimes {\bf 1}_N \;.
\end{equation}
For $Q-$dimensional Hilbert space, the density operator is represented by a $Q$ by $Q$ matrix, but it is more convenient to work in the Liouville space, where it is represented as a vector with $Q^2$ elements,  $\vec{\rho} = \left(\rho_{11},\rho_{12},\rho_{13},...,\rho_{1Q}, ...,\rho_{21},\rho_{22},...,\rho_{2Q},...,\rho_{QQ}\right)$. In fact, the selected order of the elements is not important but has to be preserved once chosen. The main idea here is to reformulate Eq.~(\ref{eqn:Lindblad}) to take the matrix equation form 
\begin{equation}
\vec{\dot{\rho}}(t) = (\hat{\mathcal{L}}^H+\hat{\mathcal{L}}^D)\vec{\rho} = \hat{\mathcal{L}}\vec{\rho}\;,
\label{eqn:matrixform}
\end{equation}
where $\hat{\mathcal{L}}^H$ and $\hat{\mathcal{L}}^D$ are superoperators acting on the vector $\rho$ in the Liouville space, where the first one represents the unitary evolution due to the free Hamiltonian while the second represents the dissipation process. The matrix elements of $\dot{\rho}$  are defined as
\begin{equation}
\dot{\rho}_{jl}(t) = -i \sum_{m,n} \;\;(\mathcal{L}_{jl,mn}^H+\mathcal{L}_{jl,mn}^D)\;\; \rho_{mn} \;,
\end{equation}
where the tetrahedral matrices $\mathcal{L}^H$ and $\mathcal{L}^D$ are given by
\begin{equation}
\mathcal{L}_{jl,mn}^H = H_{jm} \delta_{ln} - \delta_{jm} H_{nl} \;,
\end{equation}
and
\begin{equation}
\mathcal{L}_{jl,mn}^D = \frac{i}{2} \sum_{k} [2 (L_{k}^{\dagger})_{nl} (L_{k})_{jm} - (L_{k}^{\dagger} L_{k})_{jm} \delta_{ln} - \delta_{jm} (L_{k}^{\dagger} L_{k})_{nl}] \;.
\end{equation}
Now the whole problem of evaluating the time evolution of the density matrix has been reduced to seeking the solution of the standard matrix equation (\ref{eqn:matrixform}) which can be achieved once we find the set of all eigenvalues $\left\{\lambda_1,\lambda_2,\lambda_3, ... , \lambda_{Q^2}\right\}$ and eigenvectors $\left\{\vec{\eta}_1, \vec{\eta}_2, \vec{\eta}_3, ..., \vec{\eta}_{Q^2}\right\}$ of the $Q^2 \times Q^2$ tetrahedral matrix $\mathcal{L}$, and as a result the density vector becomes
\begin{equation}
\vec{\rho}(t)= \sum_{i=1}^{Q^2} A_i \; \vec{\eta}_i \; e^{\lambda_i \; t} \;,
\end{equation}
where the coefficients $A_i$ are determined from the initial conditions of the evolution process. Once the density (vector) matrix has been calculated as a function of time, we can evaluate the entanglement in the chain as explained below. 
For a one-dimensional chain with $N$ spin-1/2 particles, the dimension of the Hilbert space is $2^N$ and the dimension of the tetrahedral matrices is $2^{2N}$ which, even for a small number of spins, is extremely large and requires a huge computational storage that is more than what can be handled by most of the available computing systems and represents a real challenge in such type of problems.

For the Heisenberg spin chain described by the Hamiltonian (\ref{eqn:H}), the effect of the dissipative and thermal environment is given by the local Lindblad operator \cite{Breuer2002, Tsomokos2007_NJP9_79, Mintert2005} 
\begin{equation}
\label{Lindblad_operator}
L_k = \Gamma \; \left\{ \frac{(\bar{n}+1)}{2} \; S_{k}^{-} + \bar{n}\; S_{k}^{\dagger}\right\}\;,
\end{equation}
Where $S^{+}$ and $S^{-}$ are the spin raising and lowering operators, $S^{\pm}$=$S^{x}$ $\pm$ $iS^{y}$. $\Gamma$ is a phenomenological parameter that determines the strength of the coupling between the environment and the system and is assumed to be the same for all spins. The thermal parameter $\bar{n}$ is proportional to the temperature of the environment. Obviously, in Eq.~(\ref{Lindblad_operator}), the first term induces the dissipation process whereas the second one causes excitation. As mentioned before, for Eq.~(\ref{eqn:Lindblad}) to represent a good approximation for the time evolution of the system, certain restrictions have to apply to the system parameters, the coupling parameter between the system and the environment $\Gamma$ as well as the relaxation time scale of the environment dynamics should be small compared to that of the system dynamics manifested by the parameter $\omega$ representing the spin precession frequency around the $z$-axis. As a result, we consider values of $\Gamma$ and $J$ such that $\Gamma$ and $J << \omega$.

We adopt the concurrence as a measure of the bipartite entanglement in the system, where Wootters \cite{Wootters1998} has shown that for a pair of two-state systems $i$ and $j$, the concurrence $C_{i, j}$, which varies between $0$ to $1$, can be used to quantify the entanglement between them and is defined by
\begin{equation}
\label{concurrence}
C_{i, j}(\rho_{i, j})=max\{0,\epsilon_1-\epsilon_2-\epsilon_3-\epsilon_4\} \;,
\end{equation}
where $\rho_{i, j}$ is the reduced density matrix of the two spins under consideration, $\epsilon_i$'s are the eigenvalues of the Hermitian matrix
$R\equiv\sqrt{\sqrt{\rho_{i, j}}\tilde{\rho_{i, j}}\sqrt{\rho_{i, j}}}$ with
$\tilde{\rho_{i, j}}=(\sigma^y \otimes
\sigma^y)\rho_{i, j}^*(\sigma^y\otimes\sigma^y)$ and $\sigma^y$ is the
Pauli matrix of the spin in the $y$-direction. We use the one-tangle $\tau_1= 4 \; det \; \rho_1$ to quantify the entanglement between a single spin and the rest of the system in a pure state, where $\rho_1$ is the single site reduced density matrix \cite{Coffman2000_PRA_61_052306, Amico2004_PRA_69_022304}. On the other hand, the sum of the squared of pairwise concurrences, between the spin $i$ and the rest of the spins in the system, $\sum_{j\neq i} C_{i,j}^{2}$, defines another quantity $\tau_2$ representing the overall pairwise entanglement in the system. The ratio $R = \tau_2/\tau_1$ was introduced as a measure of the fraction of the total entanglement attributed to the pairwise correlations within the system \cite{Coffman2000_PRA_61_052306, Amico2004_PRA_69_022304}. Of course, $\tau_1$ and $R$ are evaluated only in pure states of the system (at $T=0$), otherwise they are not defined. We study the time evolution of the system using the standard basis
$\left\{\left|\uparrow\uparrow\cdots\uparrow\right\rangle,\left|\uparrow\uparrow\cdots\downarrow\right\rangle,\cdots,\left|\uparrow\downarrow\cdots\downarrow\right\rangle,\cdots,\left|\downarrow\downarrow\cdots\downarrow\right\rangle\right\}$ and starting from different initial typical states: a separable (disentangled) state, $\left|\psi_s\right\rangle=\left|\uparrow\uparrow\cdots\uparrow\right\rangle$; a partially entangled ($W$-state), $\left|\psi_w\right\rangle$=$\frac{1}{\sqrt{N}}\left(\left|\uparrow\downarrow\cdots\downarrow\right\rangle+\left|\downarrow\uparrow\cdots\downarrow\right\rangle+\cdots+\left|\downarrow\downarrow\cdots\uparrow\right\rangle\right)$ and a maximally entangled state,  $\left|\psi_m\right\rangle=\frac{1}{\sqrt{2}}\left( \left| \uparrow\downarrow\right\rangle +\left| \downarrow\uparrow\right\rangle \right) \left|\downarrow\downarrow\cdots\downarrow\right\rangle$.
\section{Dynamics of entanglement in closed boundary spin chains}
\subsection{The Free System}
It is very enlightening to start our study by considering the entanglement dynamics in the free (isolated) Heisenberg spin chains before considering the environment effect, which is described by the Hamiltonian (\ref{eqn:H}). In general, for convenience we consider the time evolution of the system in terms of the dimensionless time $T=\omega \; t$.
\begin{figure}[htbp]
\begin{minipage}[c]{\textwidth}
 \centering
   \subfigure{\includegraphics[width=8cm]{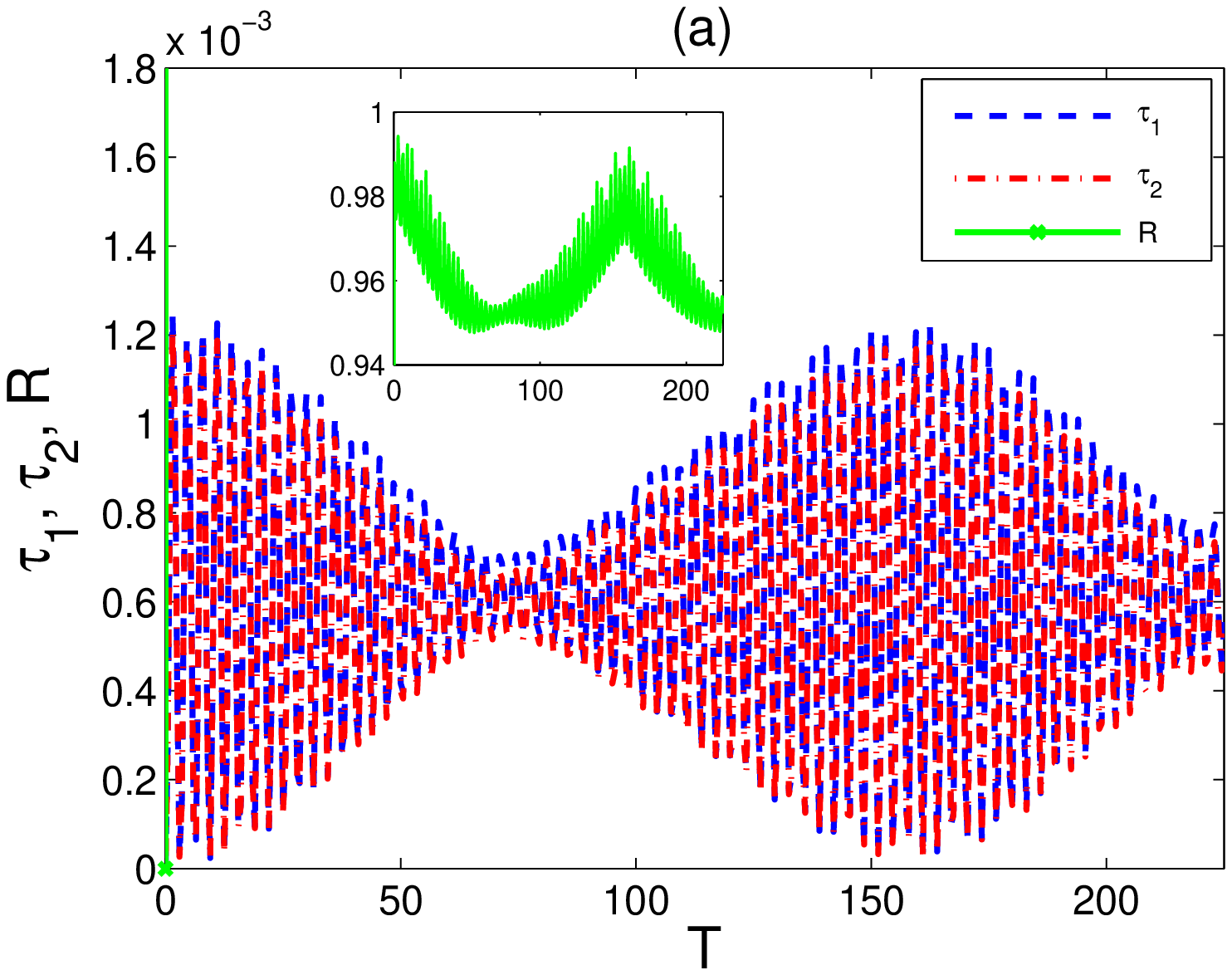}}\quad
   \subfigure{\includegraphics[width=8cm]{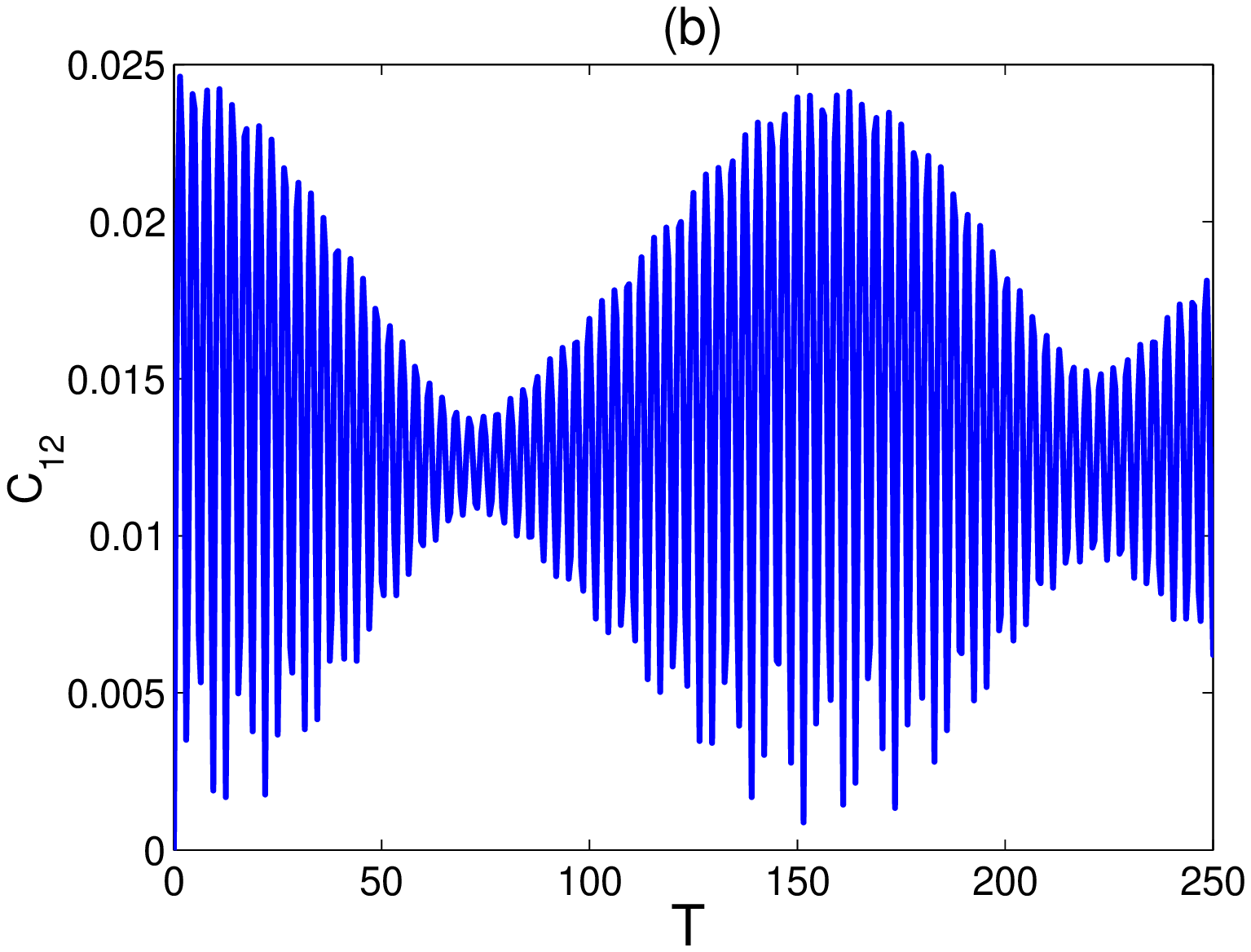}}\\
   \subfigure{\includegraphics[width=8cm]{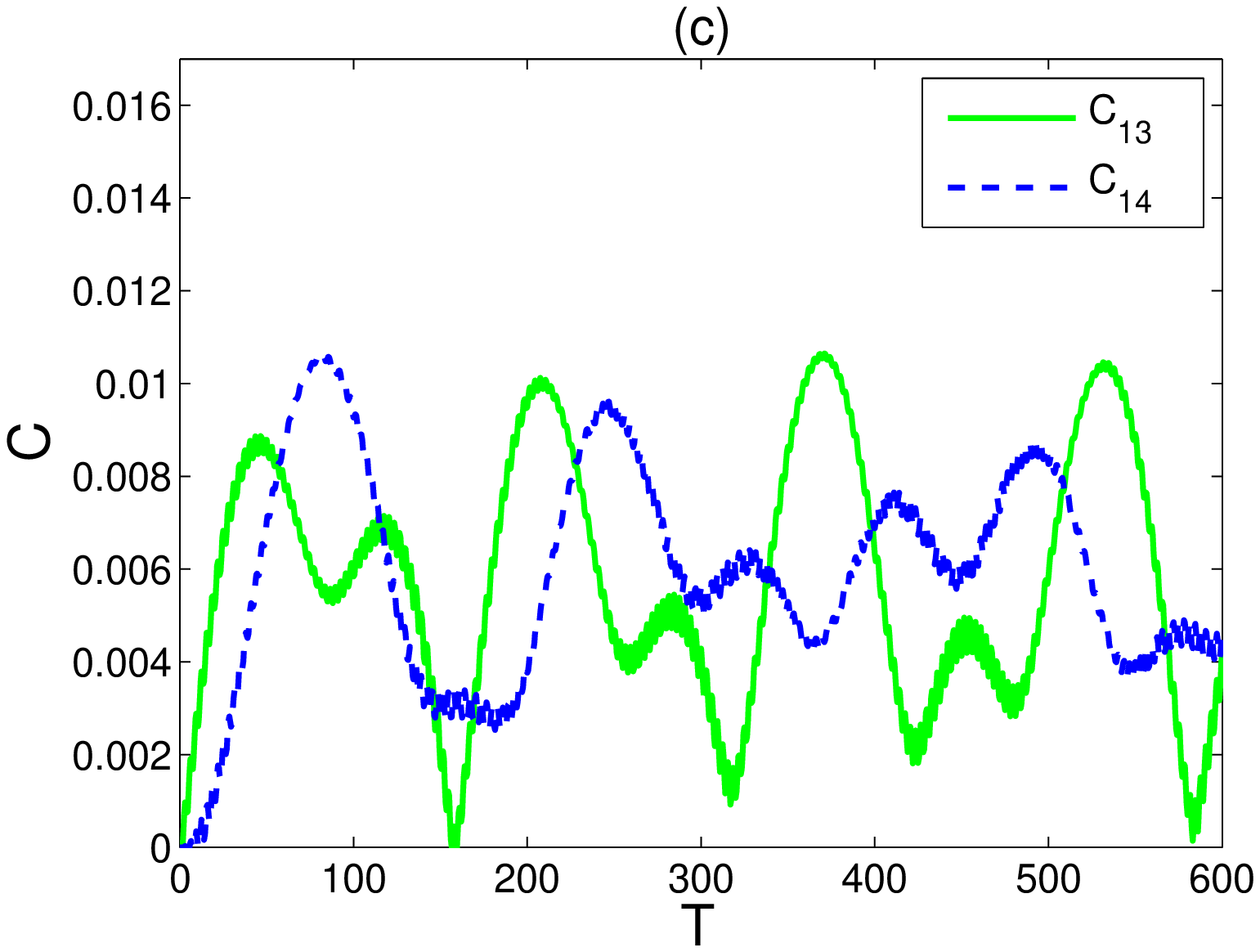}}\quad
  \caption{{\label{N7_close_Ising_dis_G0} Time evolution of (a) $\tau_1$, $\tau_2$ and $R$; (b) $C_{12}$; (c) $C_{13}$ and $C_{14}$ in the free ($\Gamma=0$) Ising system starting from an initially disentangled state, where $N=7$.}}
 \end{minipage}
\end{figure}
\begin{figure}[htbp]
\begin{minipage}[c]{\textwidth}
 \centering
   \subfigure{\includegraphics[width=8cm]{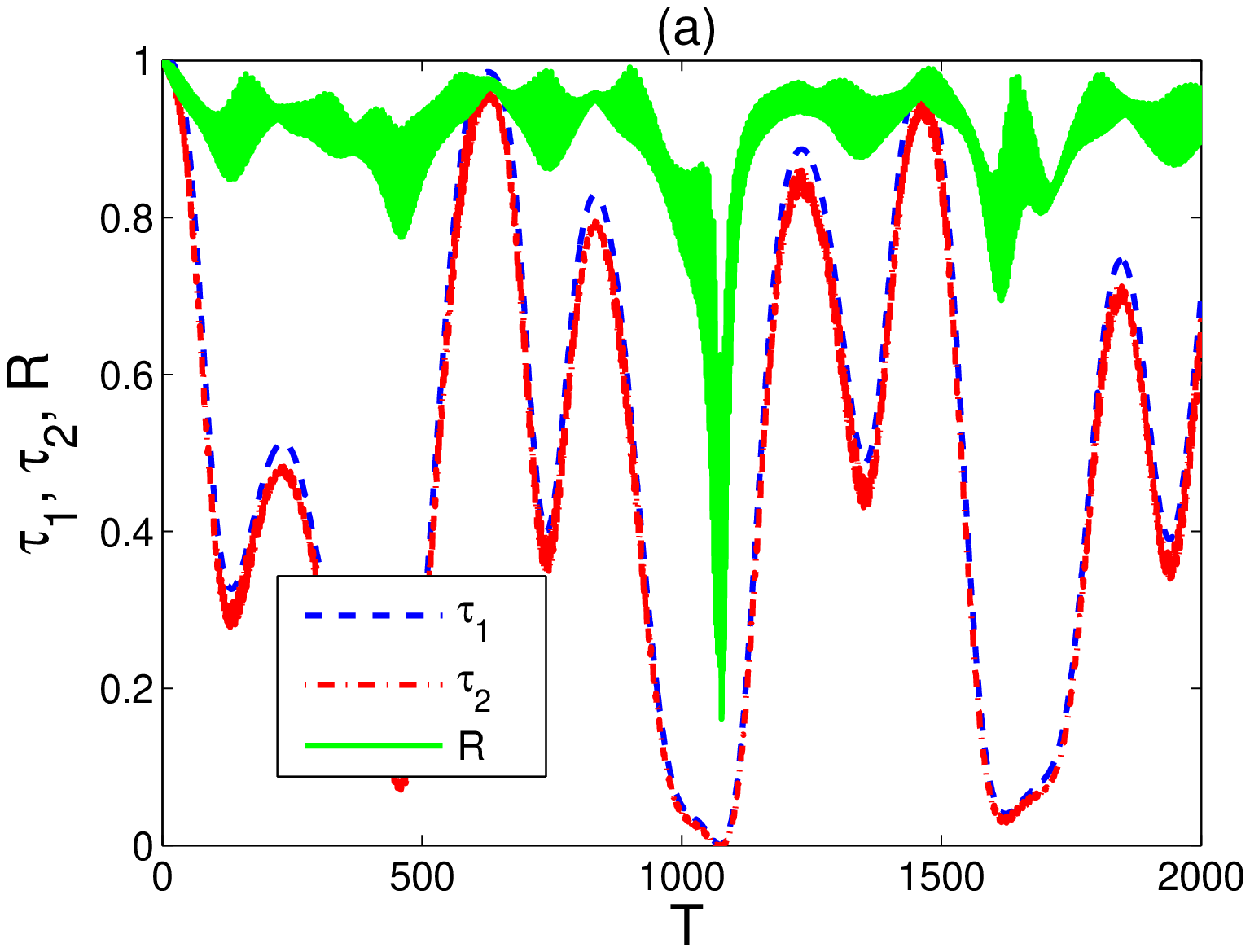}}\quad
   \subfigure{\includegraphics[width=8cm]{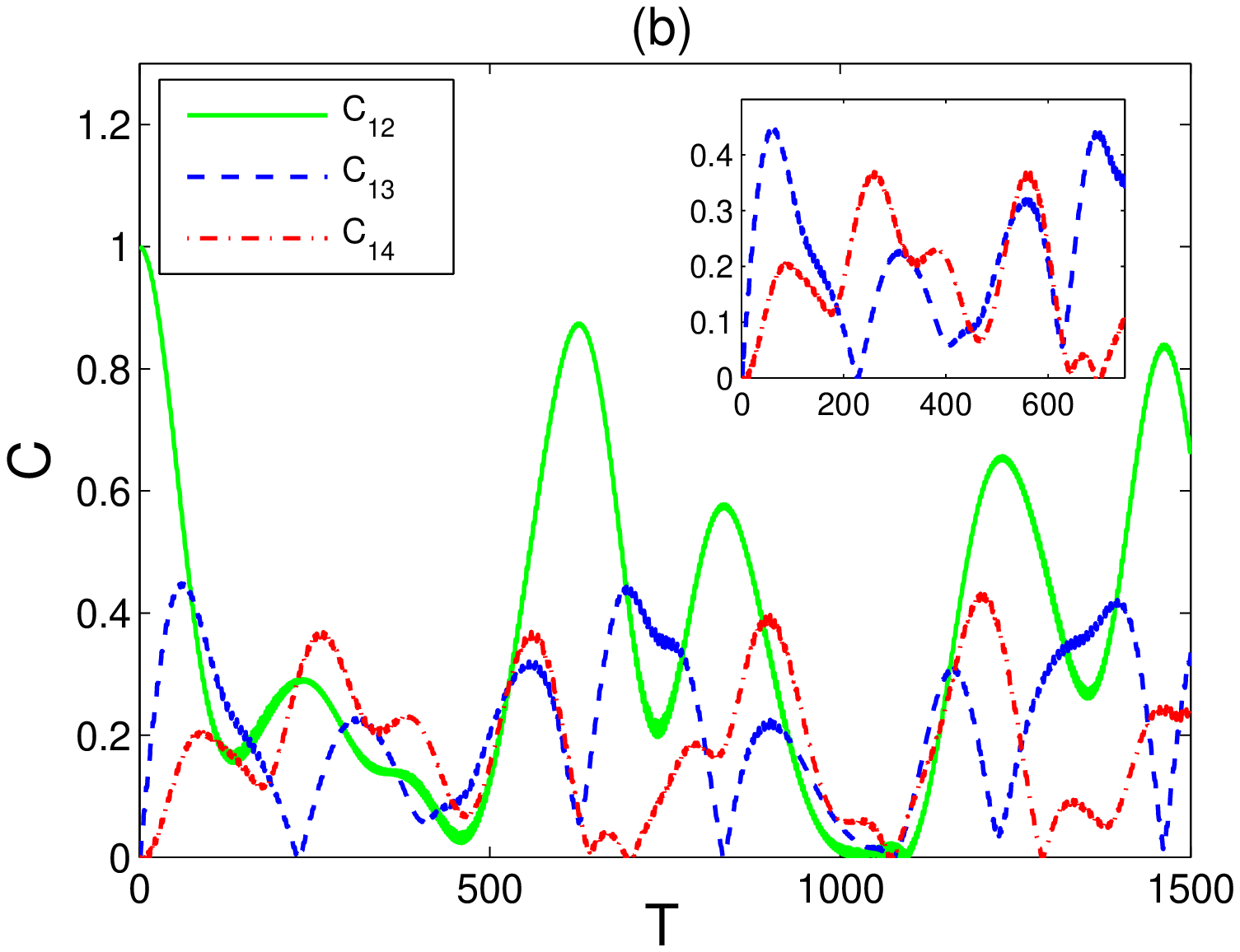}}\\
   \subfigure{\includegraphics[width=8cm]{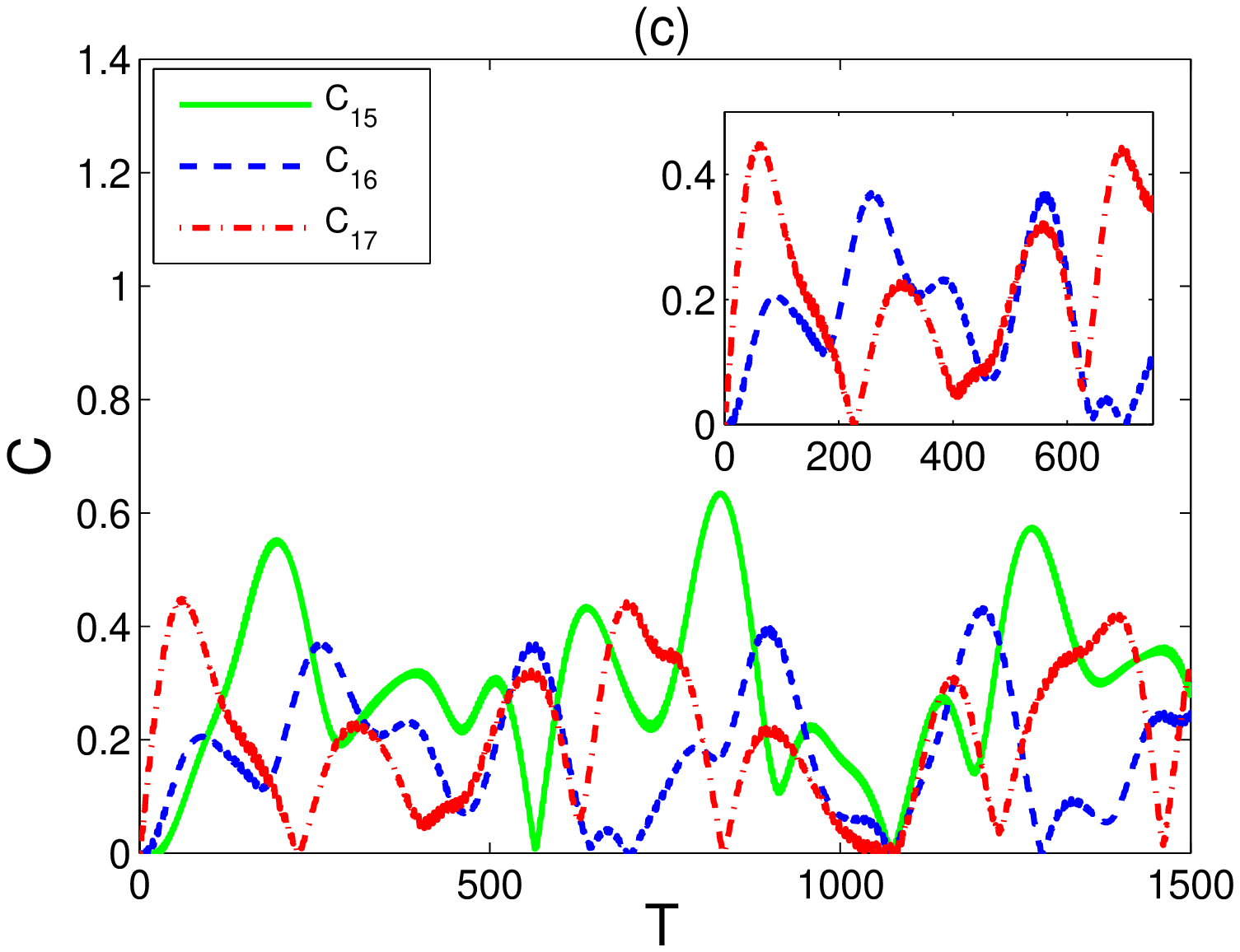}}\quad
  \caption{{\label{N7_close_Ising_max_G0} Time evolution of (a) $\tau_1$, $\tau_2$ and $R$; (b) $C_{12}$, $C_{13}$ and $C_{14}$; (c) $C_{15}$, $C_{16}$ and $C_{17}$ in the free ($\Gamma=0$) Ising system starting from an initial maximally entangled state, where $N=7$.}}
 \end{minipage}
\end{figure}
\begin{figure}[htbp]
\begin{minipage}[c]{\textwidth}
 \centering
   \subfigure{\includegraphics[width=8cm]{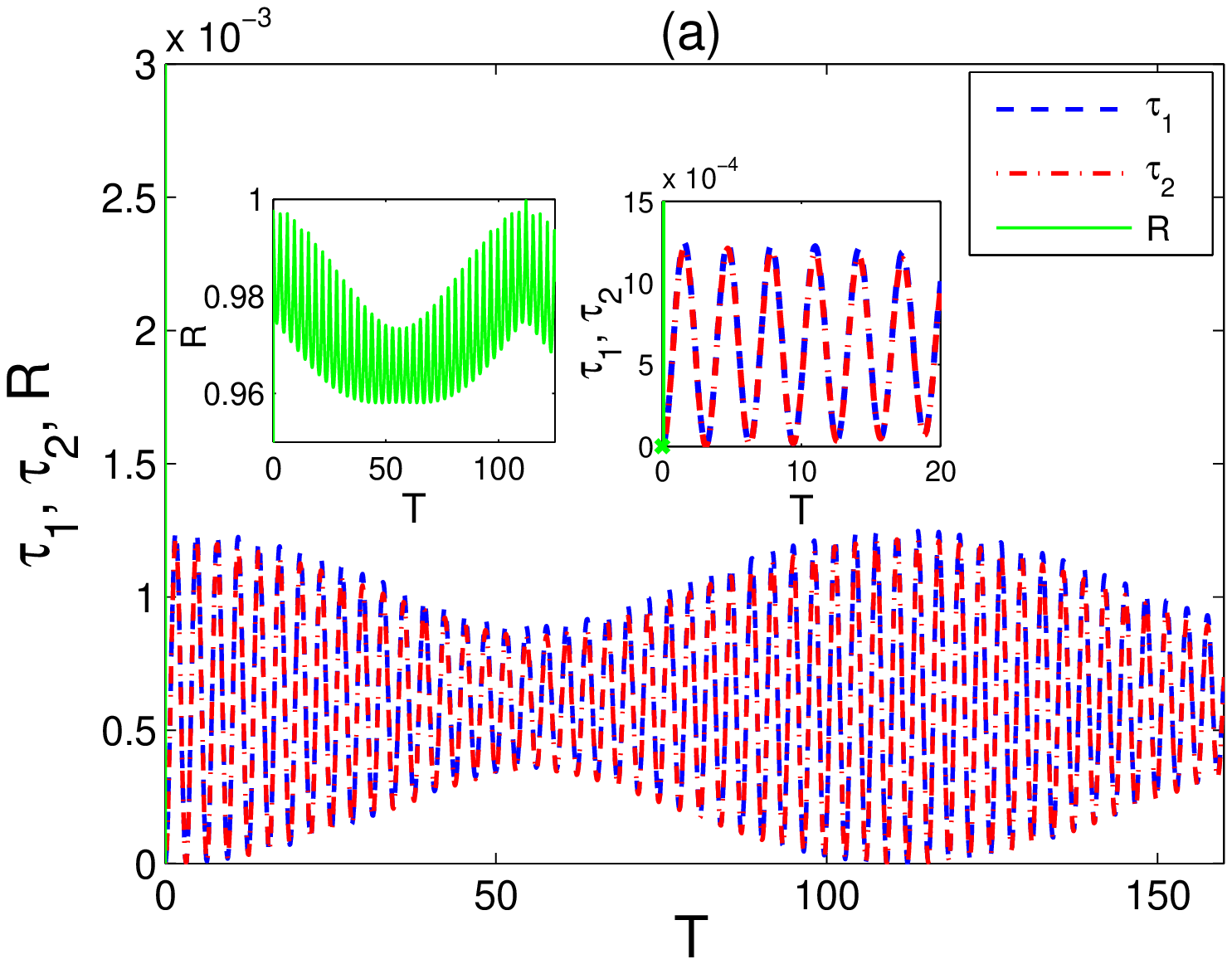}}\quad
   \subfigure{\includegraphics[width=8cm]{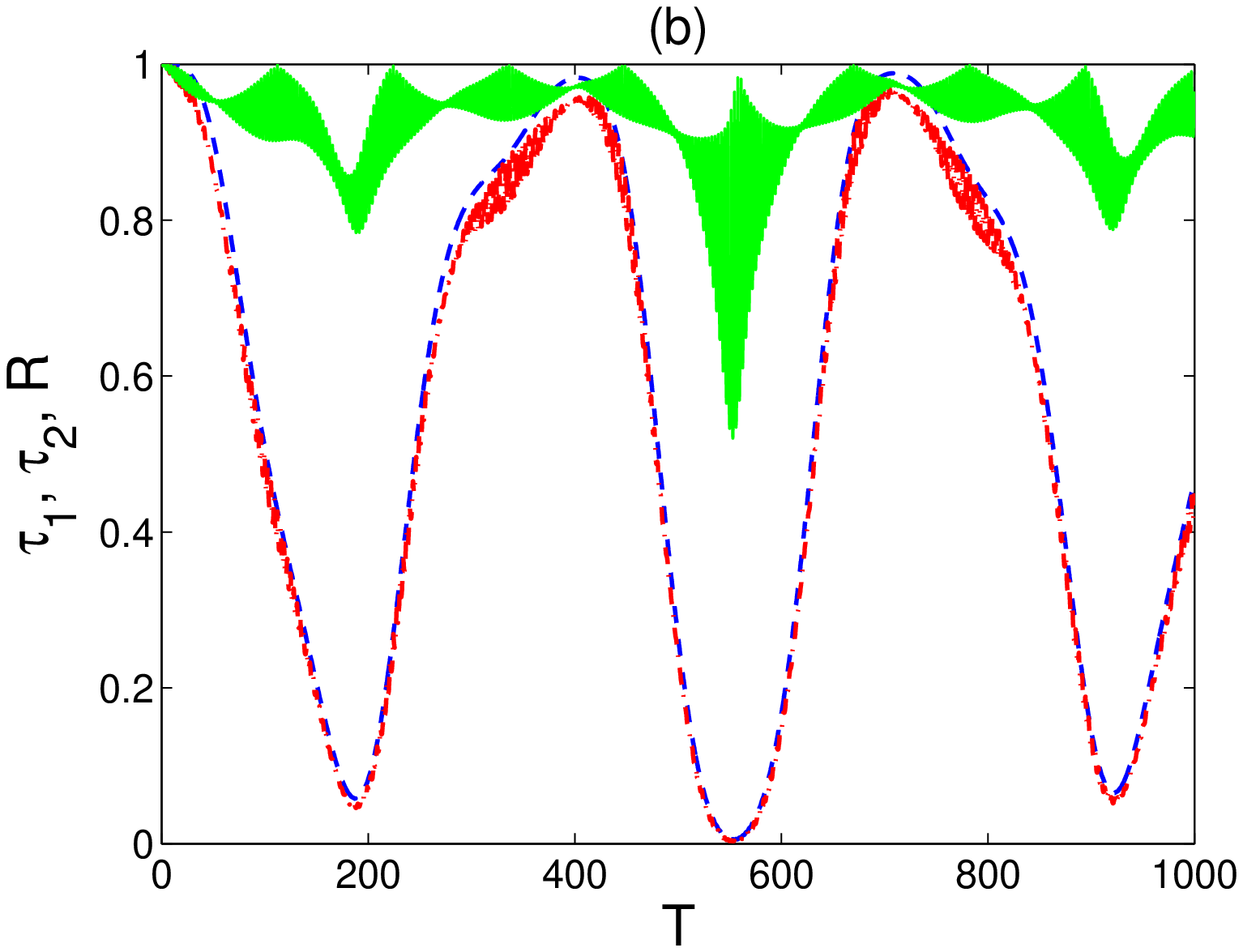}}\\
   \subfigure{\includegraphics[width=8cm]{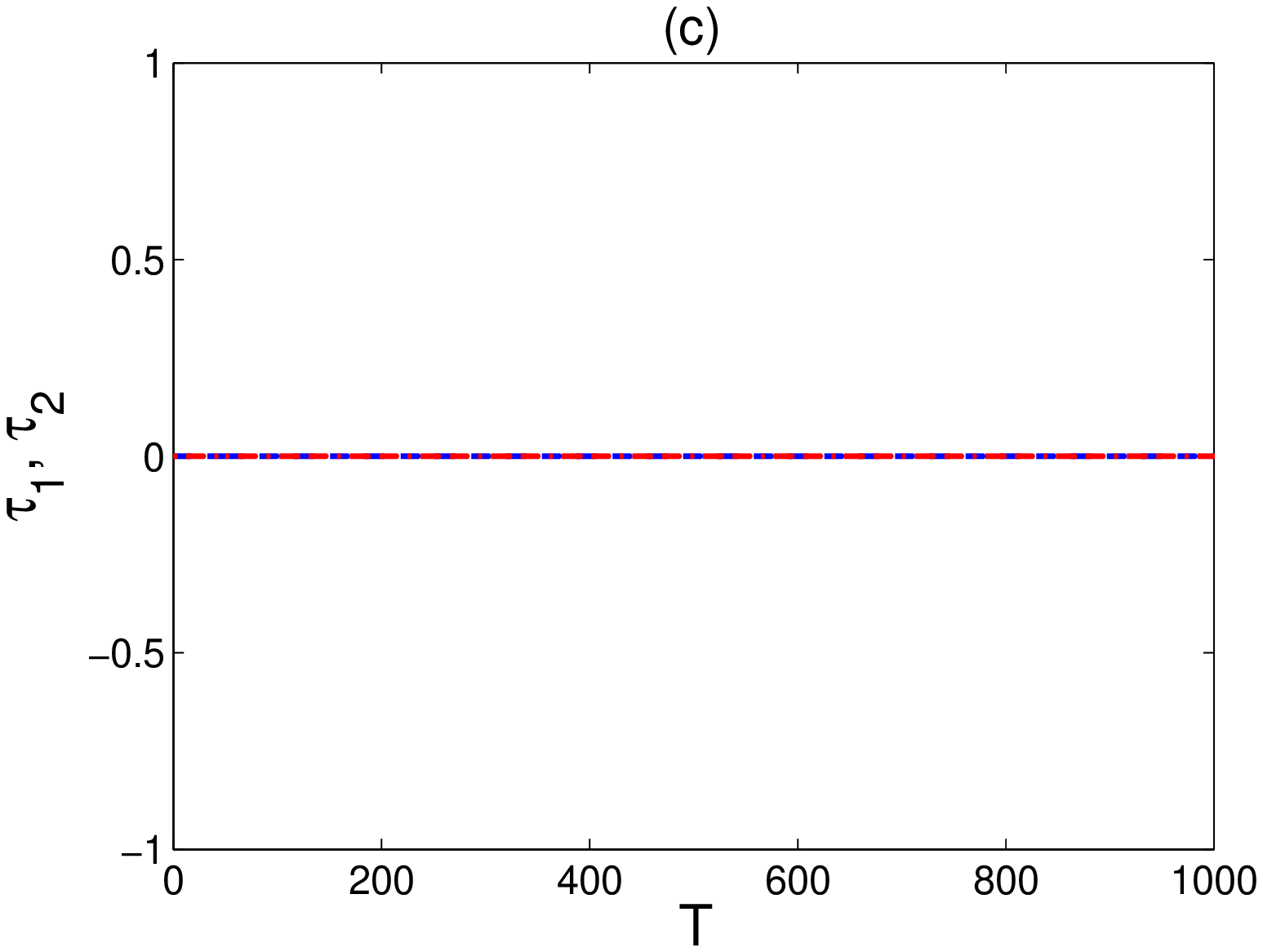}}\quad
   \subfigure{\includegraphics[width=8cm]{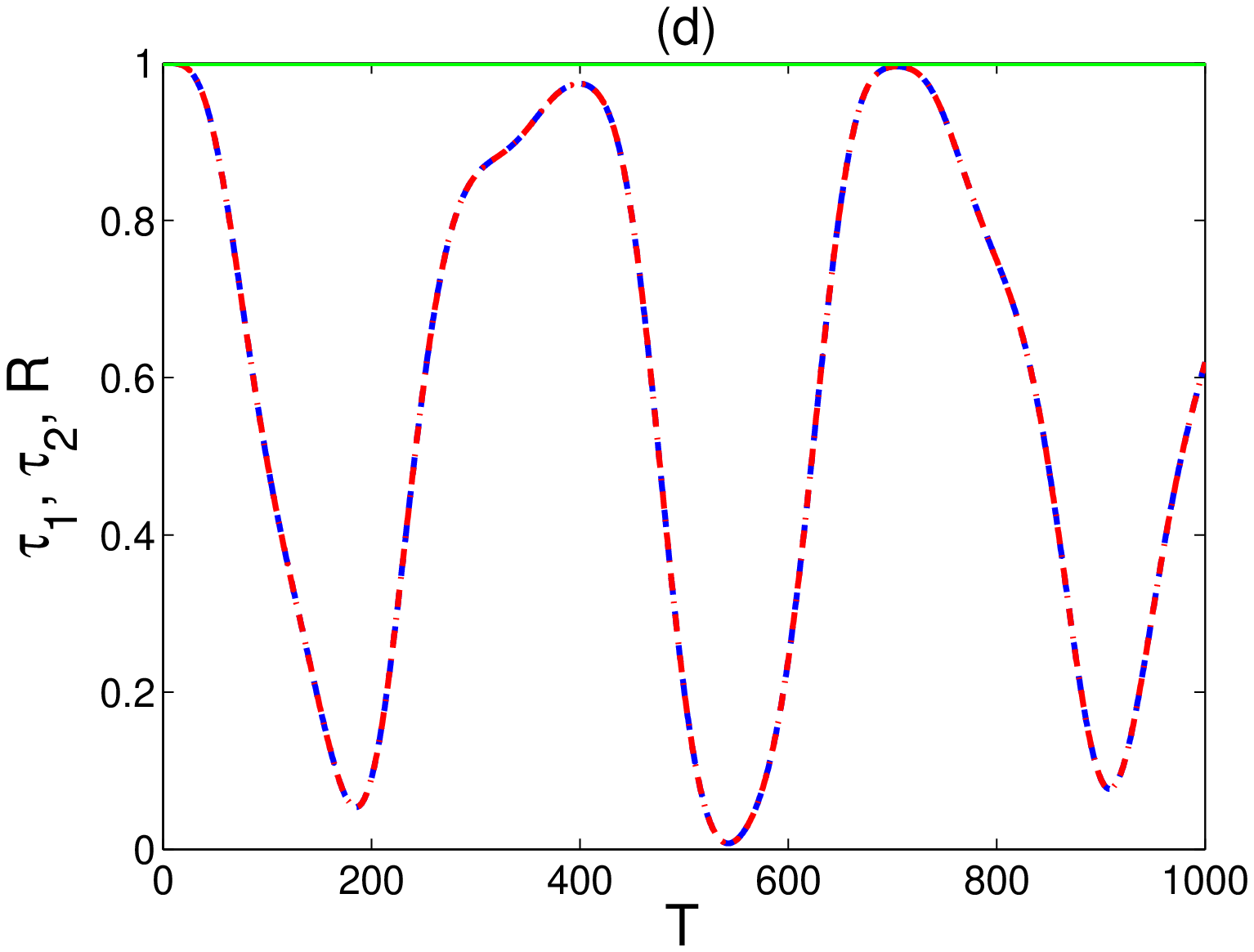}}\\
  \caption{{\label{N5_close_Ising_XX_G0} Time evolution of the free Ising system ($\Gamma=0$) starting from an initial (a) disentangled state; (b) maximally entangled state; and the free XX system starting from an initial (c) disentangled state; (d) maximally entangled state, where $N=5$. The legend is as shown in subfig. (a).}}
 \end{minipage}
\end{figure}
In fig.~\ref{N7_close_Ising_dis_G0}, we depict the time evolution of the entanglement in the closed boundary 7-spins free Ising system starting from a completely disentangled (separable) state. The one tangle $\tau_1$ and the overall bipartite entanglement $\tau_2$ between spin 1 and the rest of the chain are illustrated in fig.~\ref{N7_close_Ising_dis_G0}(a), where they show beat-like oscillatory behavior with very close magnitudes compared to each other and this is why their ratio $R=\tau_2/\tau_1$ is limited between about $0.95$ and $0.99$ as can be seen in the inner panel of the figure. The time evolution of the bipartite entanglement $C_{12}$ is very similar to that of $\tau_1$ and $\tau_2$ but with a bigger amplitude as expected as shown in fig.~\ref{N7_close_Ising_dis_G0}(b). The time evolution of the longer range entanglements $C_{13}$ and $C_{14}$ are illustrated in fig.~\ref{N7_close_Ising_dis_G0}(c), where they show a simple nonuniform oscillatory behavior with about half the amplitude of $C_{12}$. The entanglements $C_{15}$, $C_{16}$ and $C_{17}$ where found to show the same exact behavior of $C_{14}$, $C_{13}$ and $C_{12}$ respectively as expected in a closed boundary chain.
The closed boundary free Ising chain starting from a maximally entangled state is considered in fig.~\ref{N7_close_Ising_max_G0}, where it shows a different behavior from the previous case. The entanglement functions $\tau_1$ and $\tau_2$ show sustainable nonuniform oscillatory behavior, with no beating, that is very close for the two except when their magnitudes decrease significantly and their ratio $R$ changes over wider range between about $0.2$ and $1$ as shown in fig.~\ref{N7_close_Ising_max_G0}(a).
In fig.~\ref{N7_close_Ising_max_G0}(b) and (c) we plot the bipartite entanglements $C_{12}$, $C_{13}$, $C_{14}$, $C_{15}$, $C_{16}$ and $C_{17}$. They all show nonuniform oscillatory behavior, where interestingly the (nnn) entanglement $C_{13}$ profile looks exactly like that of $C_{17}$ but not like $C_{16}$ as one may have expected for a closed boundary chain. The same applies to $C_{14}$ which exactly the same as $C_{16}$ (not $C_{15}$ as one would expect). This means that the maximum entanglement that was initially created between spins 1 and 2 is propagating through the chain in both direction starting from spins 1 and 2 as a single source. Comparing the results in figs.~\ref{N7_close_Ising_dis_G0} and \ref{N7_close_Ising_max_G0}, one can notice how the initial state causes a great deal of difference on the entanglement dynamics through the entire spin chain. Starting from a maximally entangled state leads to much higher amplitude of entanglement oscillation among all spins and much smaller frequency. 
In fig.~\ref{N5_close_Ising_XX_G0}(a) and (b) we consider the Ising chain again but with only 5 spins to examine the size effect, where we focus on the time evolution of $\tau_1$, $\tau_2$ and $R$. The oscillation of the system entanglement starting from a disentangled state is losing much of its beat-like character although the amplitude is almost the same as for $N=7$ and the ration $R$ is closer to $1$ with narrower range as shown in fig.~\ref{N5_close_Ising_XX_G0}(a). The time evolution of the same system starting from an initial maximally entangled state is illustrated in fig.~\ref{N5_close_Ising_XX_G0}(b). As one can see, the oscillation of the entanglements $\tau_1$ and $\tau_2$ become more uniform compared with the $N=7$ case and also the range of $R$ is narrower.  
\begin{figure}[htbp]
\begin{minipage}[c]{\textwidth}
 \centering
   \subfigure{\includegraphics[width=8cm]{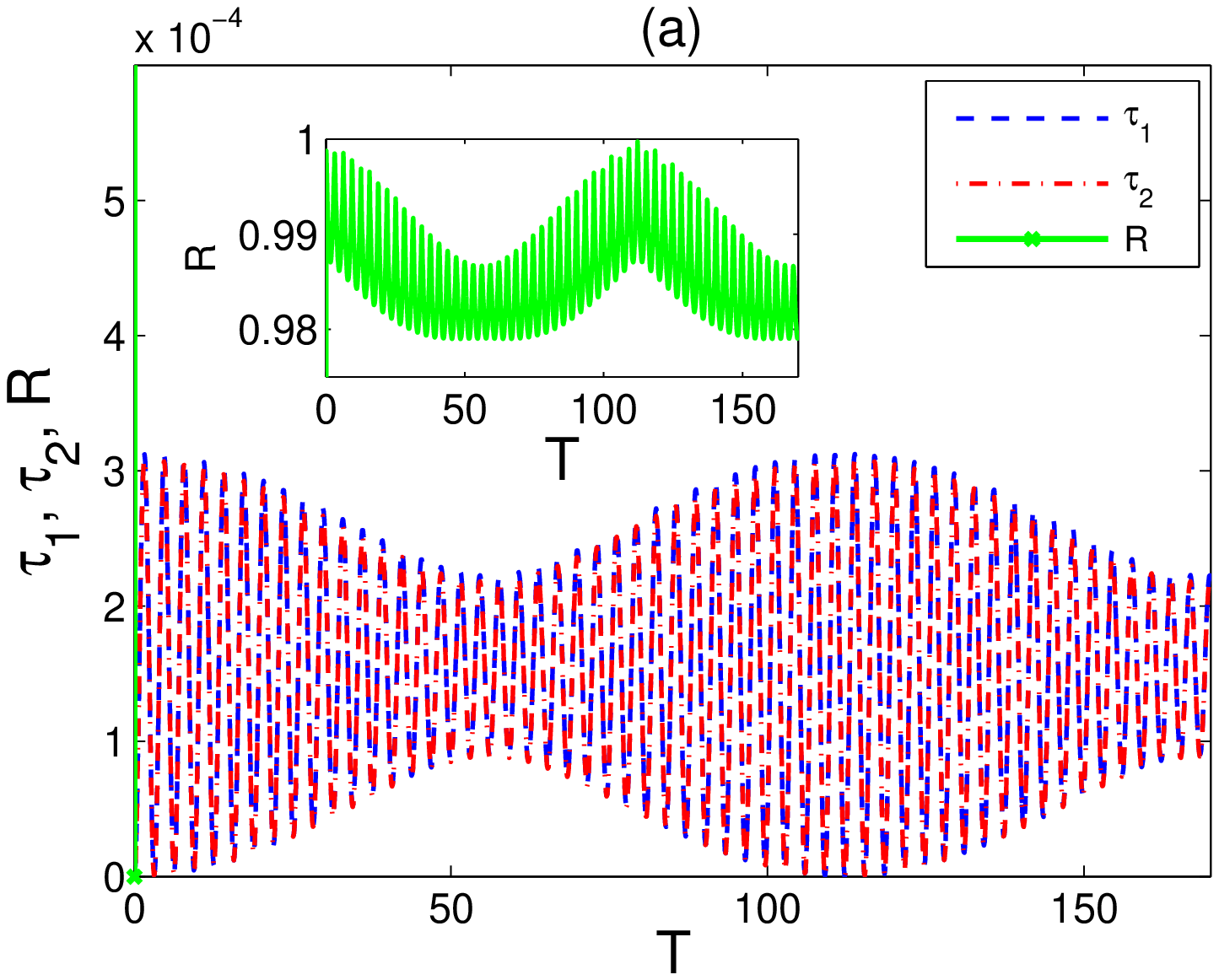}}\quad
   \subfigure{\includegraphics[width=8cm]{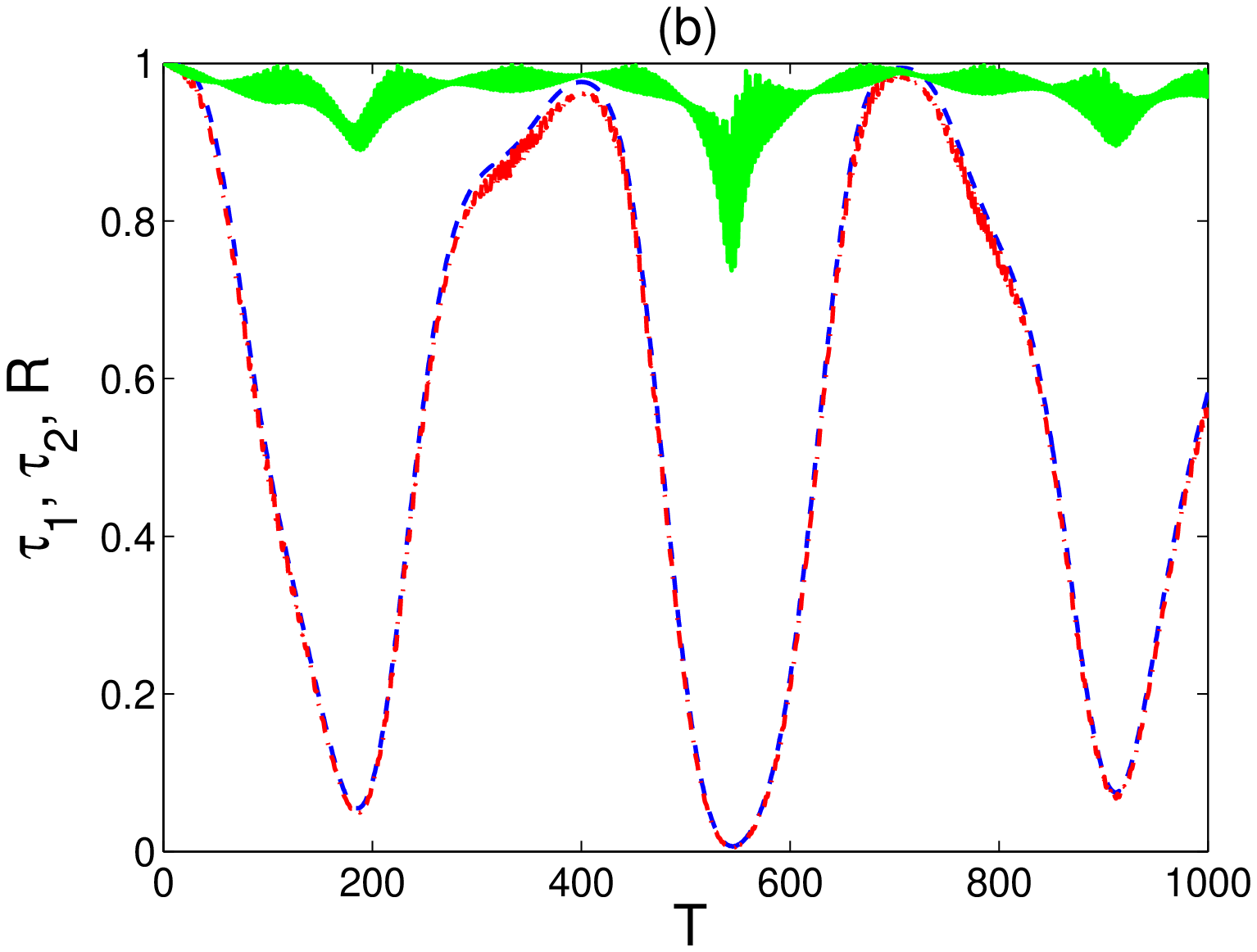}}\\
   \subfigure{\includegraphics[width=8cm]{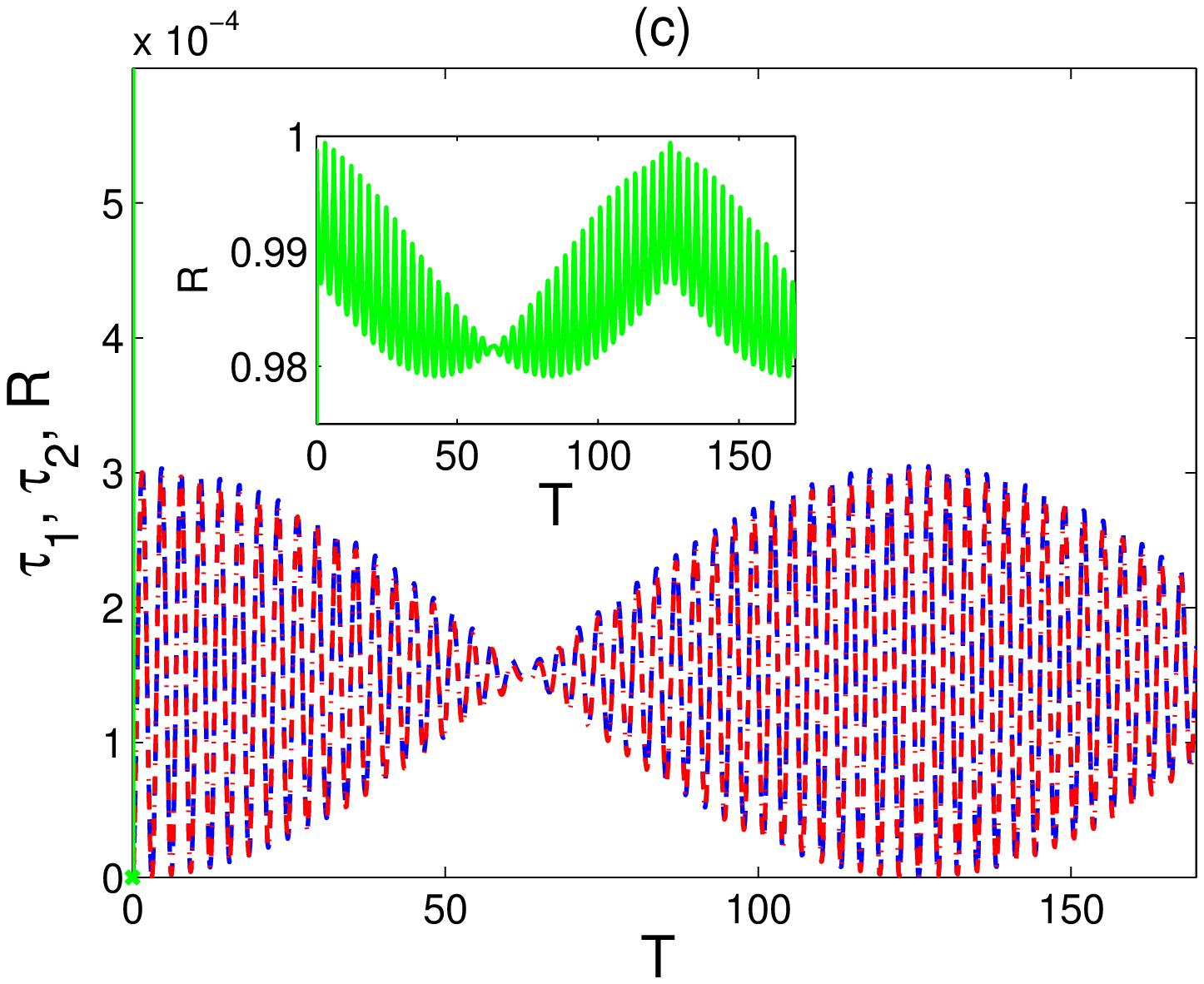}}\quad
   \subfigure{\includegraphics[width=8cm]{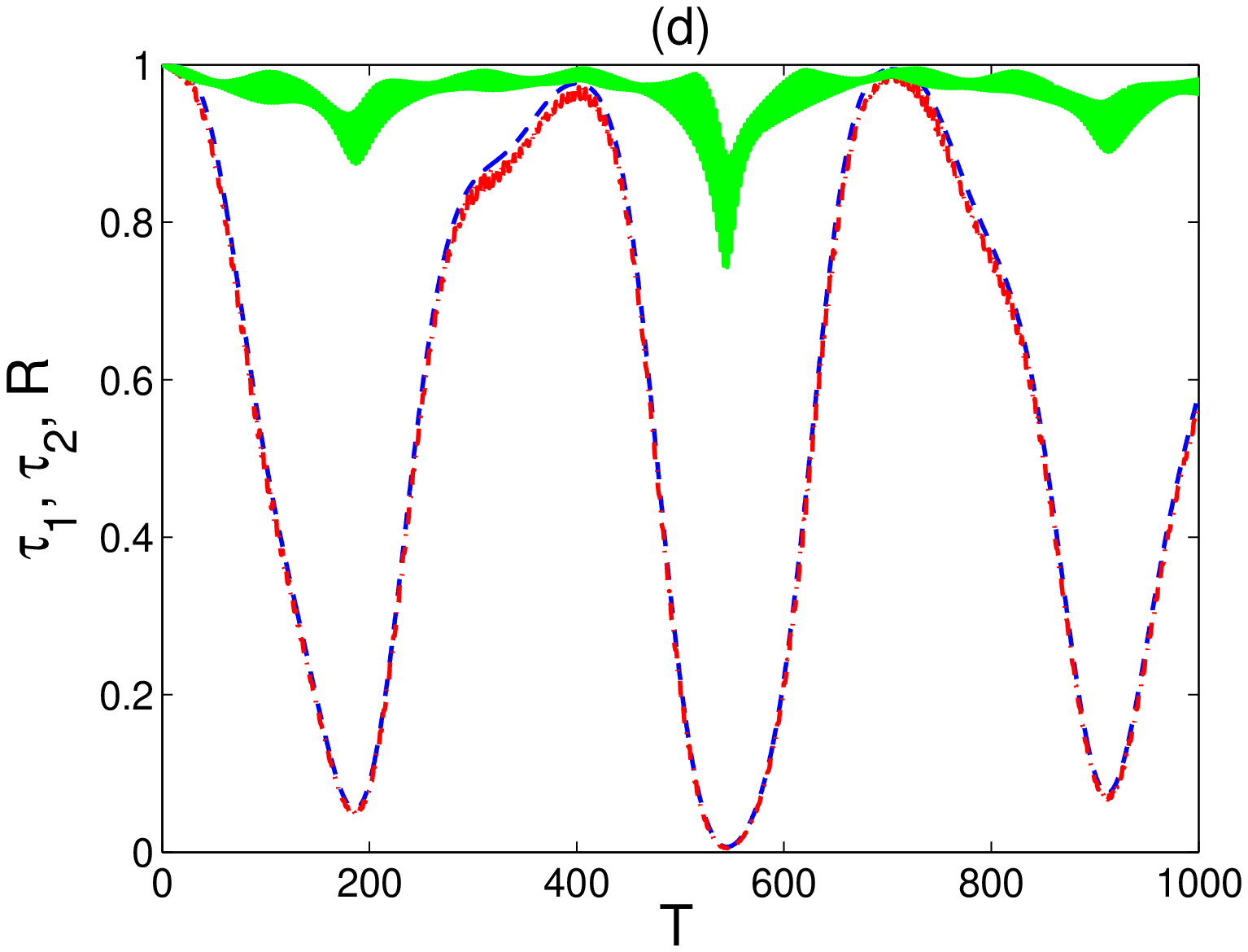}}\\
	\caption{{\label{N5_close_XY_XYZ_G0} Time evolution of the free XY system ($\Gamma=0$) starting from an initial (a) disentangled state; (b) maximally entangled state; and the free XYZ system starting from an initial (c) disentangled state; (d) maximally entangled state, where $N=5$. The legend is as shown in subfig. (a).}}
\end{minipage}
\end{figure}
In fig.~\ref{N5_close_Ising_XX_G0}(c) and (d), we test the effect of removing the anisotropy (between $X$ and $Y$) by considering the $XX$ system. The initial state of the system is significantly affecting the system dynamics where the initial separable state, as shown in fig.~\ref{N5_close_Ising_XX_G0}(c), causes the system to stay separable forever whereas the initial maximum entangled state, depicted in fig.~\ref{N5_close_Ising_XX_G0}(d), leads to an oscillation, similar to what we have seen in fig.~\ref{N5_close_Ising_XX_G0}(b) but with perfect coincidence between $\tau_1$ and $\tau_2$.
The behavior of the partial anisotropic system, $XY$, is illustrated in fig.~\ref{N5_close_XY_XYZ_G0}(a) and (b), where it looks very similar to the Ising case but with a smaller range of variation of the ratio R. In fig.~\ref{N5_close_XY_XYZ_G0}(c) and (d) we test the effect of anisotropy not only in the $X$ and $Y$-directions but also in $Z$-direction by considering the $XYZ$ system. It is clear that adding an interaction in the $z$-direction is not changing the behavior of the system significantly compared with the $XY$ model. The main change is the appearance of a node in the envelope of the oscillation in the initial separable state case. 
\subsection{Coupling to a thermal dissipative environment}
\begin{figure}[htbp]
\begin{minipage}[c]{\textwidth}
 \centering
   \subfigure{\includegraphics[width=8cm]{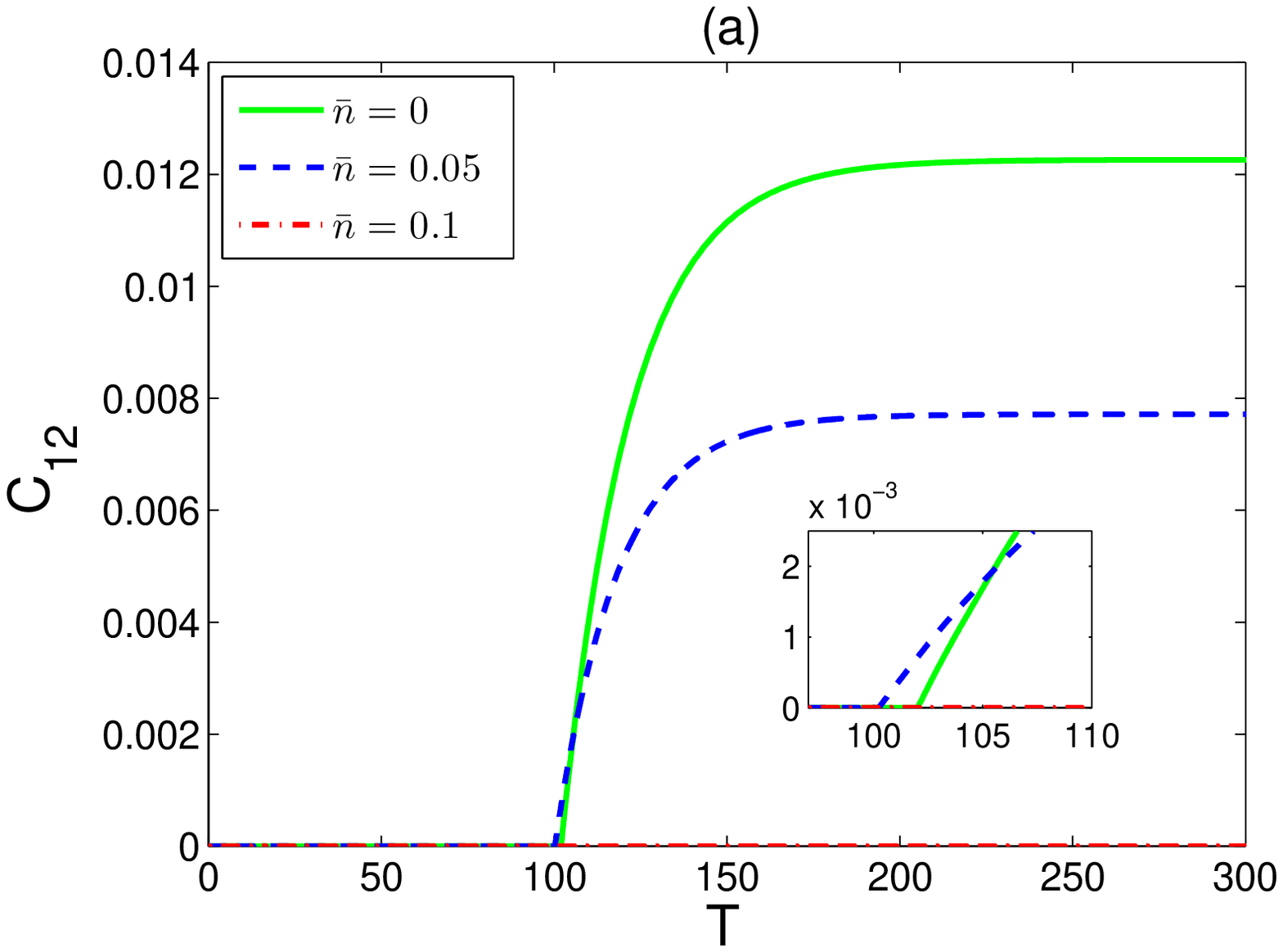}}\quad
   \subfigure{\includegraphics[width=8cm]{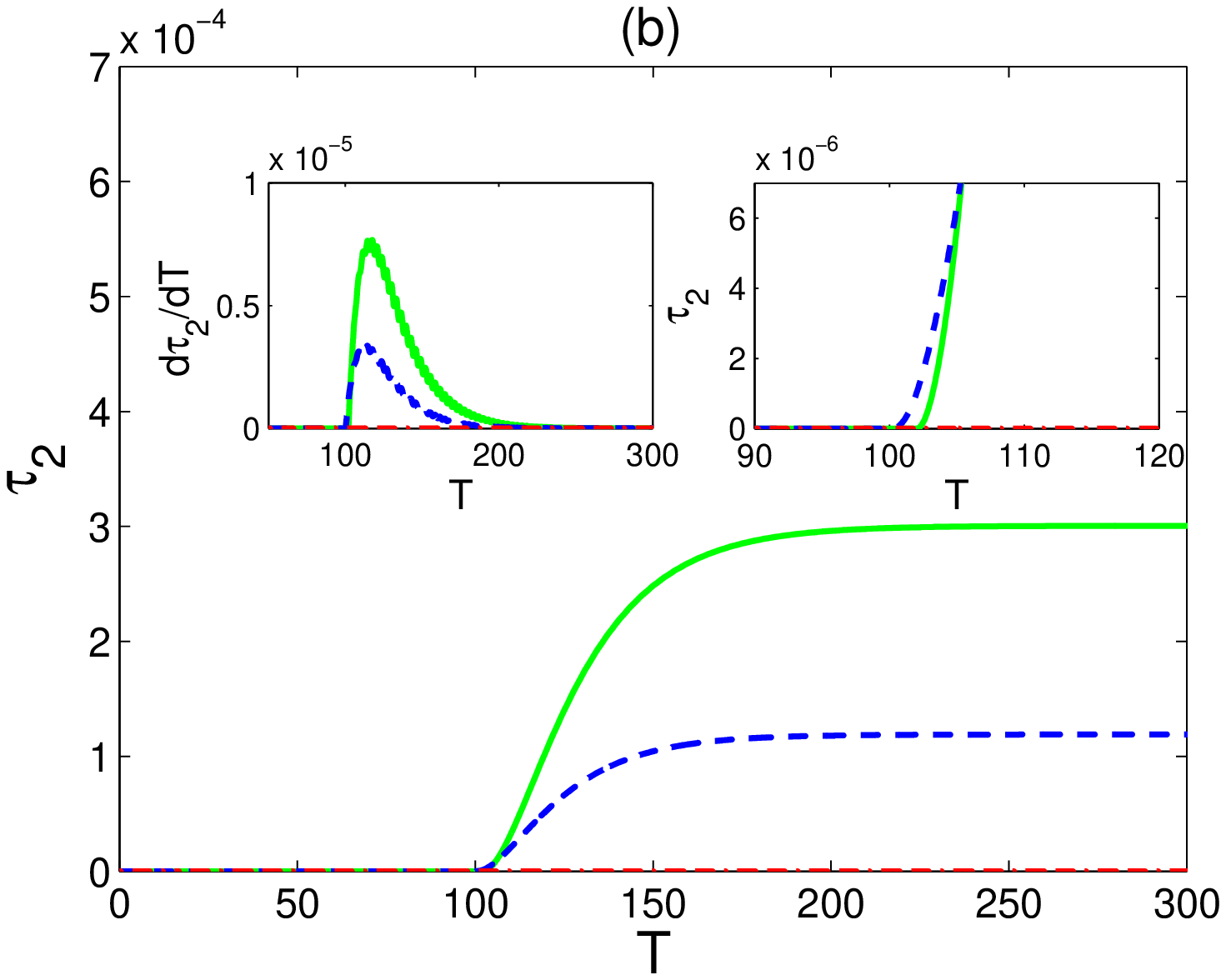}}\\
   \subfigure{\includegraphics[width=8cm]{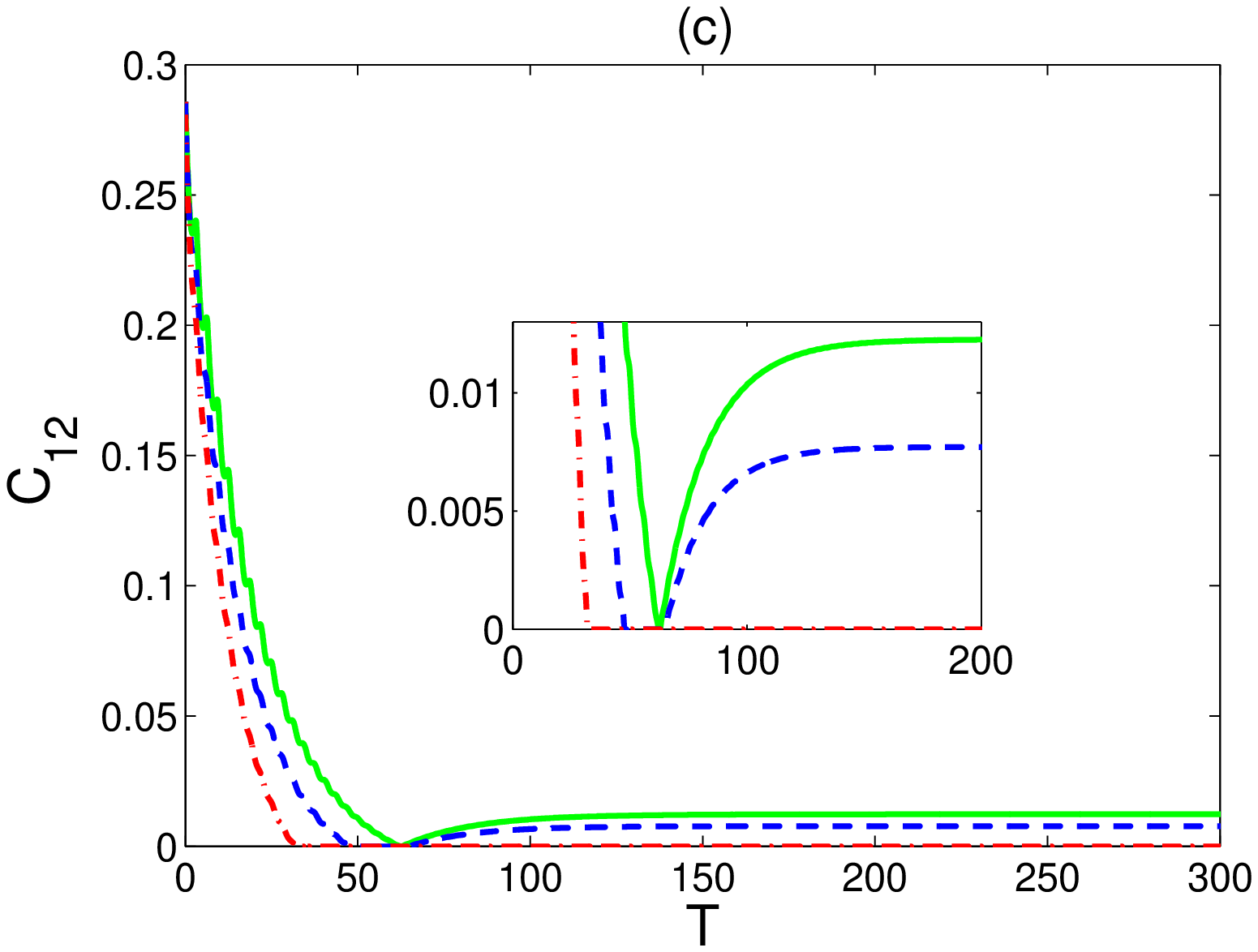}}\quad
   \subfigure{\includegraphics[width=8cm]{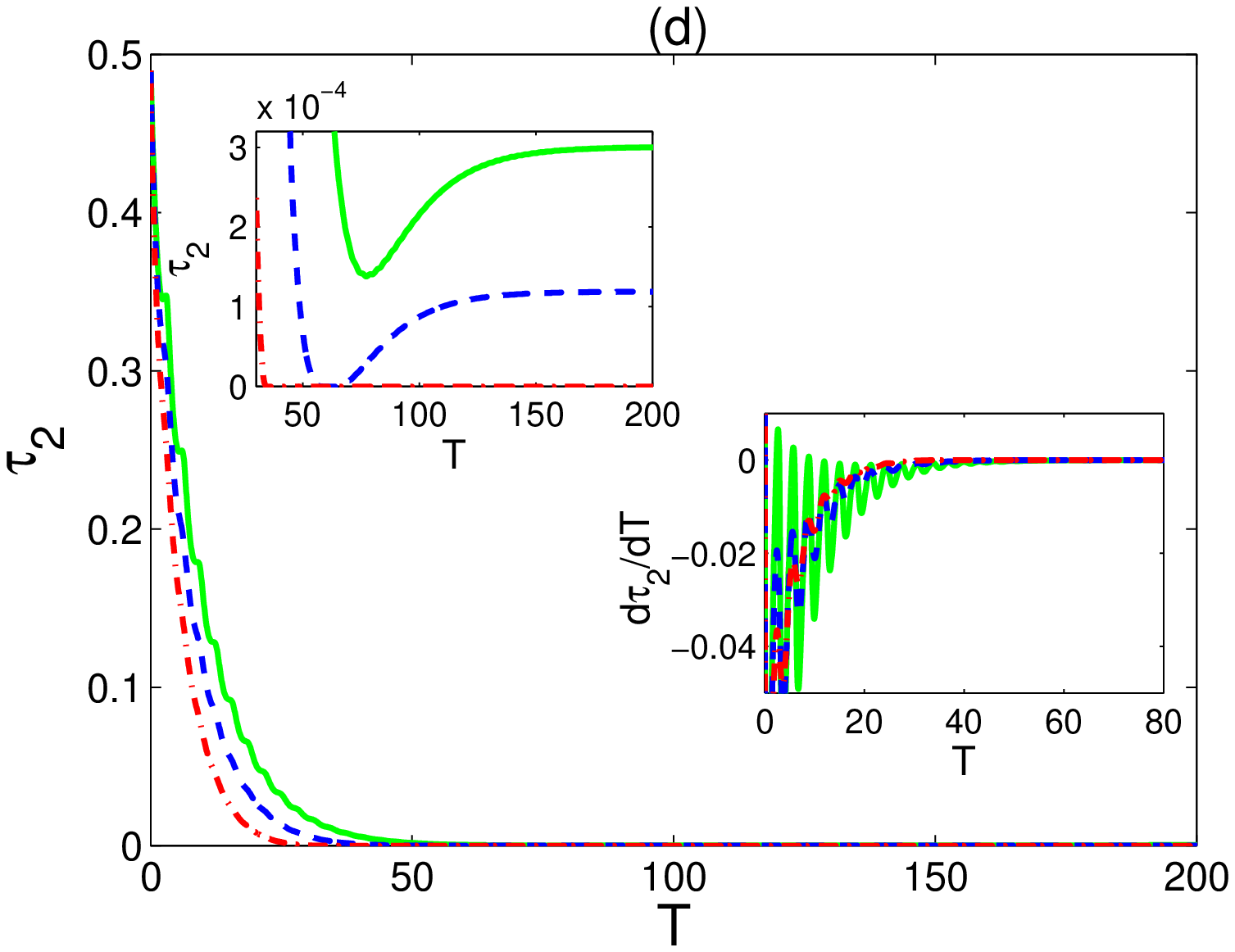}}\\
  \caption{{\label{N7_close_Ising_dis_w_G05} Time evolution of $C_{12}$ and $\tau_2$ in the Ising system in presence of the environment ($\Gamma=0.05$) starting from an initial disentangled state in (a) and (b) and an entangled W-state in (c) and (d) respectively at different temperatures $\bar{n}=0,\;0.05$ and $0.1$, where $N=7$. The legend is as shown in subfig. (a).}}
\end{minipage}
\end{figure}
In this section we study the dynamics of entanglement in different closed boundary Heisenberg spin chains, with different degrees of spatial anisotropy, coupled to a Lindblad environment at zero and finite temperatures, starting from different initial states. In this paper, we set up the system parameters such that $\omega=1$, $\Gamma=J=0.05 \; \omega$ and the temperature parameter $0 \leq\bar{n} \leq 0.1 (\sim 41 mK)$, unless otherwise stated.
We focus here on the time evolution of the nearest neighbor bipartite entanglement between the two spins 1 and 2 as well as $\tau_2$ between spin 1 and the rest of the chain, which gives a very good insight of how the the overall bipartite entanglement and the beyond nearest neighbor entanglement are behaving. We start with the Ising system, in fig.~\ref{N7_close_Ising_dis_w_G05}(a) and (b), where we show the time evolution of $C_{12}$ and $\tau_2$ respectively starting from an initially separable state. As one can see, both $C_{12}$ and $\tau_2$ start with zero initial value and stay zero for sometime before suddenly rising up and increasing monotonically to reach a steady state value. To ensure that the final state is a sustainable steady state, we plot the first derivative of $\tau_2$ versus time in the inner panel of fig.~\ref{N7_close_Ising_dis_w_G05}(b), which shows a sudden peak at around $T\approx 100$ before decaying to zero $T\approx 240$. It is very clear how devastating is the temperature effect on the steady state value of the entanglement, where having a value of $\bar{n}=0.05$ reduces the steady state value significantly compared with $\bar{n}=0$ whereas $\bar{n}=0.1$ keeps the system disentangled at all times.
\begin{figure}[htbp]
\begin{minipage}[c]{\textwidth}
 \centering
   \subfigure{\includegraphics[width=8cm]{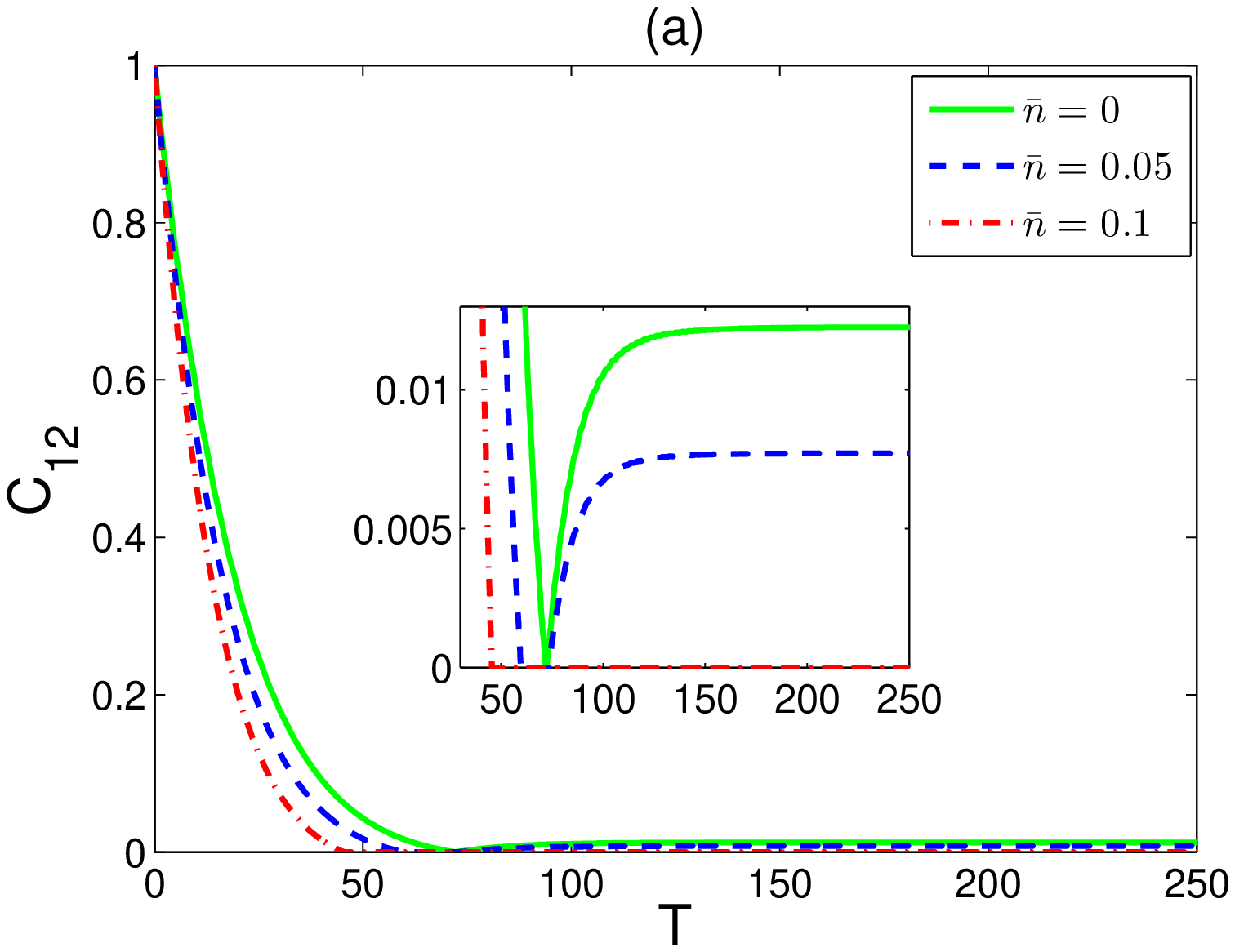}}\quad
   \subfigure{\includegraphics[width=8cm]{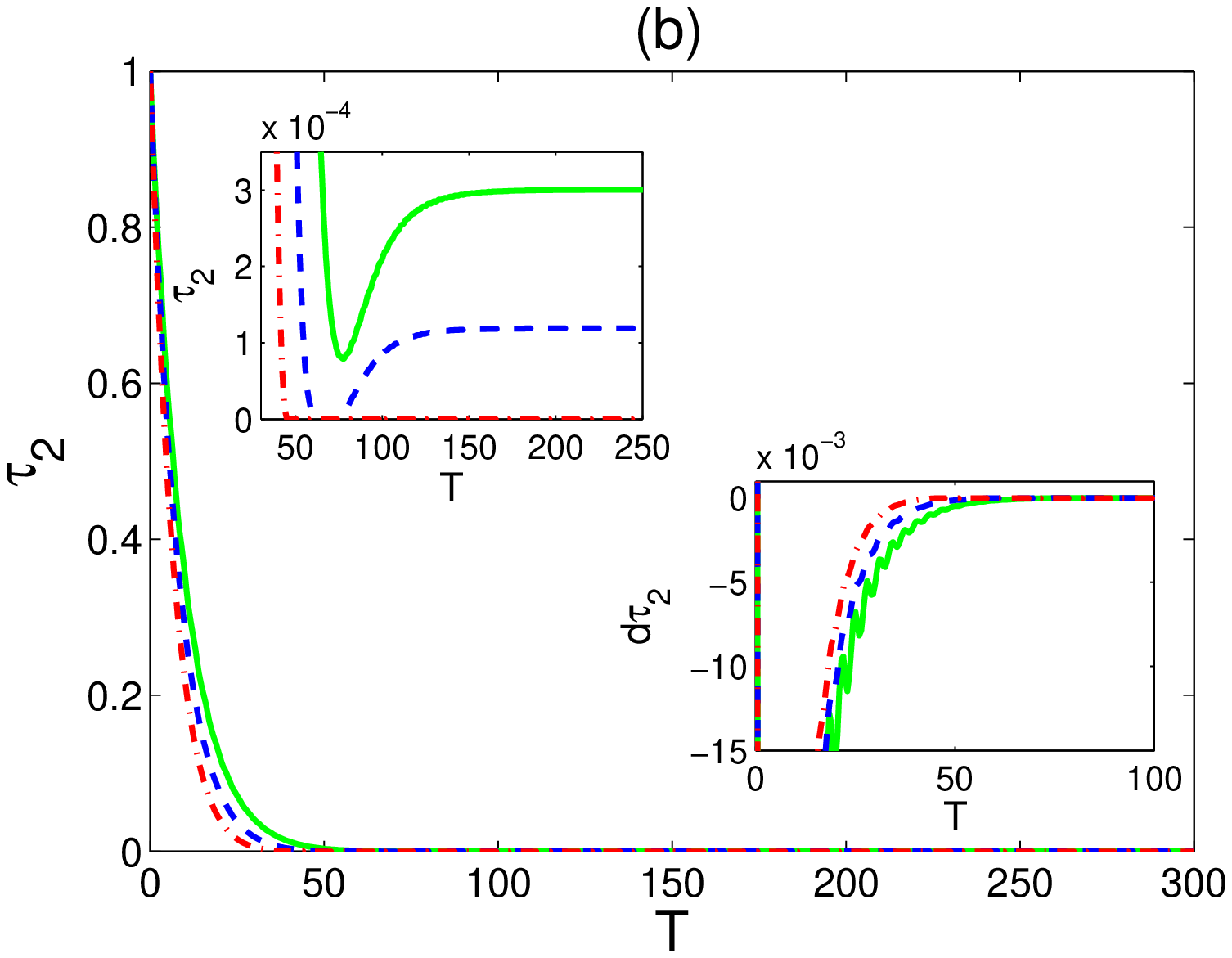}}\\
  \caption{{\label{N7_close_Ising_max_G05} Time evolution of (a) $C_{12}$ and (b)$\tau_2$ in the Ising system in presence of the environment ($\Gamma=0.05$) starting from an initial maximally entangled state at different temperatures $\bar{n}=0,\;0.05$ and $0.1$, where $N=7$. The legend is as shown in subfig. (a).}}
  \end{minipage}
\end{figure}
In fig.~\ref{N7_close_Ising_dis_w_G05}(c) and (d), the system starts from an initial partially entangled state, the w state. As a result the entanglement $C_{12}$ at zero temperature, shown in fig.~\ref{N7_close_Ising_dis_w_G05}(c), starts with an initial non-zero value but decays with time until it vanishes but immediately revives again and increases monotonically reaching a steady state. As the temperature increases, $\bar{n}=0.05$, the entanglement death period increases and the steady state value decreases. For higher temperature, $\bar{n}=0.1$, the entanglement never revive again from its zero value. Interestingly, the behavior of $\tau_2$, as illustrated in fig.~\ref{N7_close_Ising_dis_w_G05}(d), is not exactly the same as $C_{12}$, where at zero temperature $\tau_2$ decays as the system evolves, but never drop to zero, before rising up and reaching a steady state. This indicates that the beyond nearest neighbor entanglement sustains a non-zero value despite that the nearest neighbor entanglement vanishes. The effect of the finite temperature on $\tau_2$ is similar to that on $C_{12}$ as can be concluded from the inner panels. The first derivative of $\tau_2$ shows a rapid oscillation before reaching the zero value which is suppressed as the temperature is raised.
\begin{figure}[htbp]
\begin{minipage}[c]{\textwidth}
 \centering
   \subfigure{\includegraphics[width=8cm]{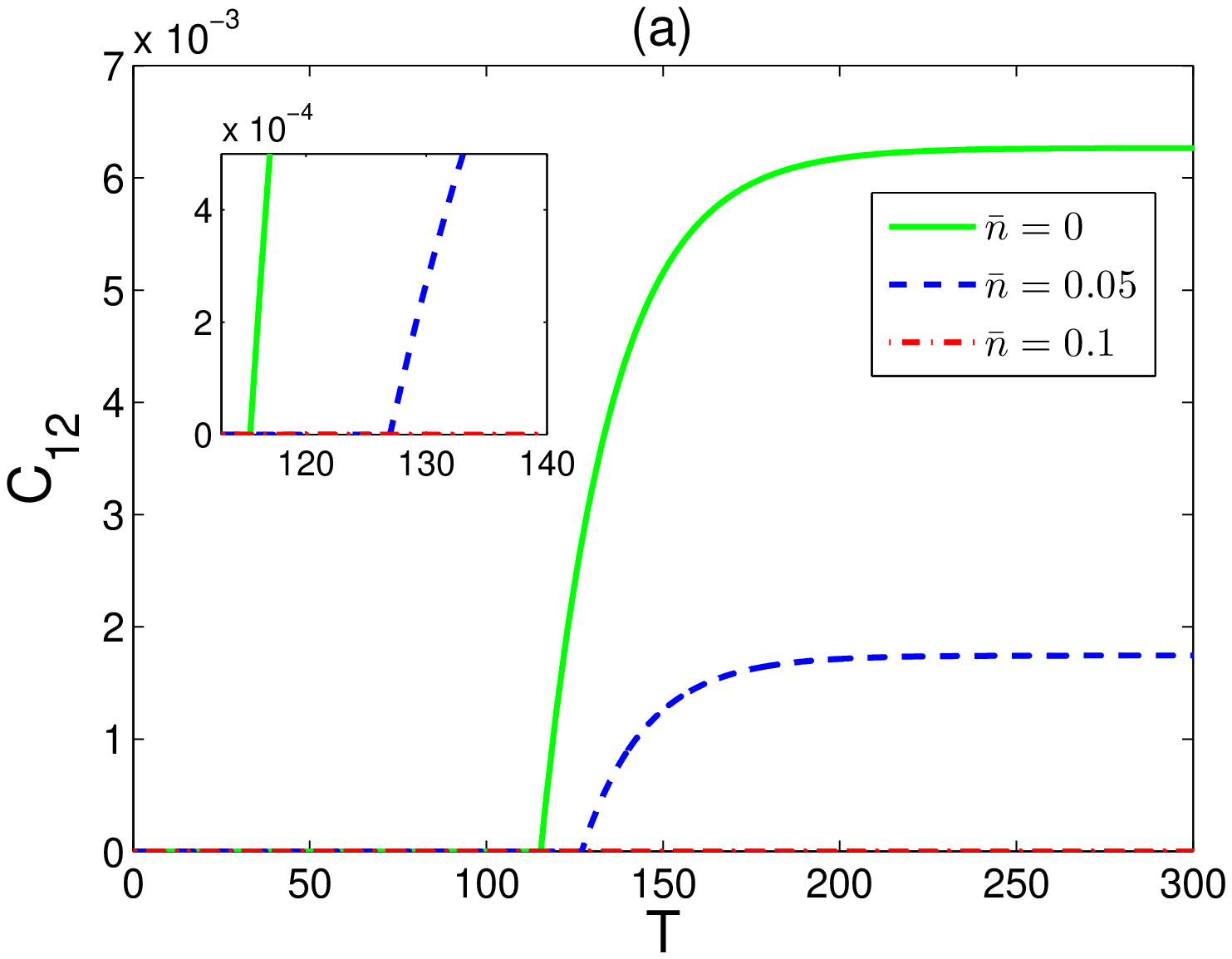}}\quad
   \subfigure{\includegraphics[width=8cm]{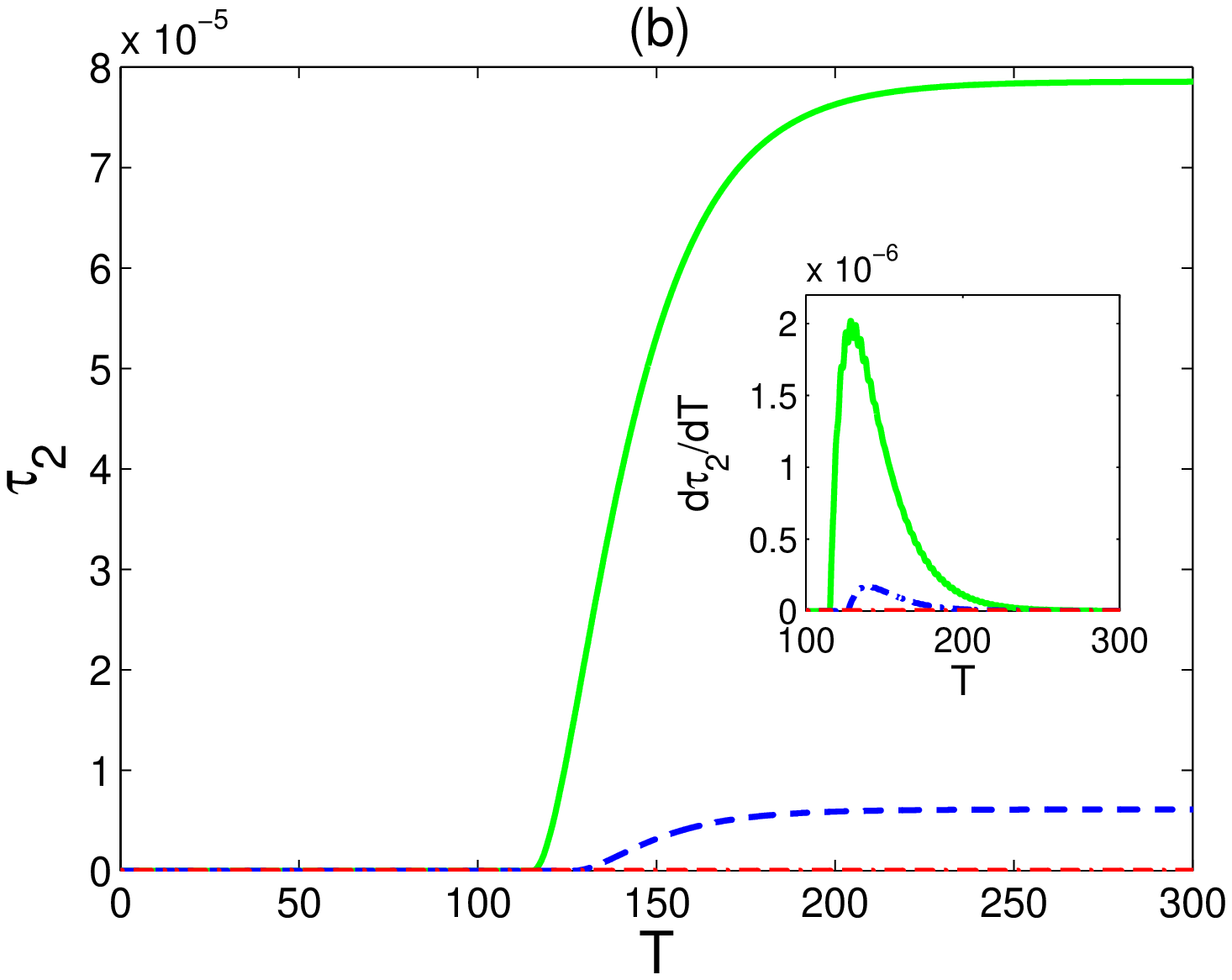}}\\
	\subfigure{\includegraphics[width=8cm]{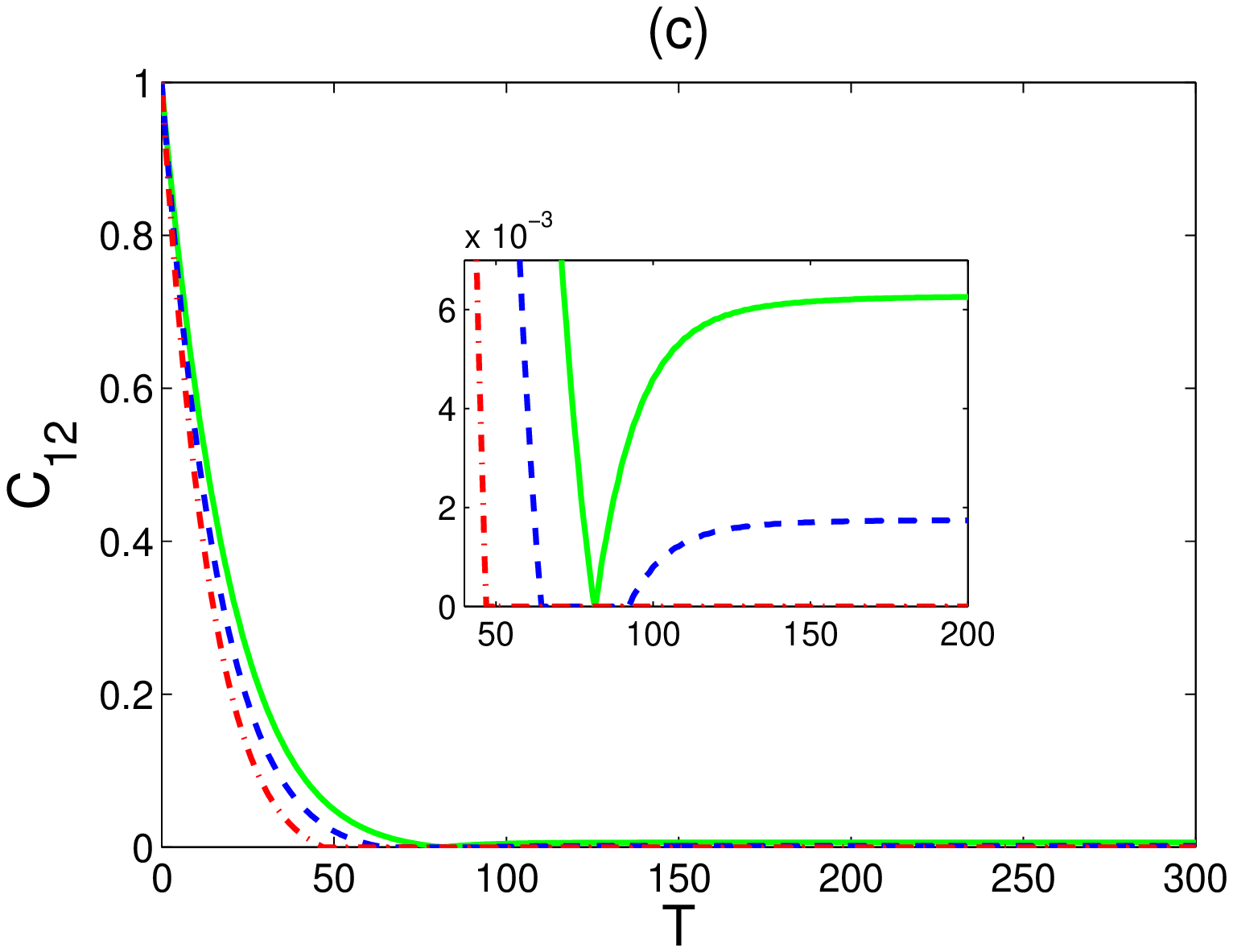}}\quad
     \subfigure{\includegraphics[width=8cm]{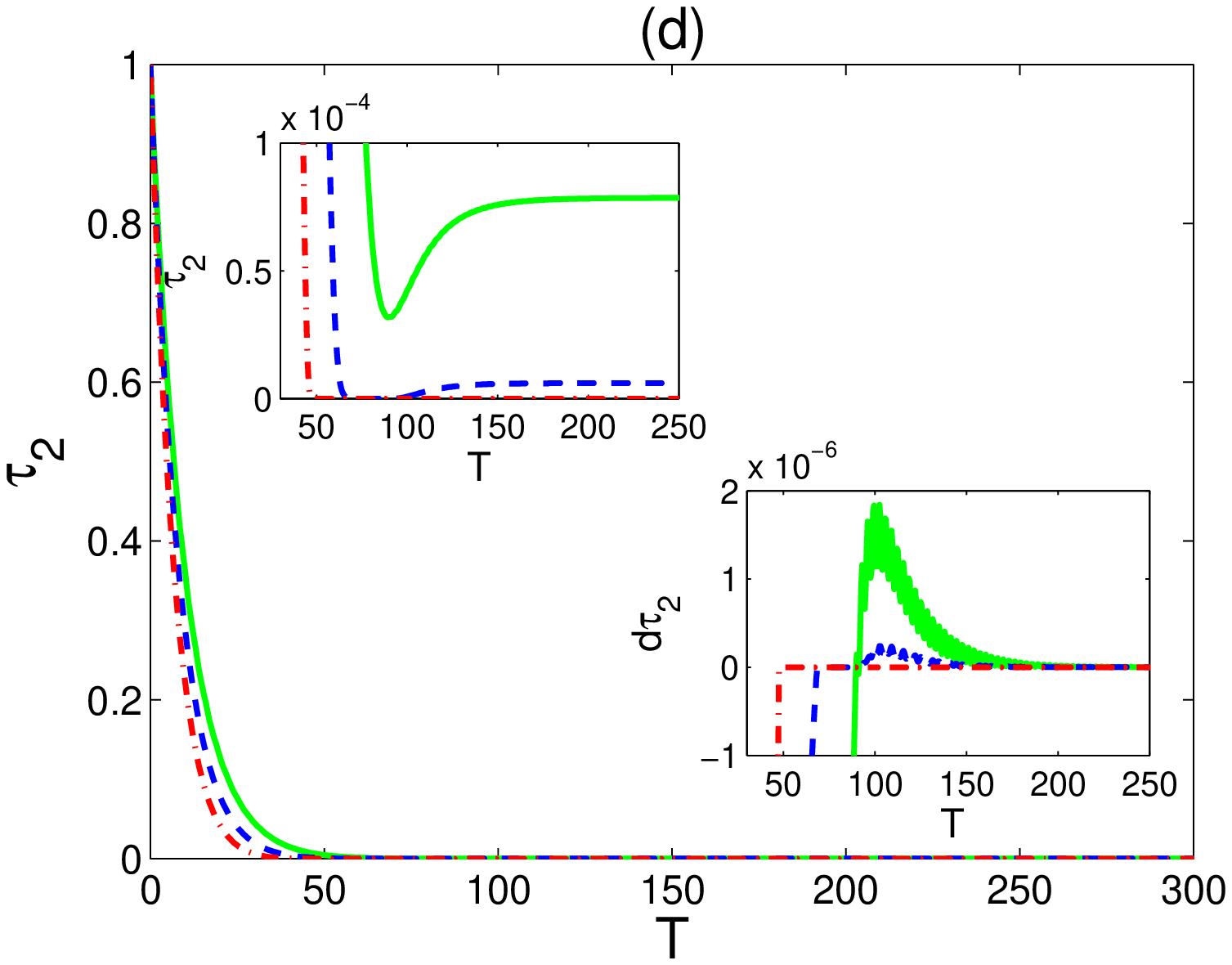}}\\
  \caption{{\label{N7_close_XYZ_XY_G05} Time evolution of $C_{12}$ and $\tau_2$ in the XYZ (or XY) system in presence of the environment ($\Gamma=0.05$) starting from an initial disentangled state in (a) and (b) and a maximally entangled state in (c) and (d) at different temperatures $\bar{n}=0,\;0.05$ and $0.1$, where $N=7$. The legend is as shown in subfig. (a).}}
 \end{minipage}
\end{figure}
In fig.~\ref{N7_close_Ising_max_G05}, we study the time evolution of the entanglement in the Ising system starting from an initial maximally entangled state. The overall dynamics of $C_{12}$ and $\tau_2$ is very close to what was observed when the system started from the W-state except that the changes are sharper and the rapid oscillation in the derivative of $\tau_2$ disappears. More importantly, the steady state values of $C_{12}$ and $\tau_2$ were found to be the same in all the three different cases of the Ising system, in figs.~\ref{N7_close_Ising_dis_w_G05} and \ref{N7_close_Ising_max_G05}, regardless of the initial state of the system.
\begin{figure}[htbp]
\begin{minipage}[c]{\textwidth}
 \centering
   \subfigure{\includegraphics[width=8cm]{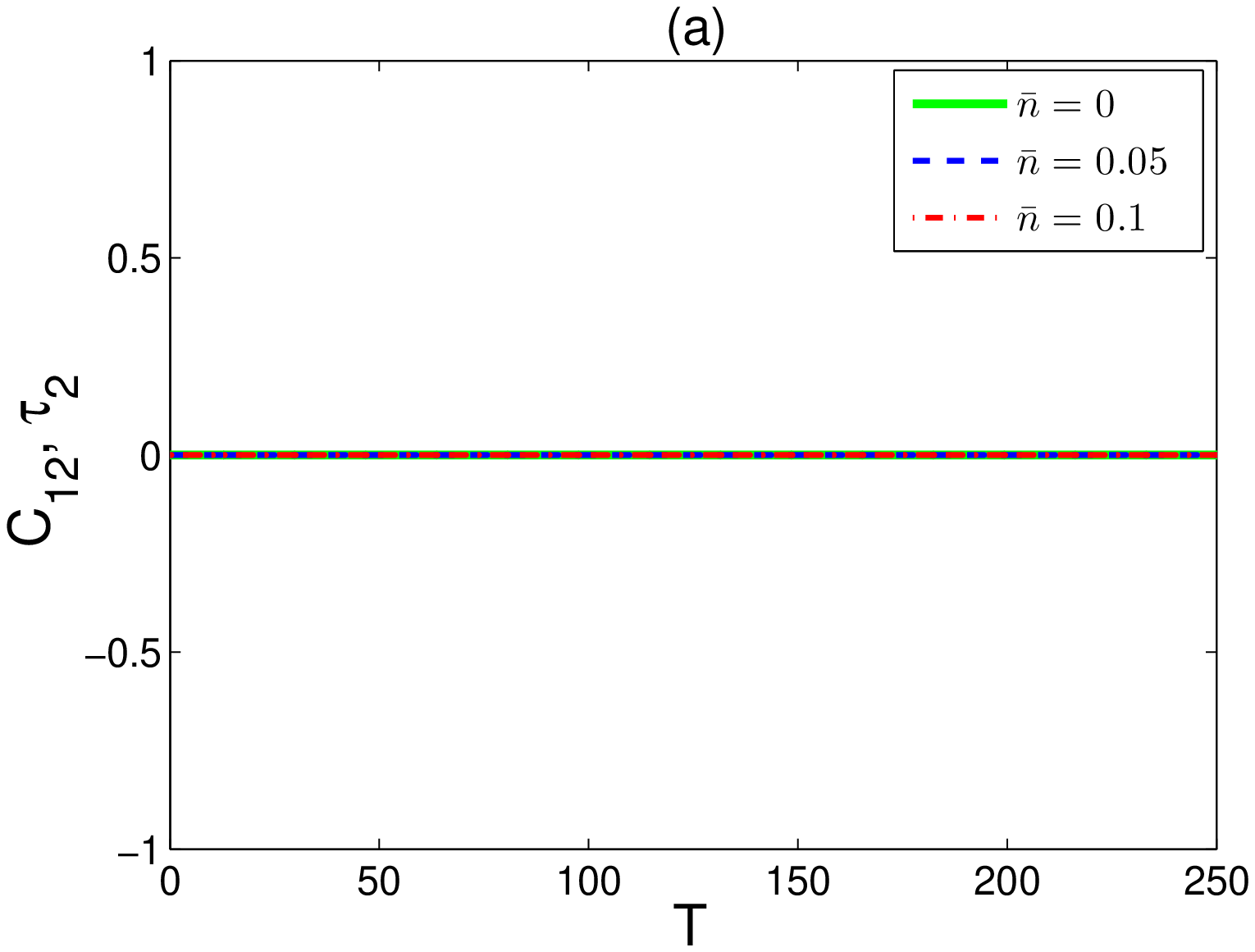}}\quad
   \subfigure{\includegraphics[width=8cm]{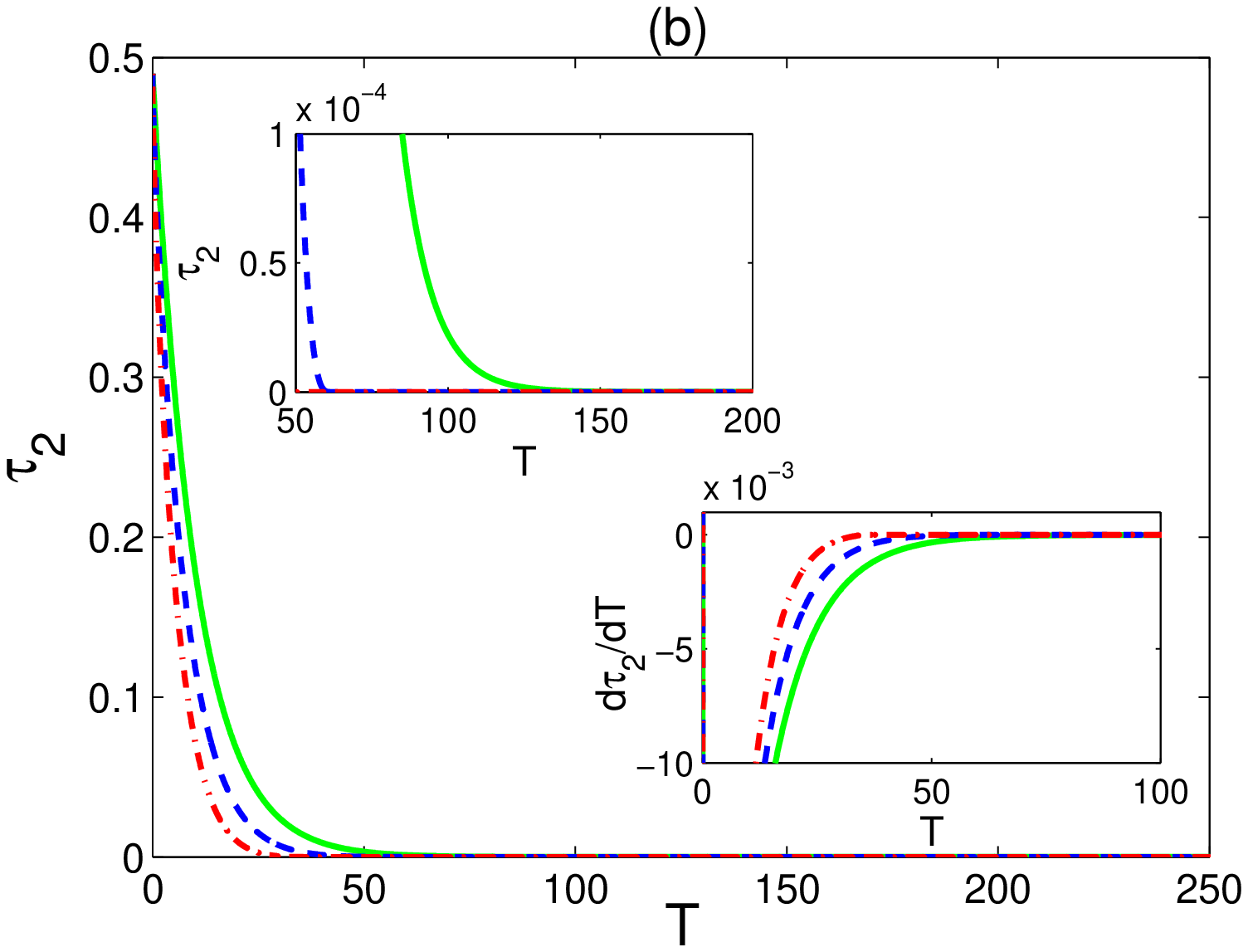}}\\
   \subfigure{\includegraphics[width=8cm]{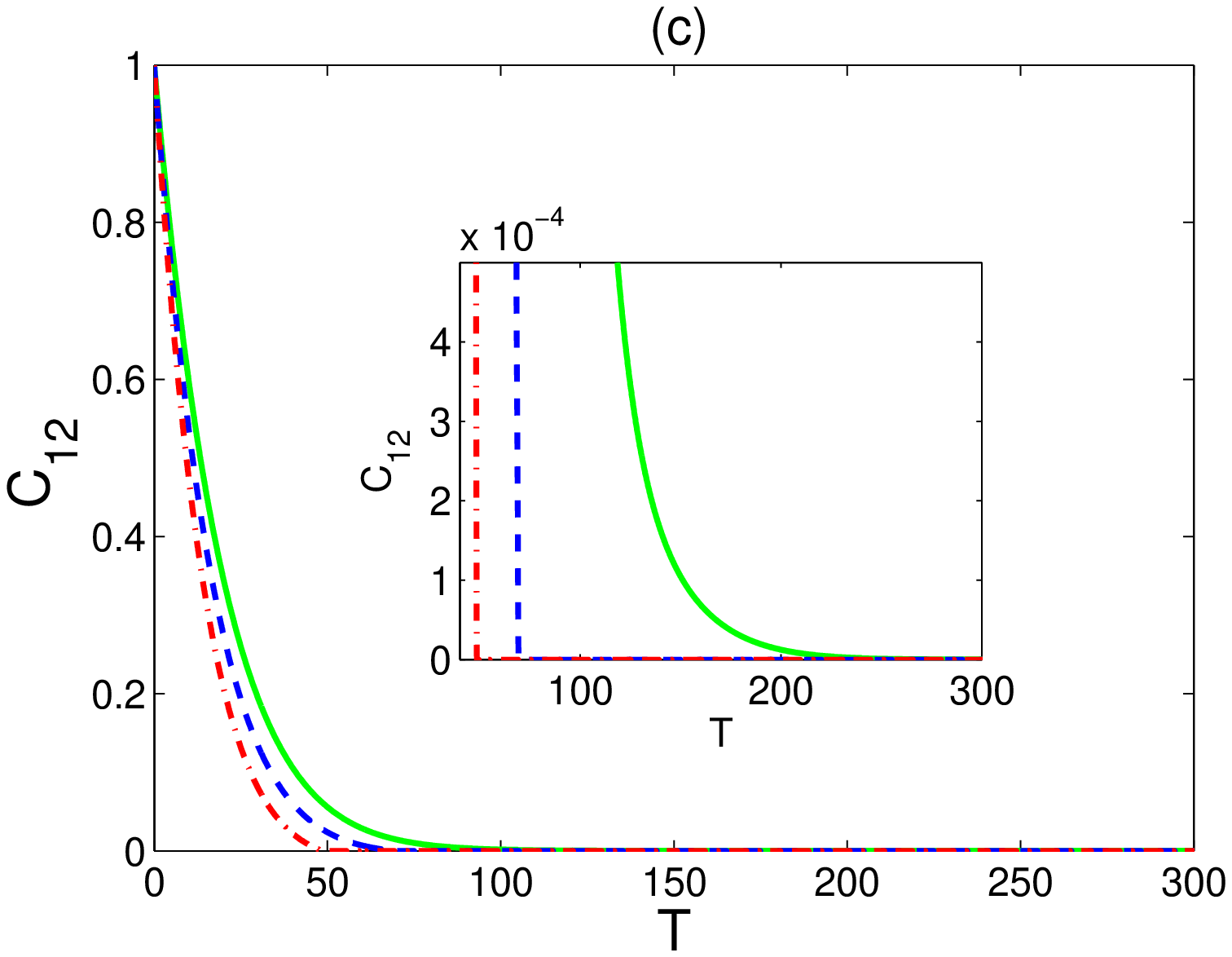}}\quad
   \subfigure{\includegraphics[width=8cm]{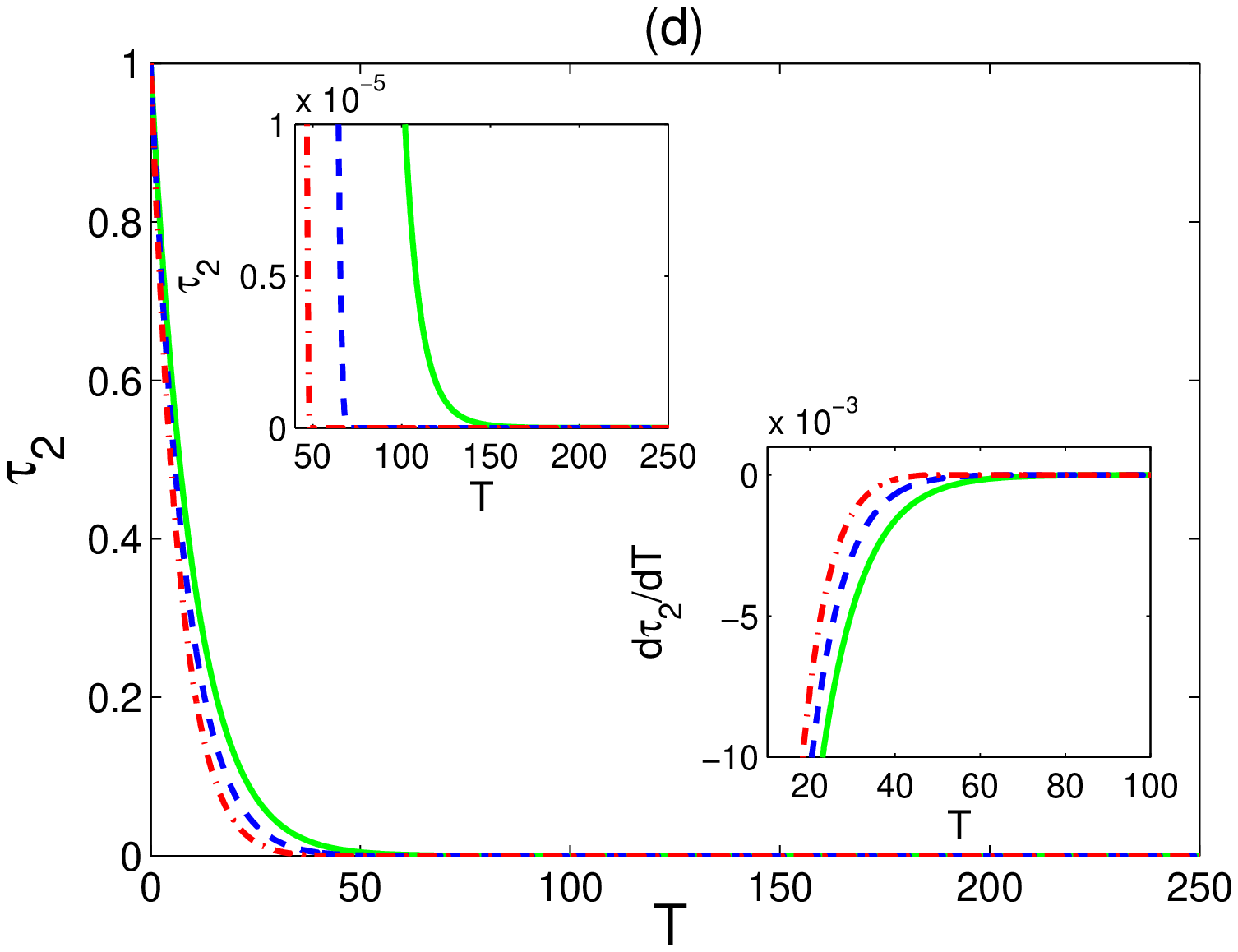}}\\
  \caption{{\label{N7_close_XX_G05} Time evolution of $C_{12}$ and $\tau_2$ in the XX (XXX or XXZ) system in presence of environment ($\Gamma=0.05$) starting from an initial (a) disentangled state; (b) entangled W-state and maximally entangled state in (c) and (d) at different temperatures $\bar{n}=0,\;0.05$ and $0.1$, where $N=7$. The legend is as shown in subfig. (a).}}
\end{minipage}
\end{figure}
In fig.~\ref{N7_close_XYZ_XY_G05}, we consider the partially anisotropic $XY$ system starting from two different initial states , separable in (a) and (b) and Maximally entangled in (c) and (d). The behavior of the entanglement $C_{12}$ and $\tau_2$ are similar to that of the Ising system with one main difference, which is a much smaller steady state values for $C_{12}$ and $\tau_2$. Also we have tested the effect of the spin coupling in the z-direction, by considering $0 < \delta \leq 1$, and particularly in the $XYZ$ system. We didn't find any noticeable change in either the dynamics of the system or the steady-state values as a result of this coupling for the set of parameter values that we are adopting here.
\begin{figure}[htbp]
\begin{minipage}[c]{\textwidth}
 \centering
   \subfigure{\includegraphics[width=8cm]{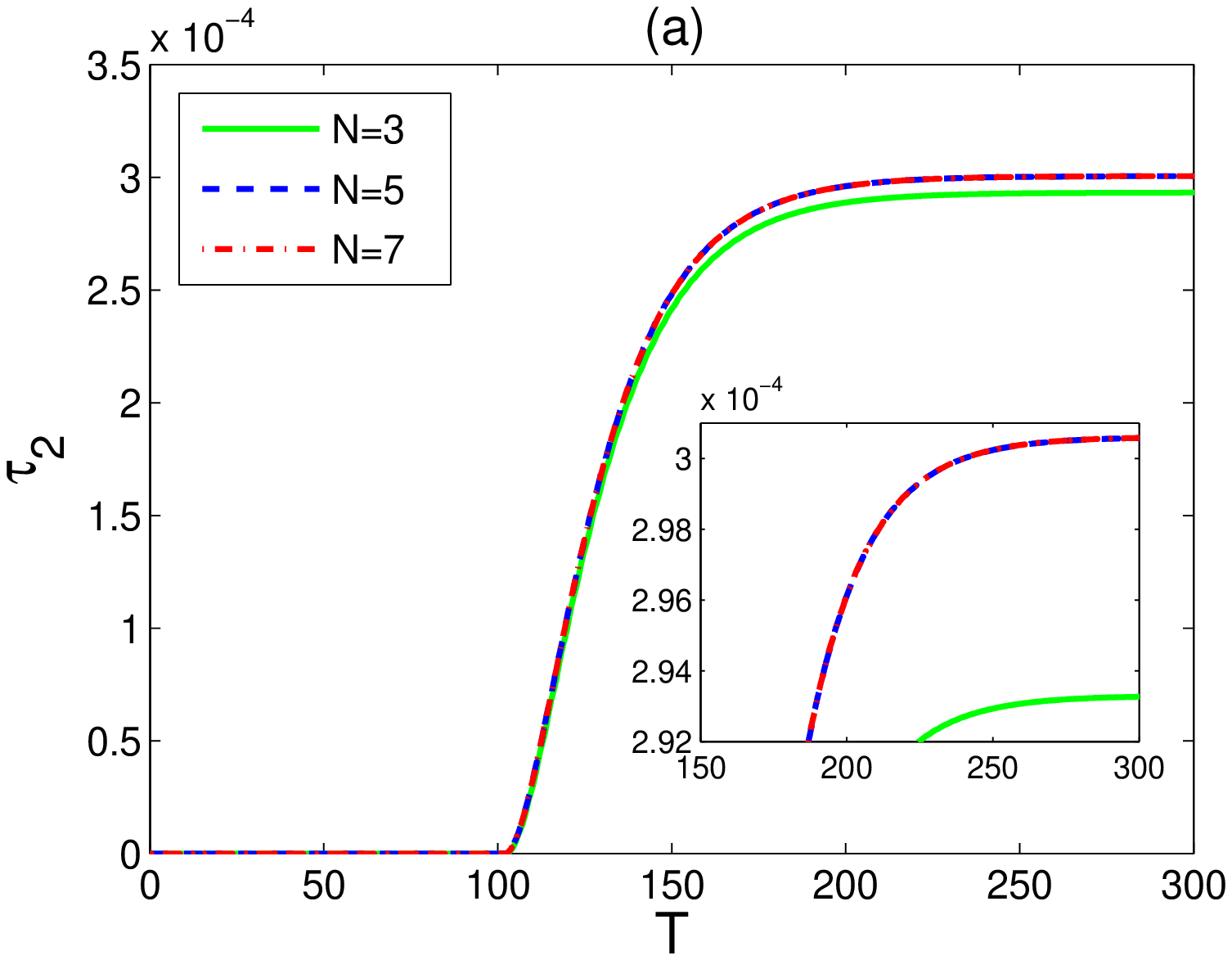}}\quad
   \subfigure{\includegraphics[width=8cm]{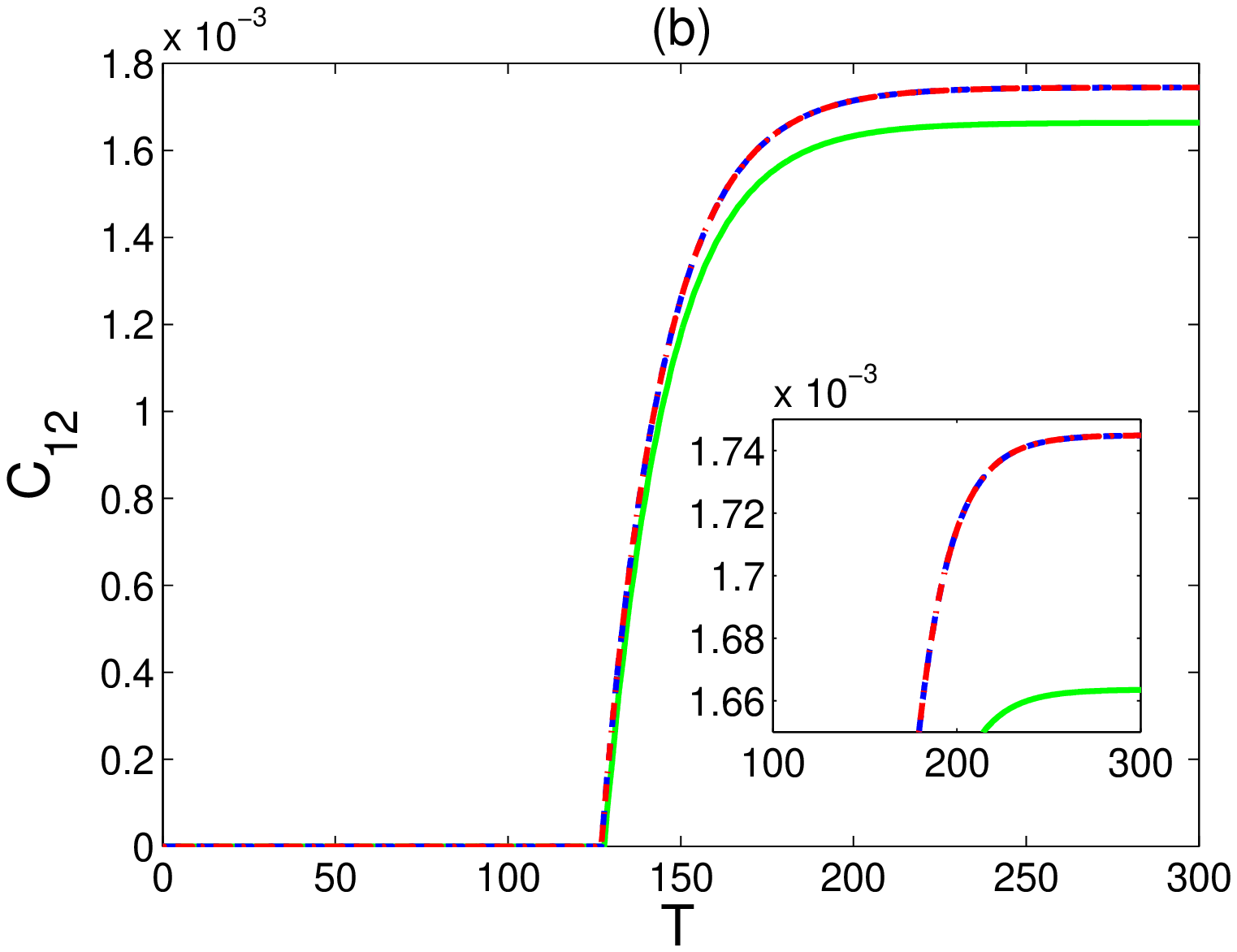}}\\
   \subfigure{\includegraphics[width=8cm]{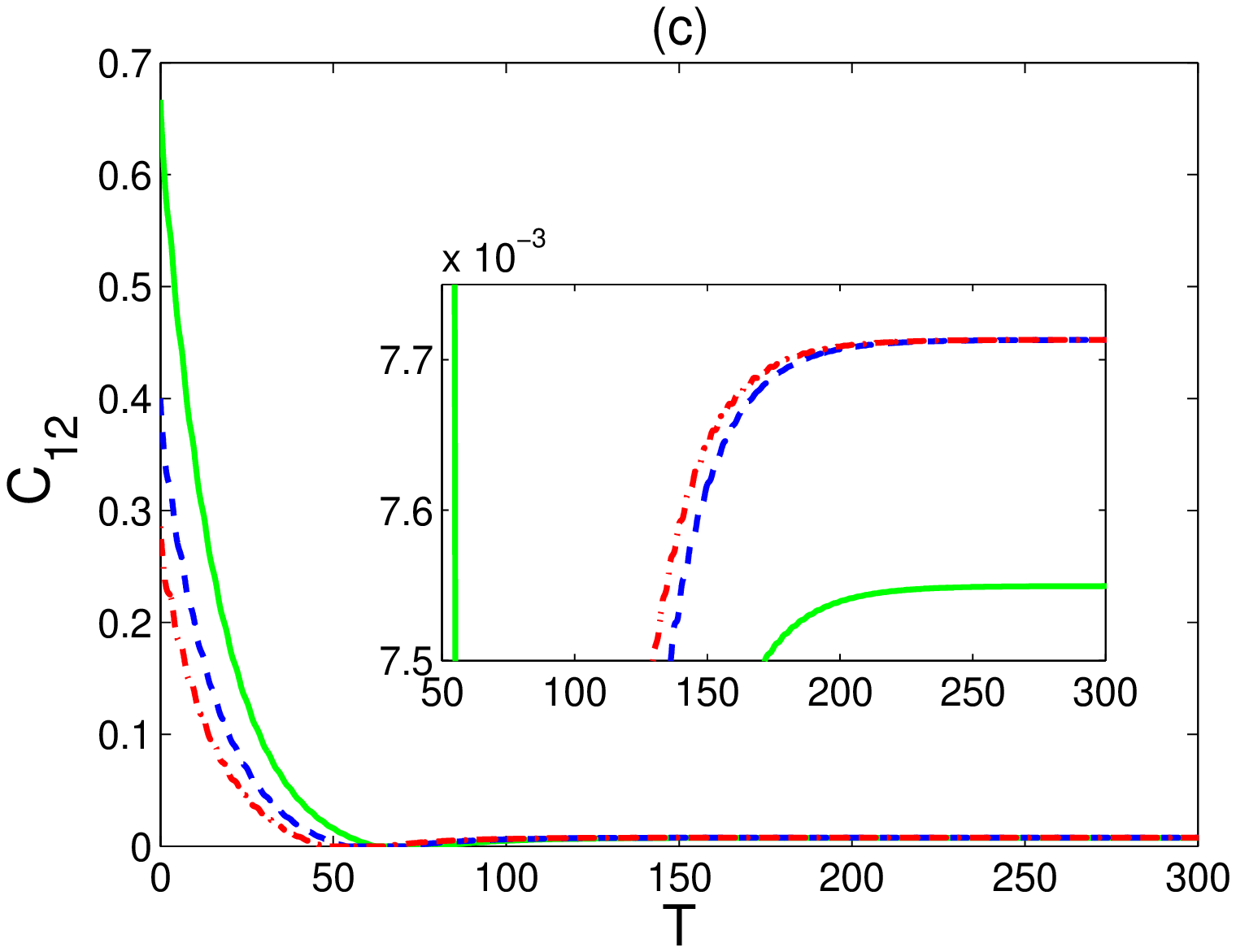}}\quad
   \subfigure{\includegraphics[width=8cm]{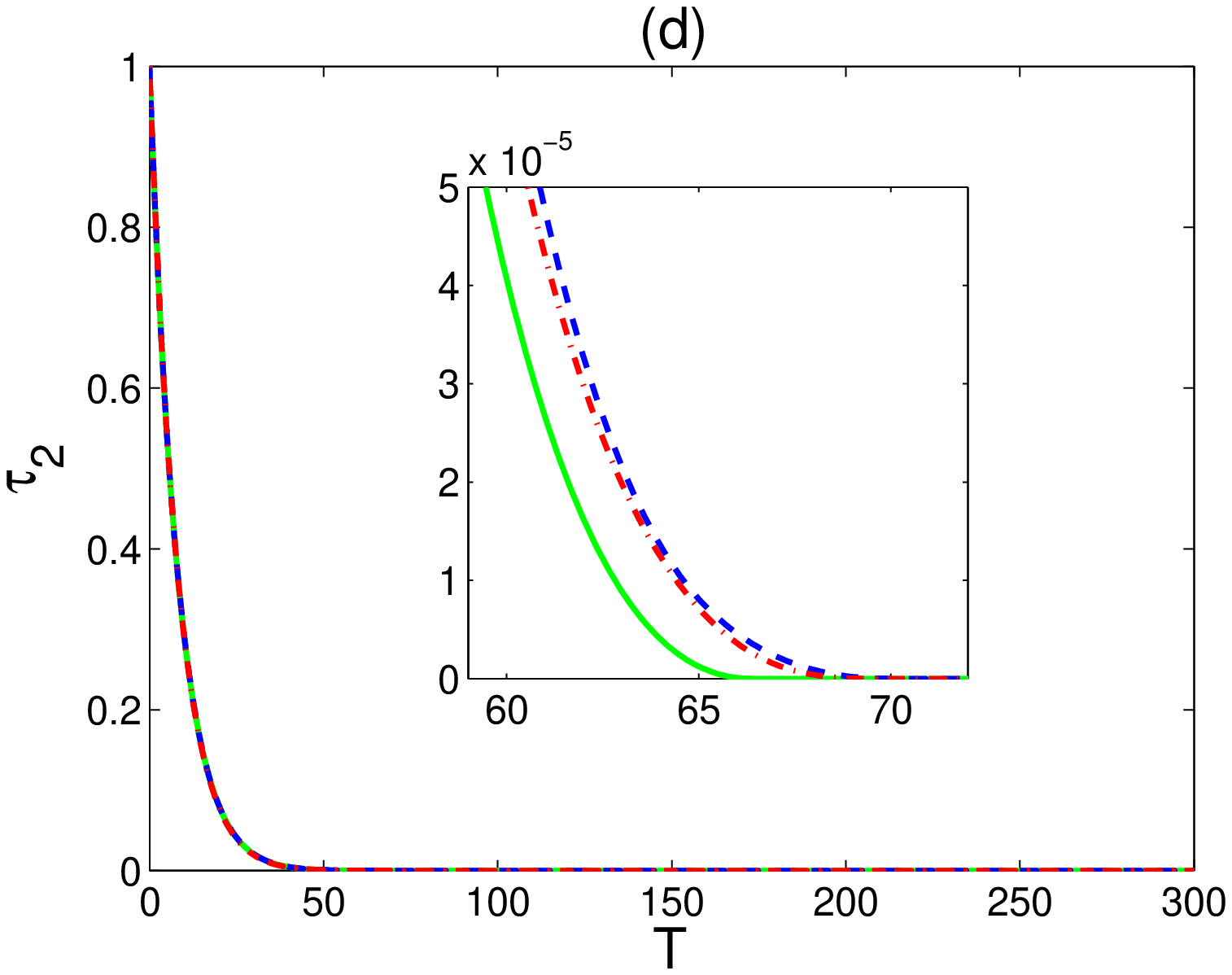}}\\
  \caption{{\label{size_effect} Time evolution at different chain sizes ($N=3, \;5$ and $7$) in presence of the environment ($\Gamma=0.05$) of (a) $\tau_2$ in the Ising system starting from an initial disentangled state at zero temperature; (b) $C_{12}$ in the $XYZ$ system starting from an initial disentangled state at temperature $\bar{n}=0.05$; (c) $C_{12}$ in the Ising system starting from an initial W-entangled state at temperature $\bar{n}=0.05$; (d) $\tau_2$ in the XX system starting from an initial maximally entangled state at zero temperature. The legend is as shown in subfig.(a).}}
 \end{minipage}
\end{figure}
The completely isotropic $XXX$ system is explored in fig.~\ref{N7_close_XX_G05}, which shows a significantly different profile from the Ising and the $XY$ systems. As one can see in fig.~\ref{N7_close_XX_G05}(a), when the system starts from an initial separable state, both $C_{12}$ and $\tau_2$ start with and sustain a zero value as the system evolves in time at zero and finite temperatures. In fig.~\ref{N7_close_XX_G05}(b), the time evolution of $\tau_2$ is monitored in the $XXX$ system starting from the W-state. As can be seen, $\tau_2$ starts with a value of about 0.5 and decays rapidly as the time elapses but ends up vanishing completely without any revival. As the temperature increases, the vanishing of entanglement becomes sharper and earlier in time as can be concluded from the inner panels in fig.~\ref{N7_close_XX_G05}(b). A very similar behavior of $C_{12}$ and $\tau_2$ is observed as the $XXX$ system starts from an initial maximally entangled state following the same dynamical behavior and ending up with a zero value, as illustrated in fig.~\ref{N7_close_XX_G05}(c) and (d). Again testing the effect of spin coupling in the z-direction, by studying the $XX$ or $XXZ$ systems, there were no noticeable changes, compared with the $XXX$ system, either in the dynamics of the systems or the asymptotic values they reach.
\begin{figure}[htbp]
\begin{minipage}[c]{\textwidth}
 \centering
   \subfigure{\includegraphics[width=8cm]{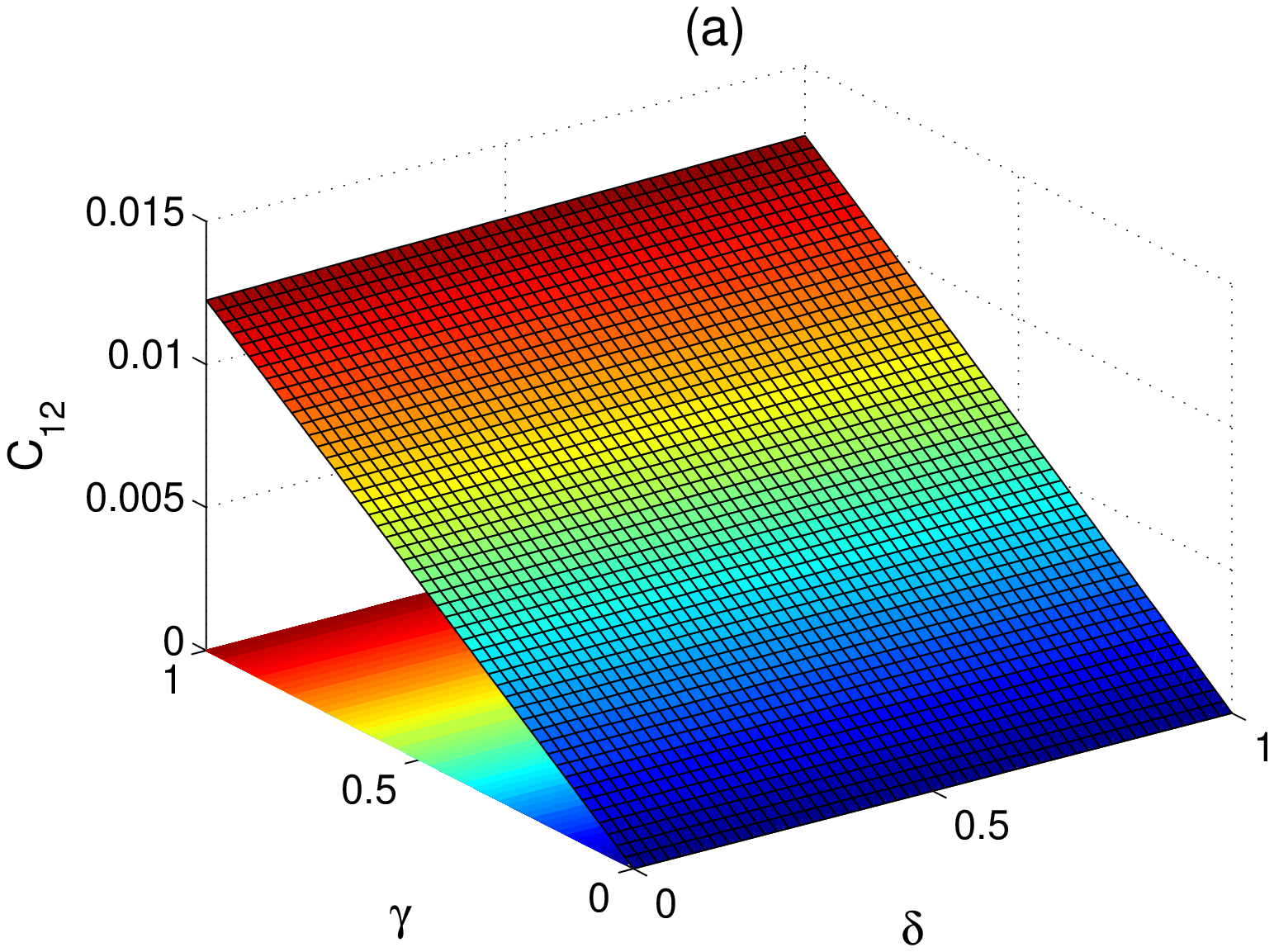}}\quad
   \subfigure{\includegraphics[width=8cm]{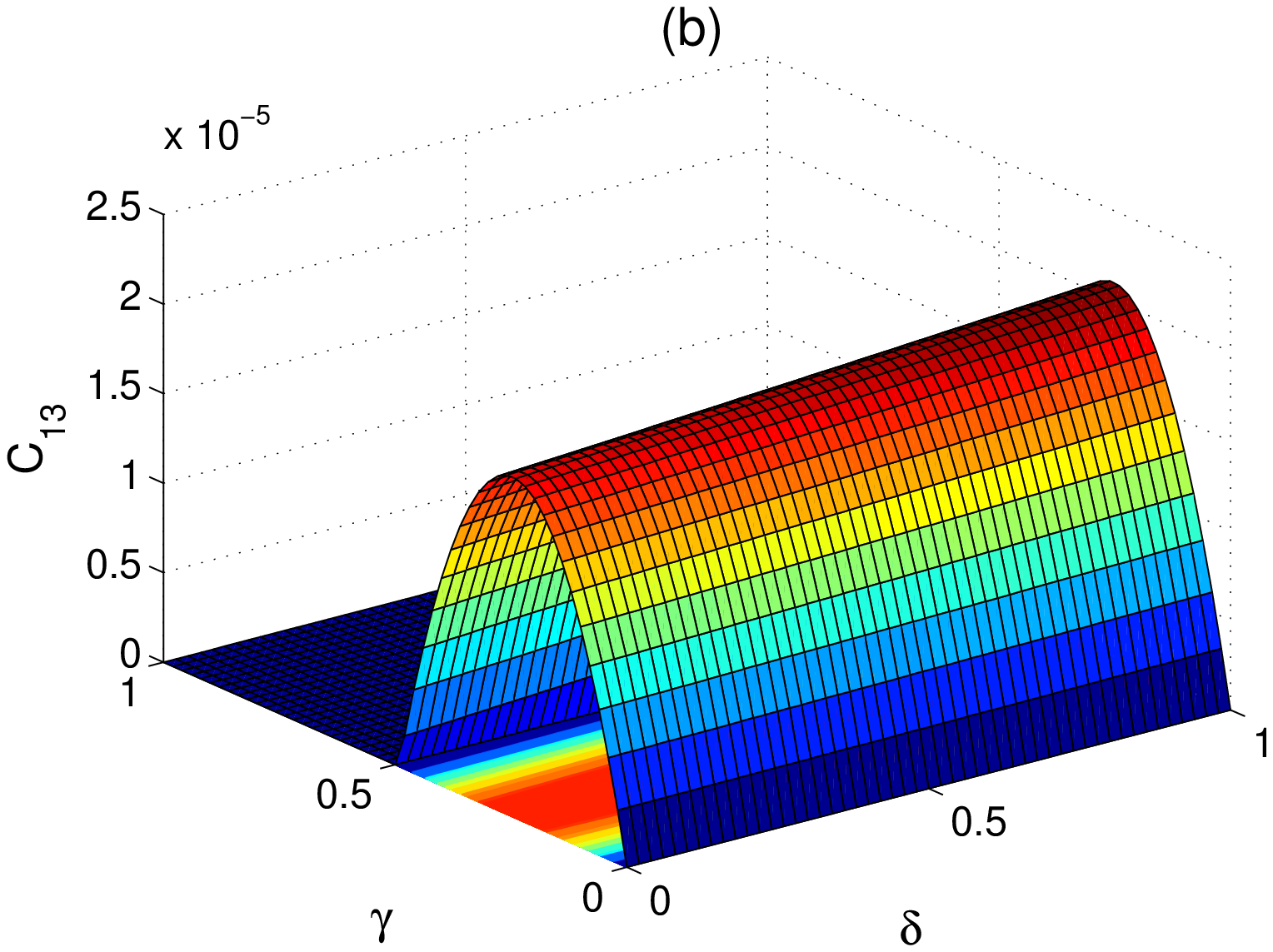}}\\
    \subfigure{\includegraphics[width=8cm]{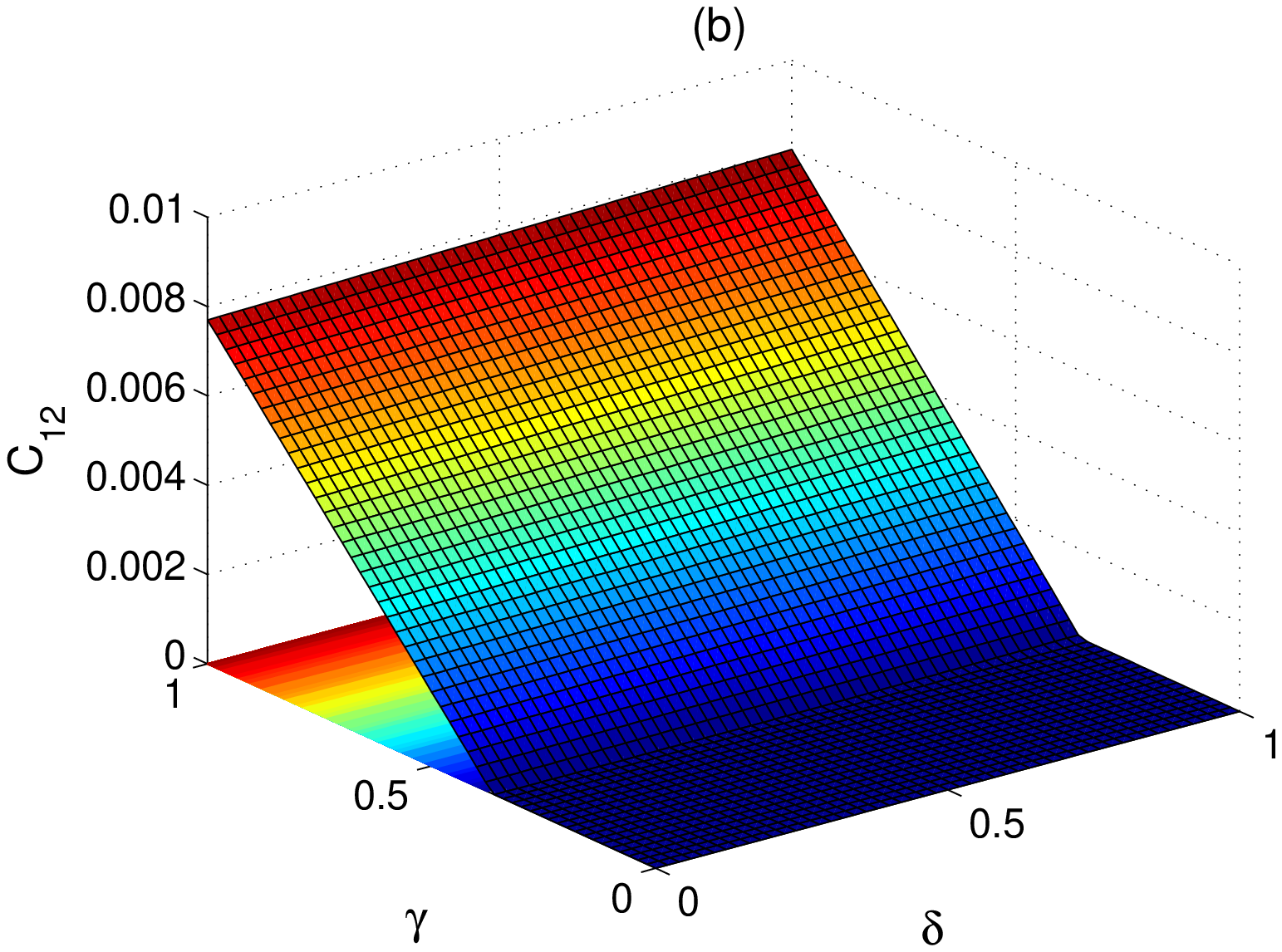}}\quad
   \subfigure{\includegraphics[width=8cm]{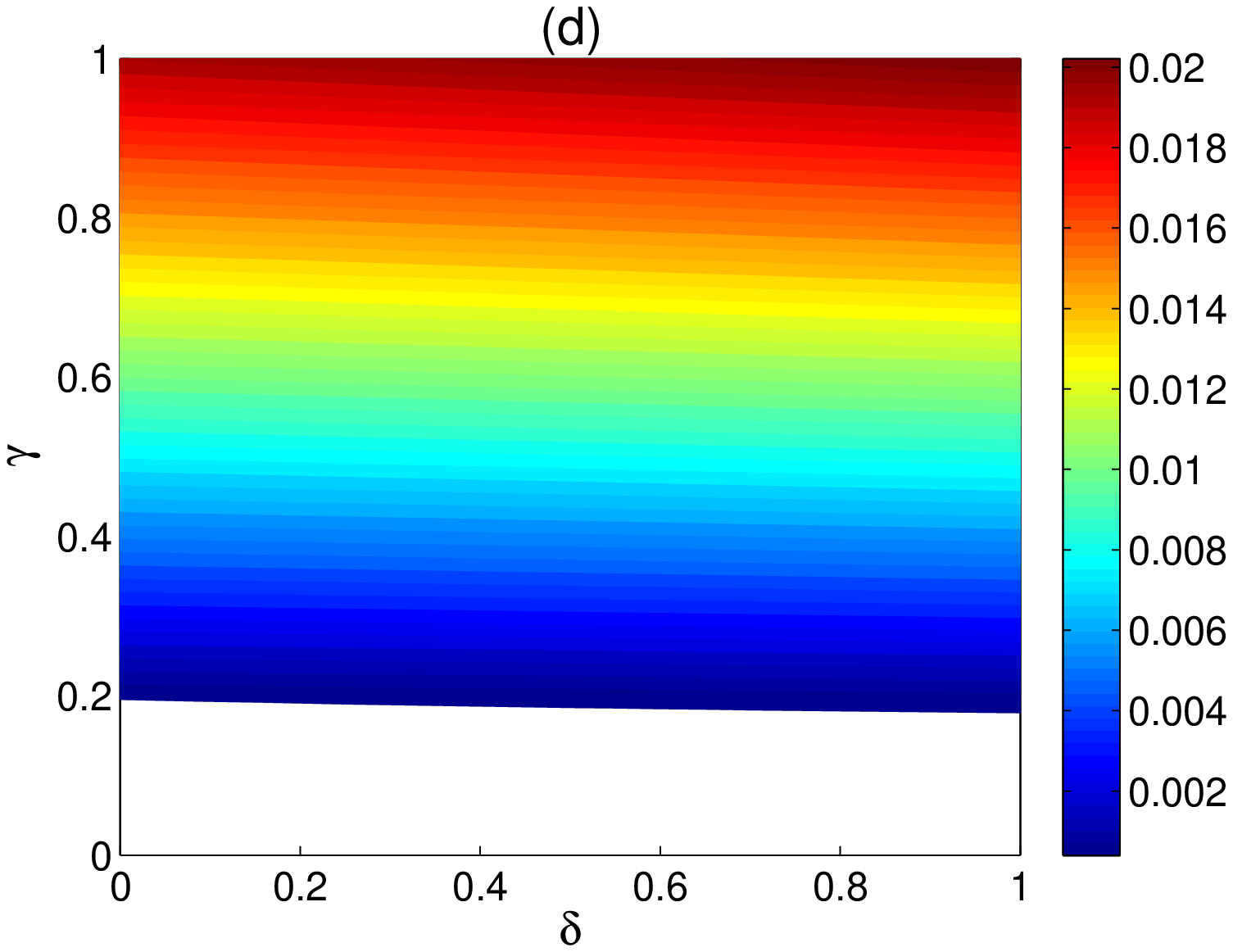}}\\
  \caption{{\label{N5_closed_Ent_vs_gamma_delta} The asymptotic behavior in the $\gamma-\delta$ space of the Heisenberg XYZ system in presence of the environment ($\Gamma=0.05$), with $J=0.05$ starting from any initial state (disentangled, entangled or maximally entangled) of (a) $C_{12}$ at $\bar{n}=0$; (b) $C_{13}$ at $\bar{n}=0$; (c) $C_{12}$ at $\bar{n}=0.05$ and (d) the contour plot of $C_{12}$ at $\bar{n}=0.05$ but for $J=0.1$.}}
\end{minipage}
\end{figure}

In fig.~\ref{size_effect}, we examine the system size effect by studying the time evolution of the entanglement in chains with different total number of spins. In fig.~\ref{size_effect}(a), we depict the time evolution of $\tau_2$ for an Ising chain starting from an initial disentangled state at zero temperature for three different chain sizes ($N=3, 5$ and 7). As can be noticed, the behavior of the entanglement dynamics converges very rabidly as $N$ increases and the difference between the two cases of ($N=5$ and 7) is quite small, which indicates a very small effect played by the system size as N becomes 5 or higher. The time evolution of $C_{12}$ in an $XYZ$ chain with different sizes starting from a disentangled state at finite temperature, $\bar{n}=0.05$, is considered in fig.~\ref{size_effect}(b). The behavior of $C_{12}$ is very similar to that of $\tau_2$, in fig.~\ref{size_effect}(a), showing a rapid convergence and an almost same steady state value for $N=5$ and 7. In fig.~\ref{size_effect}(c) we again examine the Ising system size at finite temperature but starting from partially entangled state whereas in fig.~\ref{size_effect}(d) we depict $\tau_2$ of the $XX$ model starting from a maximally entangled state at zero temperature. The behavior of the entanglements $C_{12}$ and $\tau_2$, as illustrated fig.~\ref{size_effect}(c) and (d), confirms our conclusion from fig.~\ref{size_effect}(a) and (b).
\begin{figure}[htbp]
\begin{minipage}[c]{\textwidth}
 \centering
 \subfigure{\label{fig14a}\includegraphics[width=8cm]{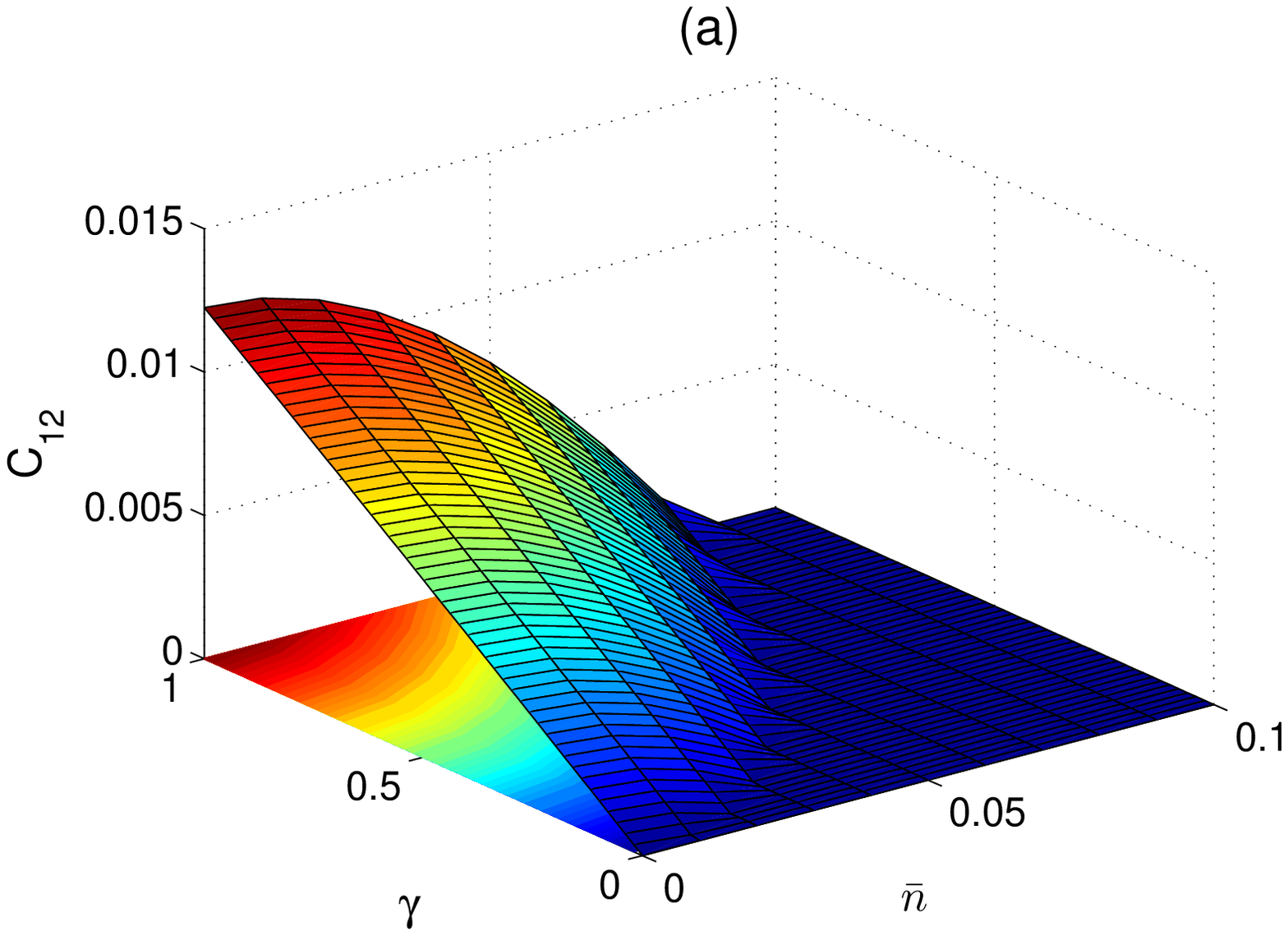}}\quad 
 \subfigure{\label{fig14b}\includegraphics[width=8cm]{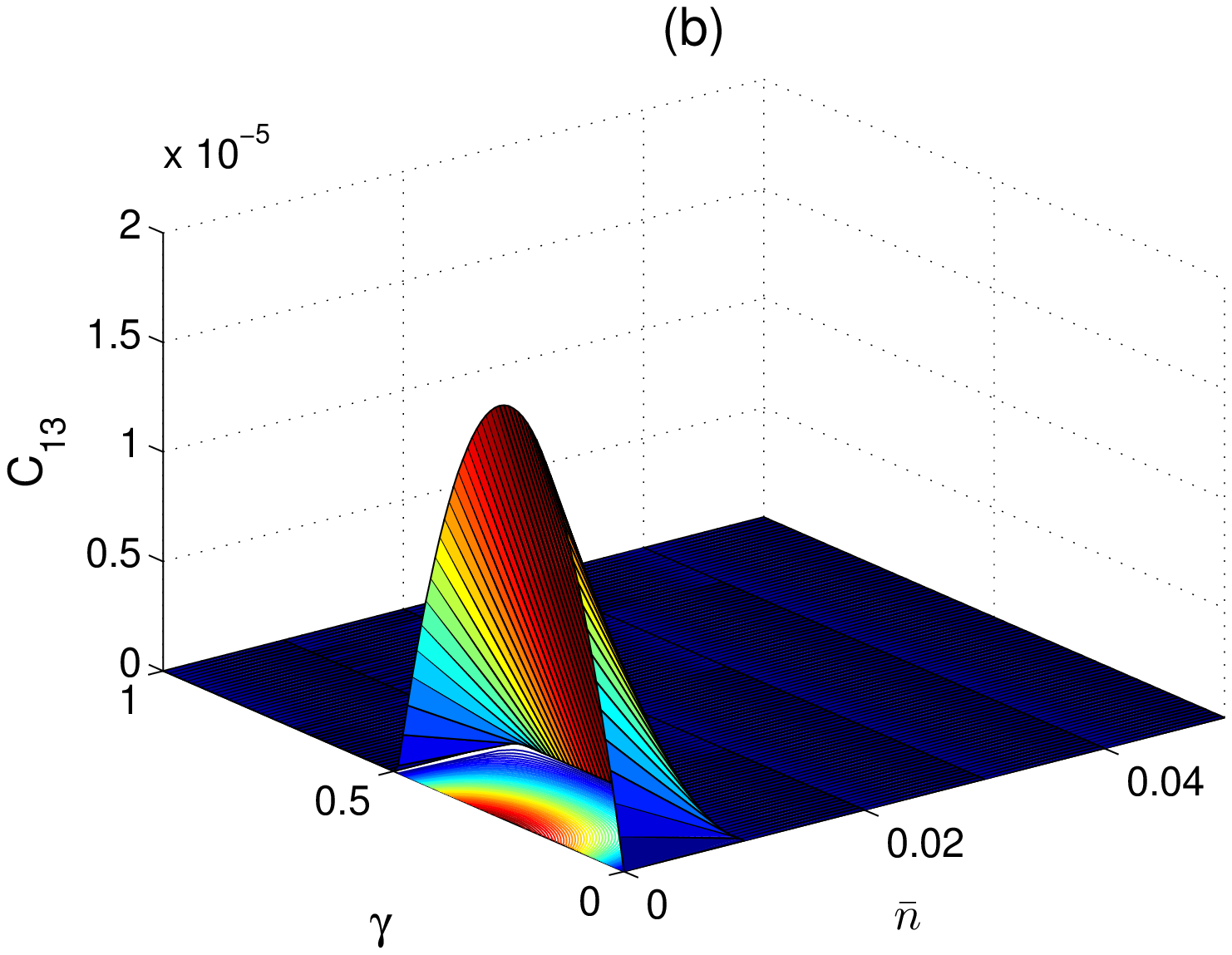}}\\
  \caption{{\label{N5_closed_Ent_vs_gamma_n}\protect\footnotesize The asymptotic behavior of $C_{12}$ and  $C_{13}$ in the Heisenberg system in presence of the environment ($\Gamma=0.05$) versus the anisotropy parameter $\gamma$ and the temperature parameter $\bar{n}$ starting from any initial state (disentangled, entangled or maximally entangled) and at any value of $0 \leq \delta \leq 1$.}}
 \end{minipage}
\end{figure}

In fig.~\ref{N5_closed_Ent_vs_gamma_delta}, the asymptotic (steady state) behavior of the entanglement in the $\gamma-\delta$ space of the $XYZ$ Heisenberg 5-spins chain is explored, where the steady state value of $C_{12}$ and $C_{13}$, at time $T=300$, is depicted versus the anisotropic parameters $\gamma$ and $\delta$. The asymptotic value was found to be independent of the initial state of the system. Interestingly, the steady state value of the entanglement, at zero temperature, shows a monotonic linear decay profile as the anisotropic parameter $\gamma$ decreases and it vanishes at $\gamma = 0$ as shown in fig.~\ref{N5_closed_Ent_vs_gamma_delta}(a), whereas the parameter $\delta$ shows no effect on the steady state value. The entanglement $C_{13}$ shows a completely different behavior, as illustrated in fig.~\ref{N5_closed_Ent_vs_gamma_delta}(b), where it sustains a value of zero for $\gamma=1$ up to $\gamma\approx0.5$ before rising up to reach a maximum value at $\gamma \approx 0.25$, then it decays again until completely vanishing at $\gamma=0$. Obviously, the robustness of the entanglement $C_{13}$ against the decohering effect of the environment is not highest at maximum anisotropy, in contrary to $C_{12}$. As the temperature increases, the entanglement $C_{12}$ decreases but chains with higher $\gamma$ (anisotropy) is more robust to thermal excitation whereas chains with low anisotropy lose their entanglement completely, as can be noticed in fig.~\ref{N5_closed_Ent_vs_gamma_delta}(c) where $\bar{n}=0.05$, but as the temperature is raised further, the Heisenberg chains become fully disentangled regardless of their degree of anisotropy. As we concluded  before and as can be noticed in fig.~\ref{N5_closed_Ent_vs_gamma_delta}(a), (b) and (c), the anisotropic parameter $\delta$ has no noticeable effect on the entanglement dynamics, the reason is the overwhelming magnetic field in the z-direction compared with the component of spin coupling $J$  in the same direction. To clarify this point, in fig.~\ref{N5_closed_Ent_vs_gamma_delta}(d), we have applied a greater value of $J$, $0.1$ instead of $0.05$, and as can be seen in the contour plot of the entanglement $C_{12}$ versus $\gamma$ and $\delta$, the entanglement steady state value slightly increases as $\delta$ is increased, which means higher $\delta$ would enhance the value of the entanglement.
To further investigate the effect of thermal excitations on the asymptotic steady state of the Heisenberg chains, we depict the asymptotic values of the entanglements $C_{12}$ and $C_{13}$ versus the anisotropy parameter $\gamma$ and the temperature parameter $\bar{n}$ in fig.~\ref{N5_closed_Ent_vs_gamma_n}. The results confirm our observations from the previous figure, where the (nn) entanglement $C_{12}$ is more robust to thermal excitation in the completely anisotropic system and less as the degree of anisotropy decreases until it becomes very fragile in the isotropic system, as shown in fig.~\ref{N5_closed_Ent_vs_gamma_n}(a). Also the (nnn) entanglement $C_{13}$, explored in fig.~\ref{N5_closed_Ent_vs_gamma_n}(b), shows robustness for approximately $0 < \gamma \leq 0.5$ with its peak at $\gamma \approx 0.25$. This indicates that while the completely anisotropic system ($\gamma=1$) enjoys a very robust nearest neighbor entanglement, its beyond nearest neighbor entanglement is not and vice versa for the partially anisotropic system ($0<\gamma<0.5$), if we ignore the role of the parameter $\delta$. The entanglements $C_{14}$ and $C_{15}$ were found to show exactly the same behavior as $C_{13}$ and $C_{12}$ respectively as would be expected for a closed boundary spin chain.
\begin{figure}[htbp]
\begin{minipage}[c]{\textwidth}
 \centering
   \subfigure{\includegraphics[width=8cm]{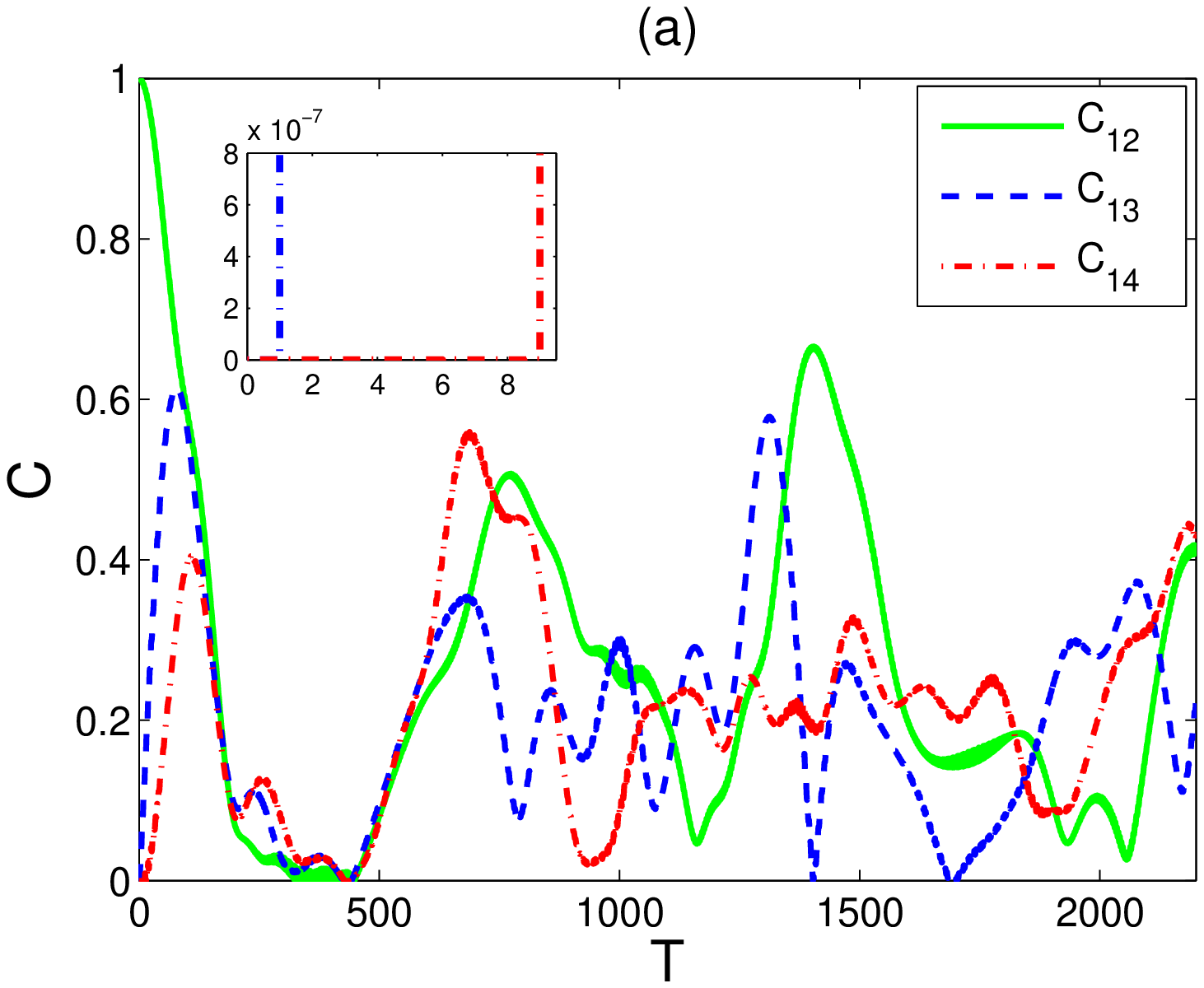}}\quad
   \subfigure{\includegraphics[width=8cm]{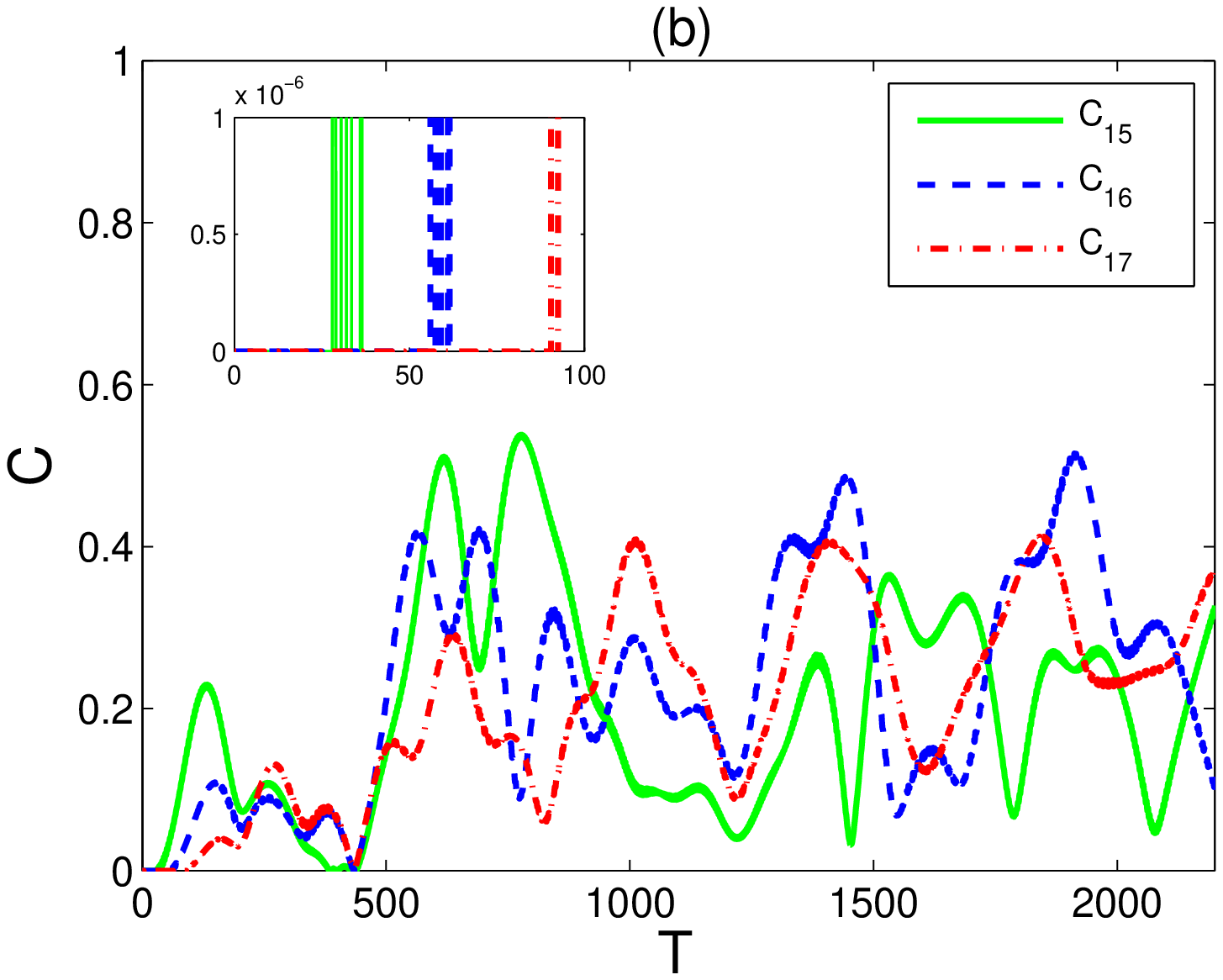}}\\
	\subfigure{\includegraphics[width=8cm]{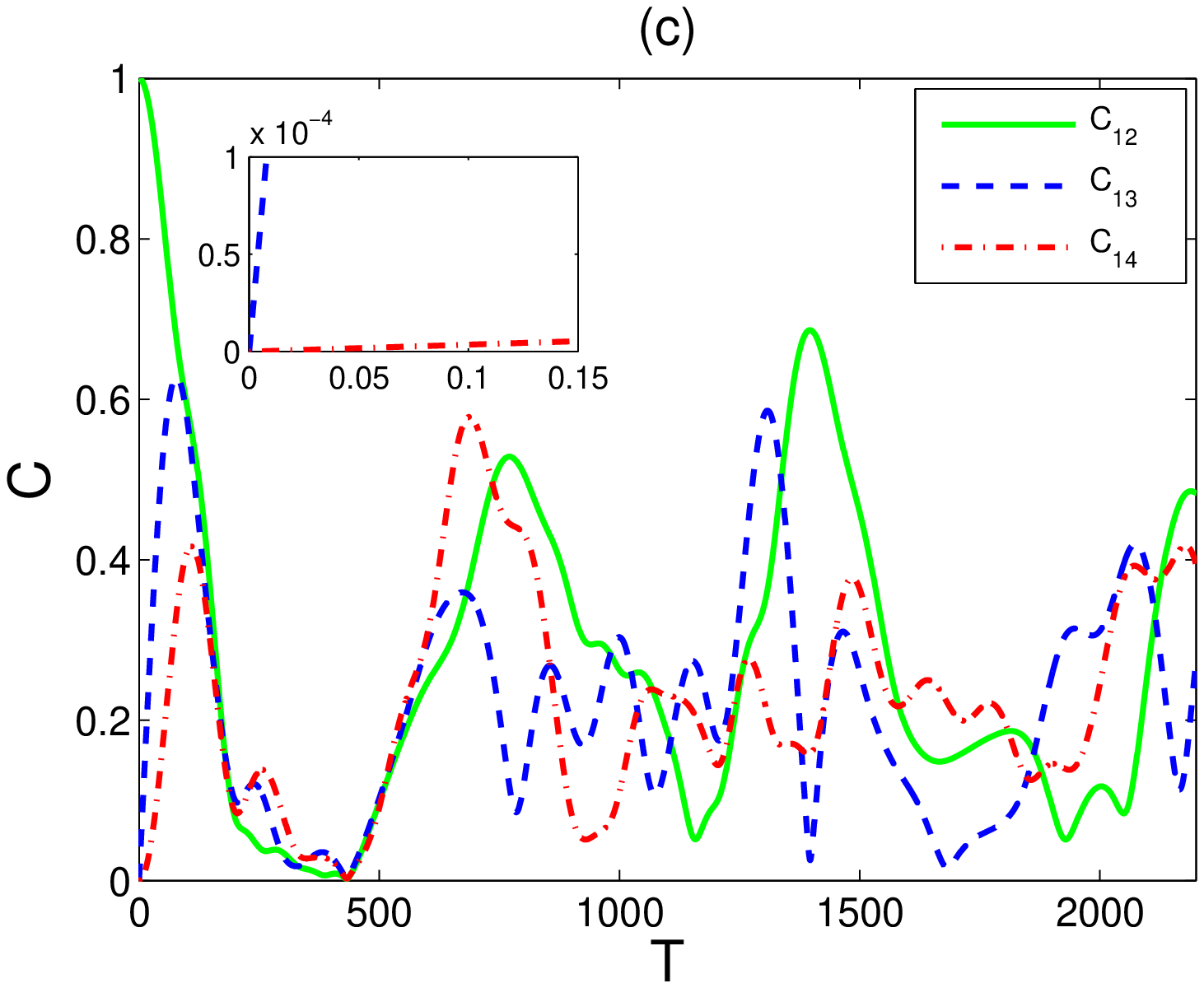}}\quad
   	\subfigure{\includegraphics[width=8cm]{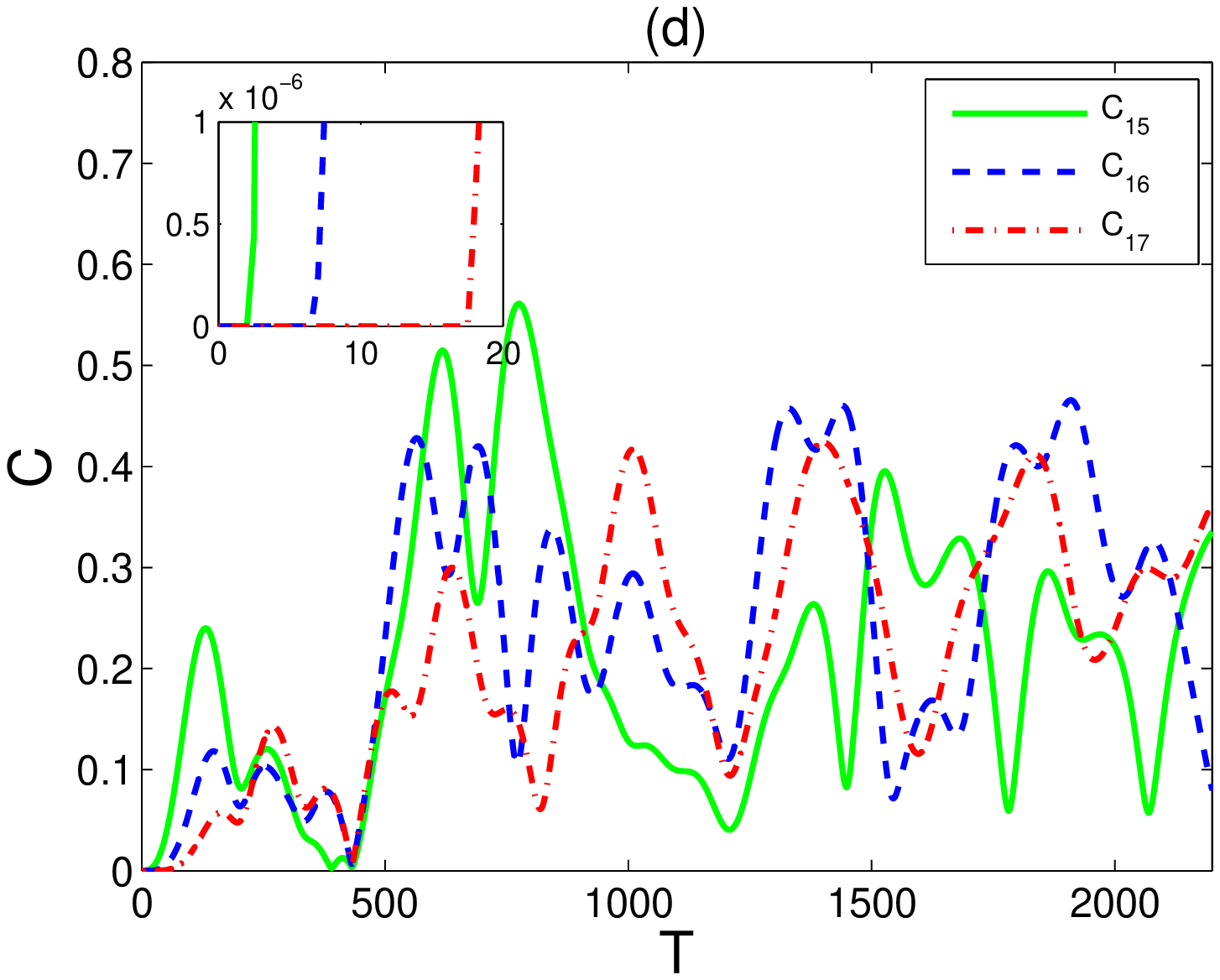}}\\
  \caption{{\label{N7_Ising_XX_open_max_C_G0} The dynamics of the entanglements $C_{12}$, $C_{13}$, $C_{14}$, $C_{15}$, $C_{16}$ and $C_{17}$ starting from a maximally entangled state in the free Ising system ($\Gamma=0$) in (a) and (b) and the free $XX$ system in (c) and (d) with open boundary condition, where N=7. The inner panels illustrate the rise up of the entanglement from zero.}}
 \end{minipage}
\end{figure}
\section{End to end entanglement transfer in open boundary spin chains}
\subsection{The Free System}
The entanglement transfer through open boundary spin systems has been always in the focus of interest as it plays an important role in implementing the different algorithms in quantum computing systems.
\begin{figure}[htbp]
\begin{minipage}[c]{\textwidth}
 \centering
   \subfigure{\includegraphics[width=8cm]{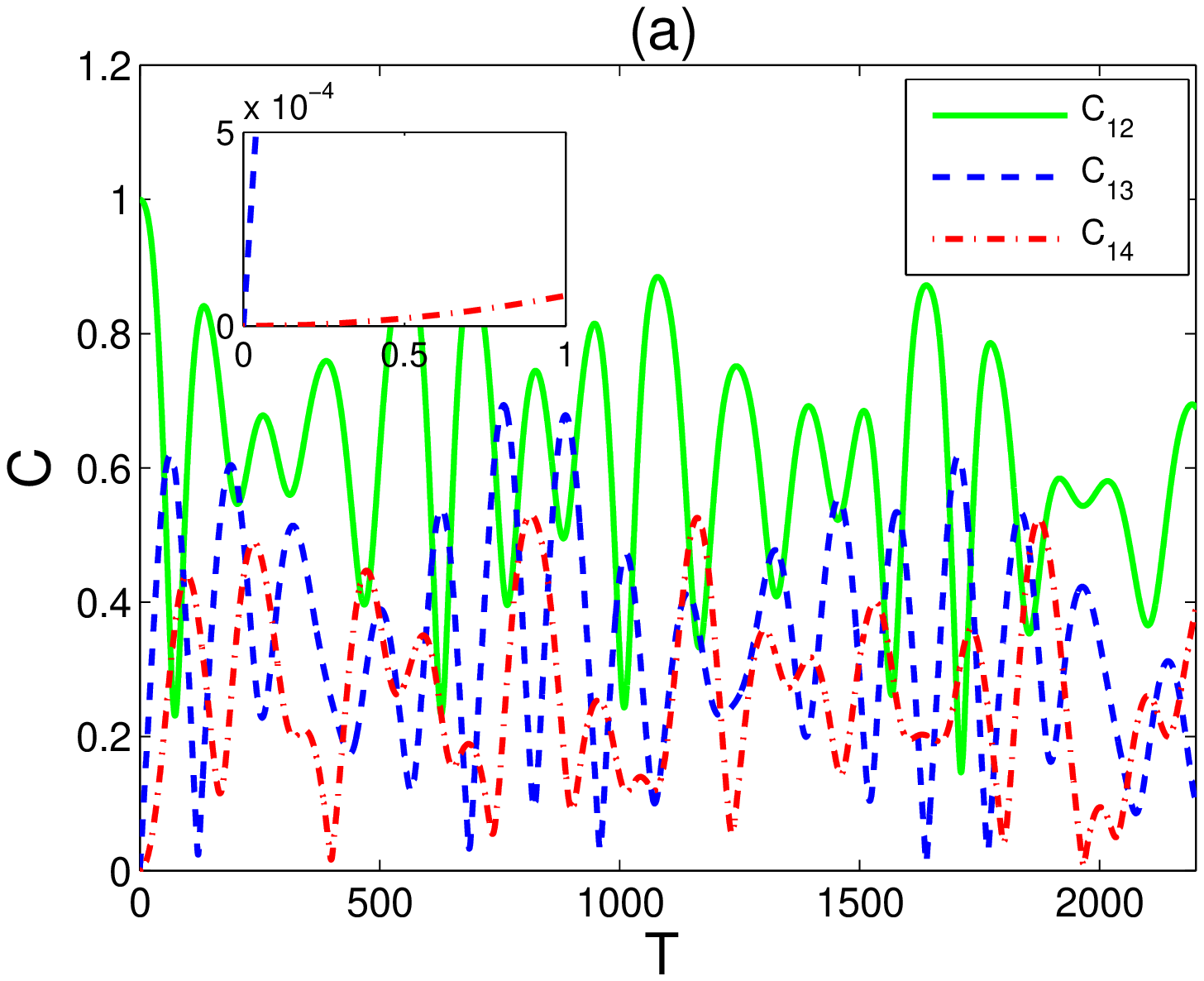}}\quad
   \subfigure{\includegraphics[width=8cm]{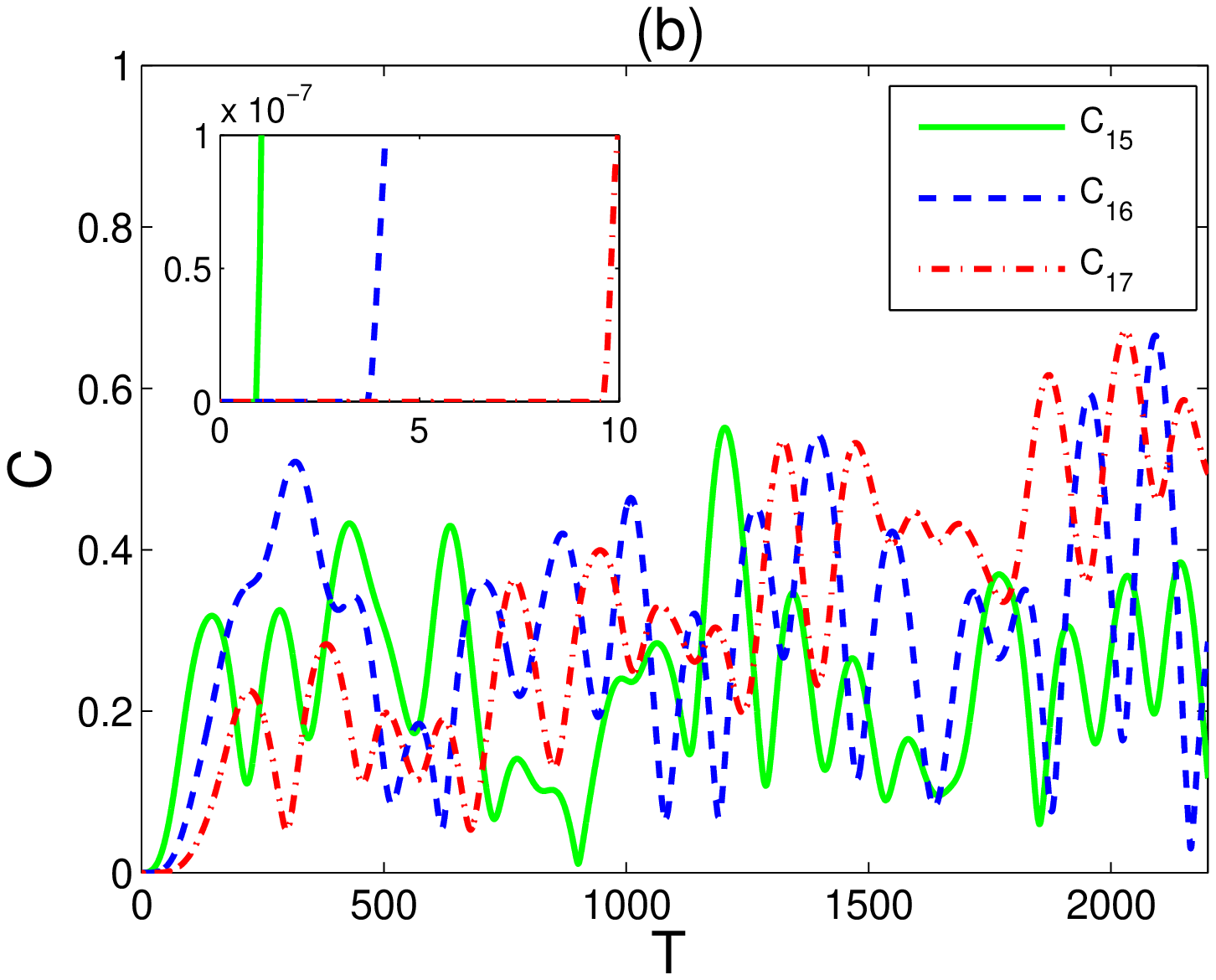}}\\
   \subfigure{\includegraphics[width=8cm]{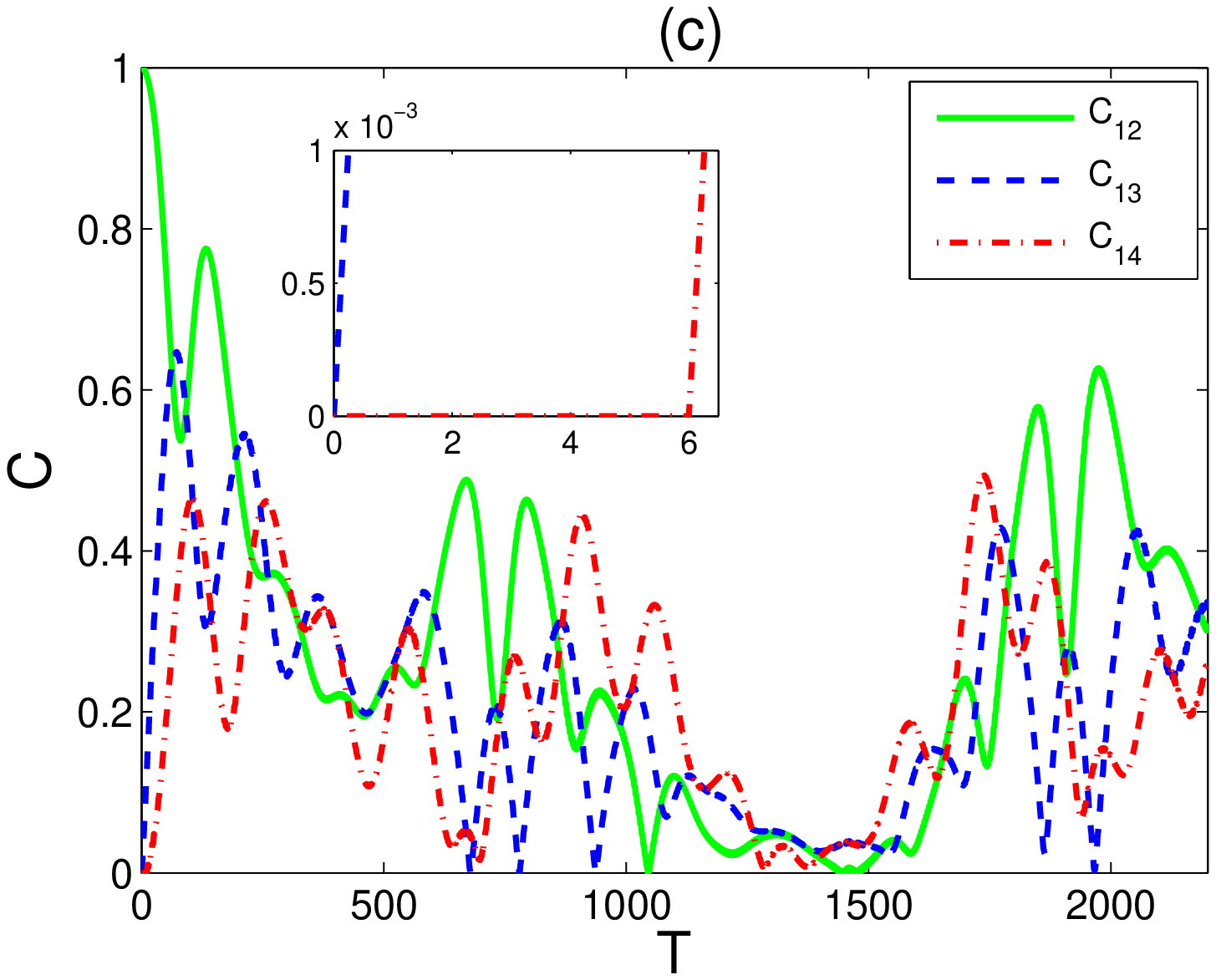}}\quad
   \subfigure{\includegraphics[width=8cm]{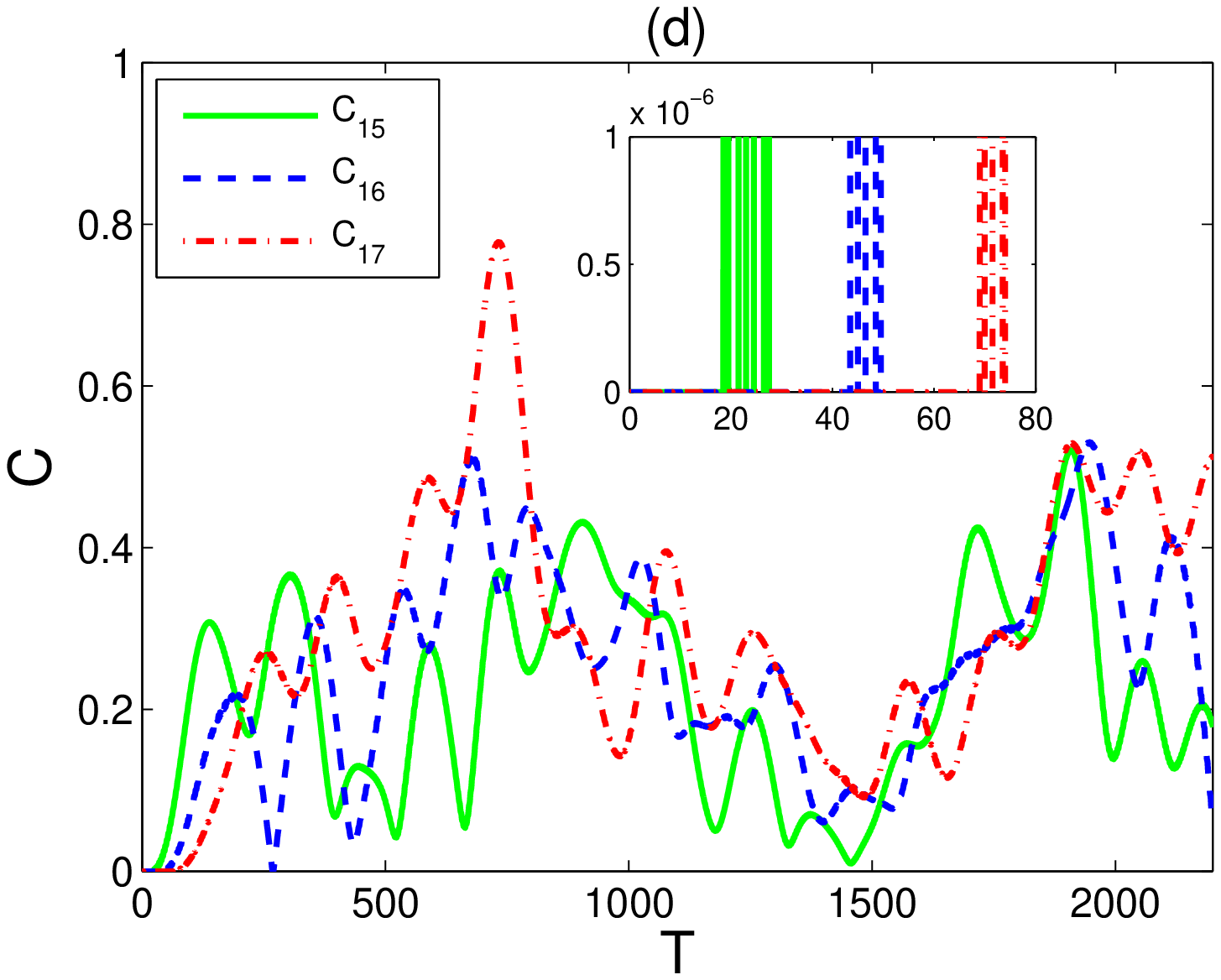}}\\
  \caption{{\label{N7_XXZ_XYZ_open_max_C_G0} The dynamics of the entanglements $C_{12}$, $C_{13}$, $C_{14}$, $C_{15}$, $C_{16}$ and $C_{17}$ starting from a maximally entangled state of the free $XXZ$ system ($\Gamma=0$) in (a) and (b) and the free $XYZ$ system in (c) and (d) with open boundary condition, where N=7. The inner panels illustrate the rise up of the entanglement from zero.}}
 \end{minipage}
\end{figure}
\begin{figure}[htbp]
\begin{minipage}[c]{\textwidth}
 \centering
   \subfigure{\includegraphics[width=8cm]{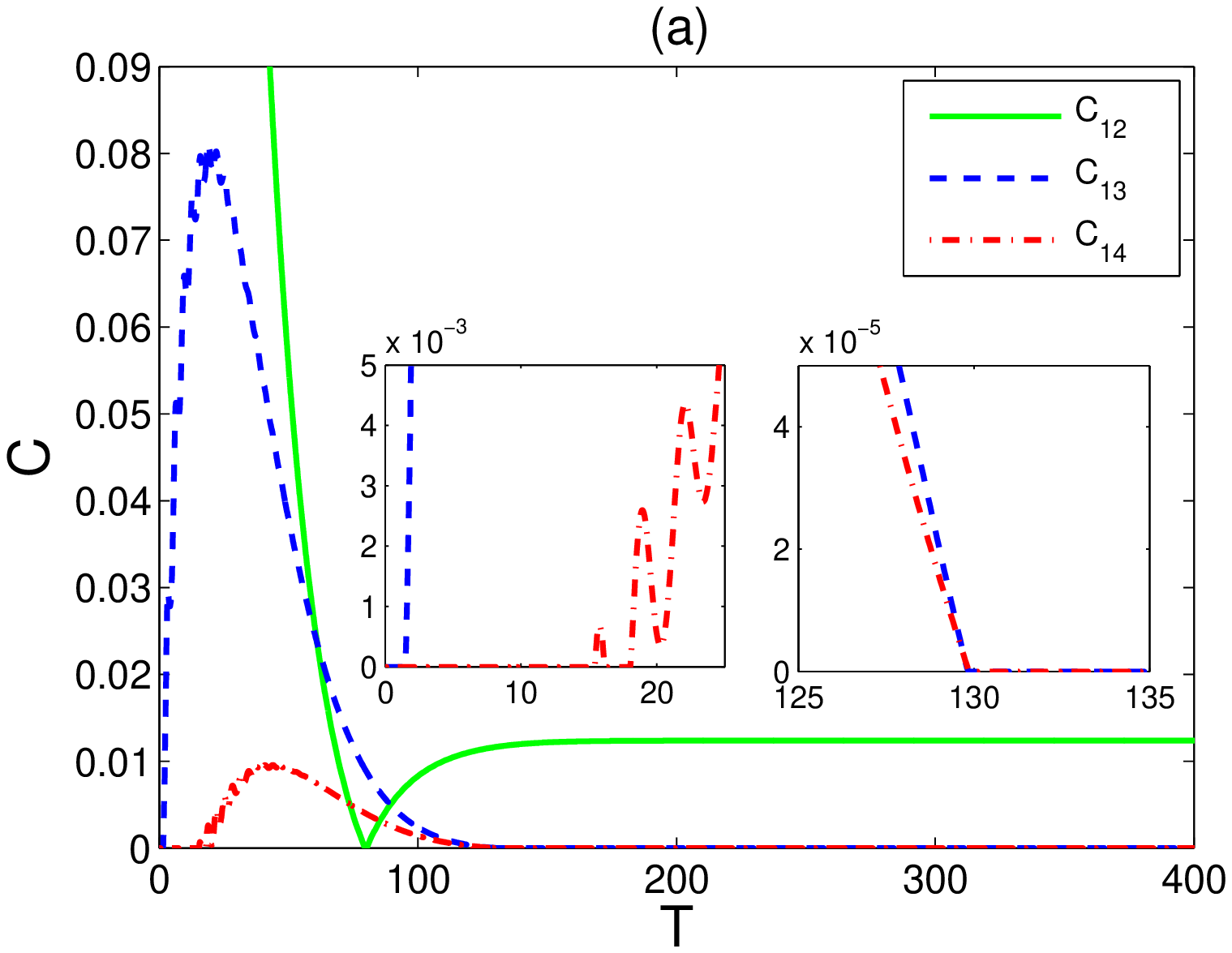}}\quad
   \subfigure{\includegraphics[width=8cm]{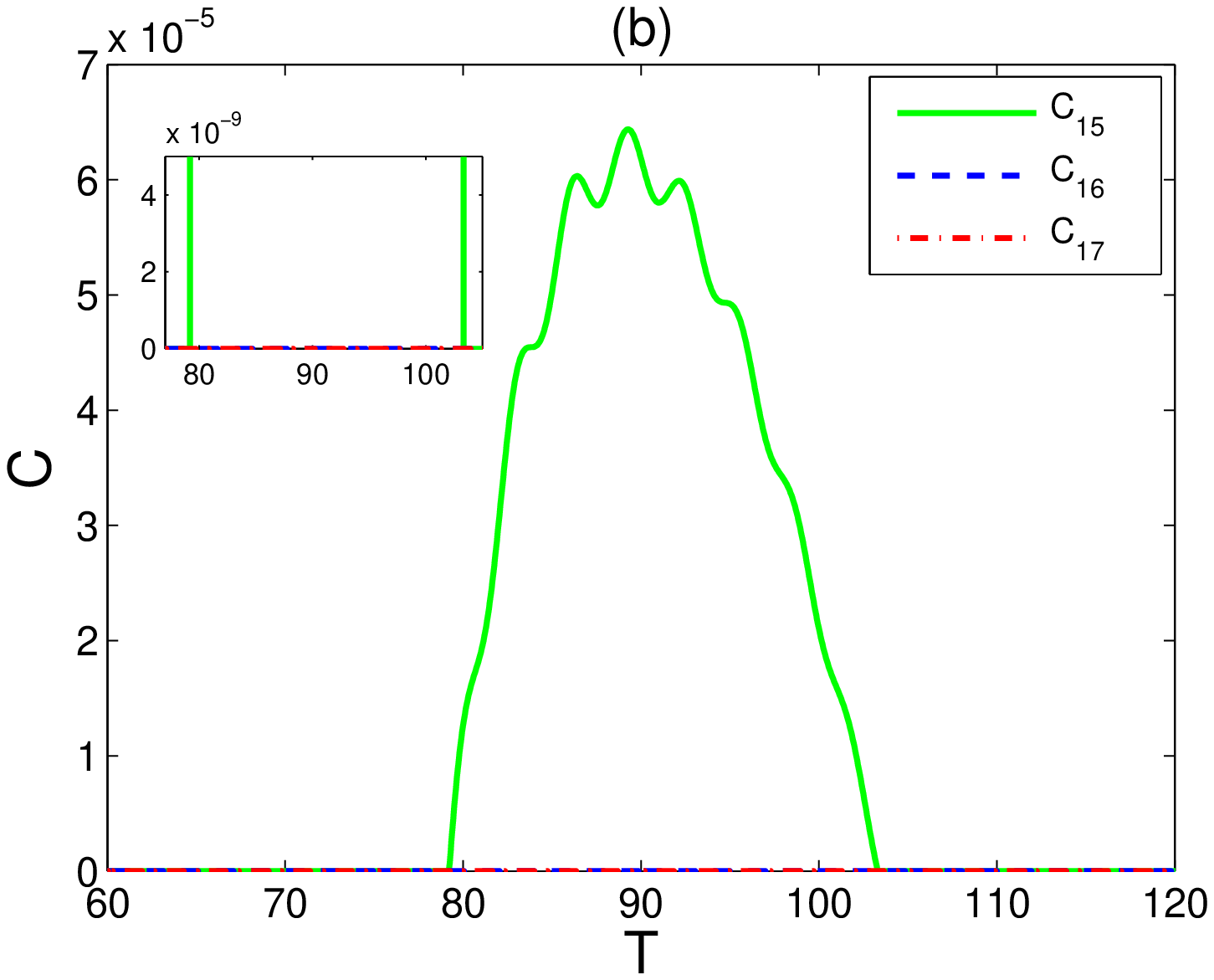}}\\
   \subfigure{\includegraphics[width=8cm]{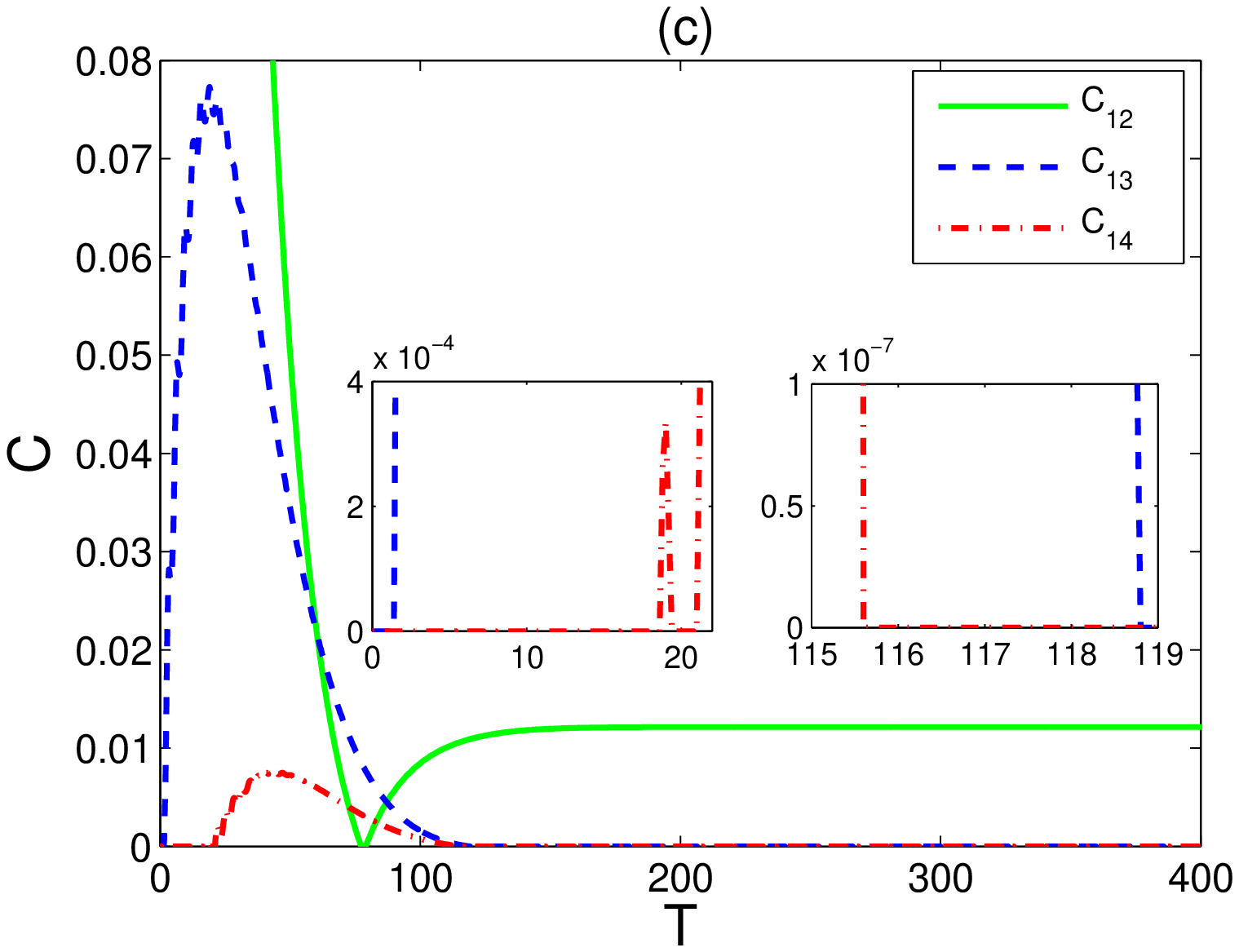}}\quad
   \subfigure{\includegraphics[width=8cm]{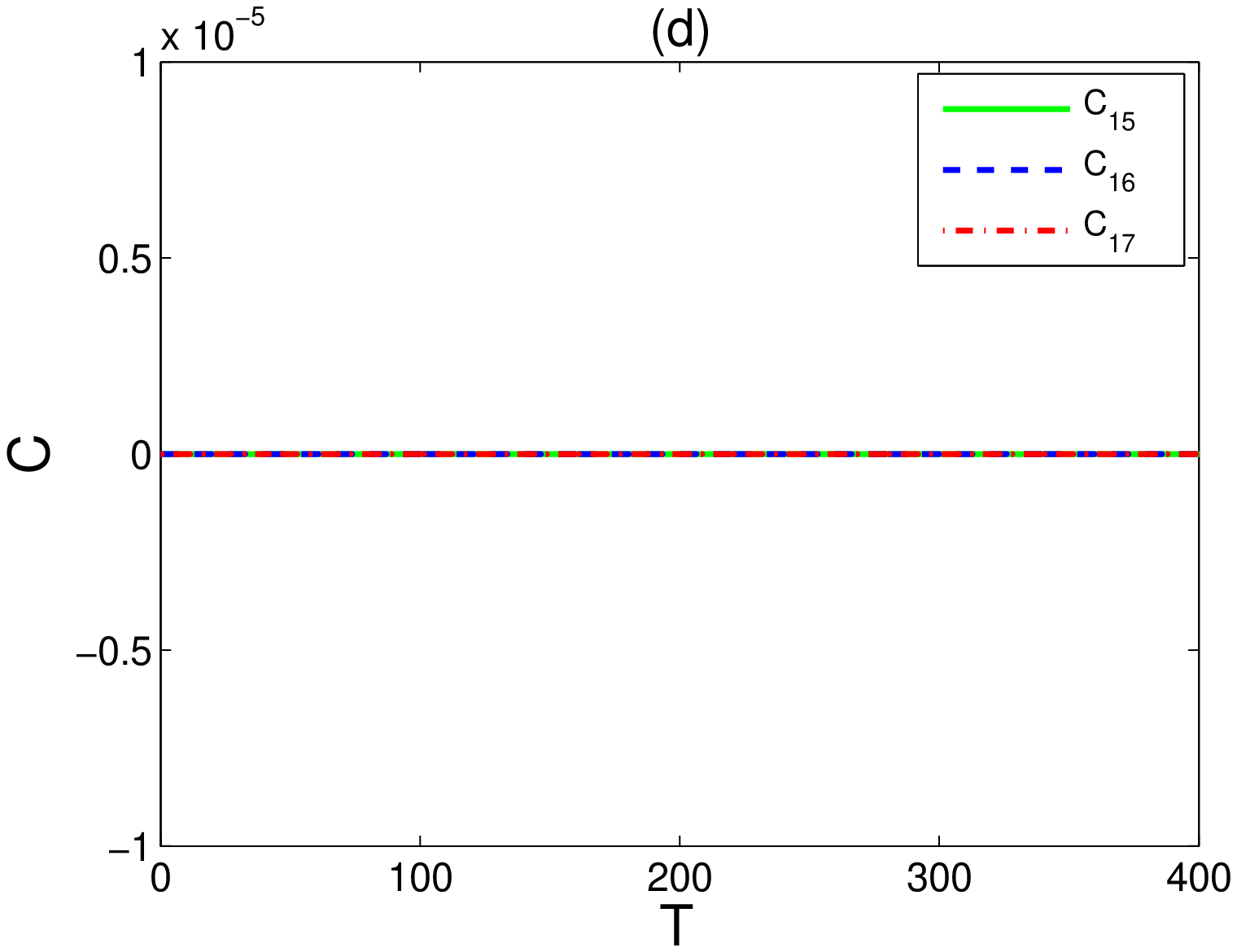}}\\
  \caption{{\label{N7_Ising_open_G05_n_0_001} The dynamics of the entanglements $C_{12}$, $C_{13}$, $C_{14}$, $C_{15}$, $C_{16}$ and $C_{17}$ starting from a maximally entangled state in the Ising system in presence of the environment ($\Gamma=0.05$) at $\bar{n}=0$ in (a) and (b) and $\bar{n}=0.01$ in (c) and (d), where N=7. In (a) and (c), the left inner panels illustrate the rise up of entanglement while the right ones illustrate its death. In (b) the inner panel shows both of the rise up and death of entanglement.}}
 \end{minipage}
\end{figure}
In this section, we start by investigating the entanglement dynamics and transfer in one-dimensional free Heisenberg spin chains with open boundary condition.
\begin{figure}[htbp]
\begin{minipage}[c]{\textwidth}
 \centering
  \subfigure{\includegraphics[width=8cm]{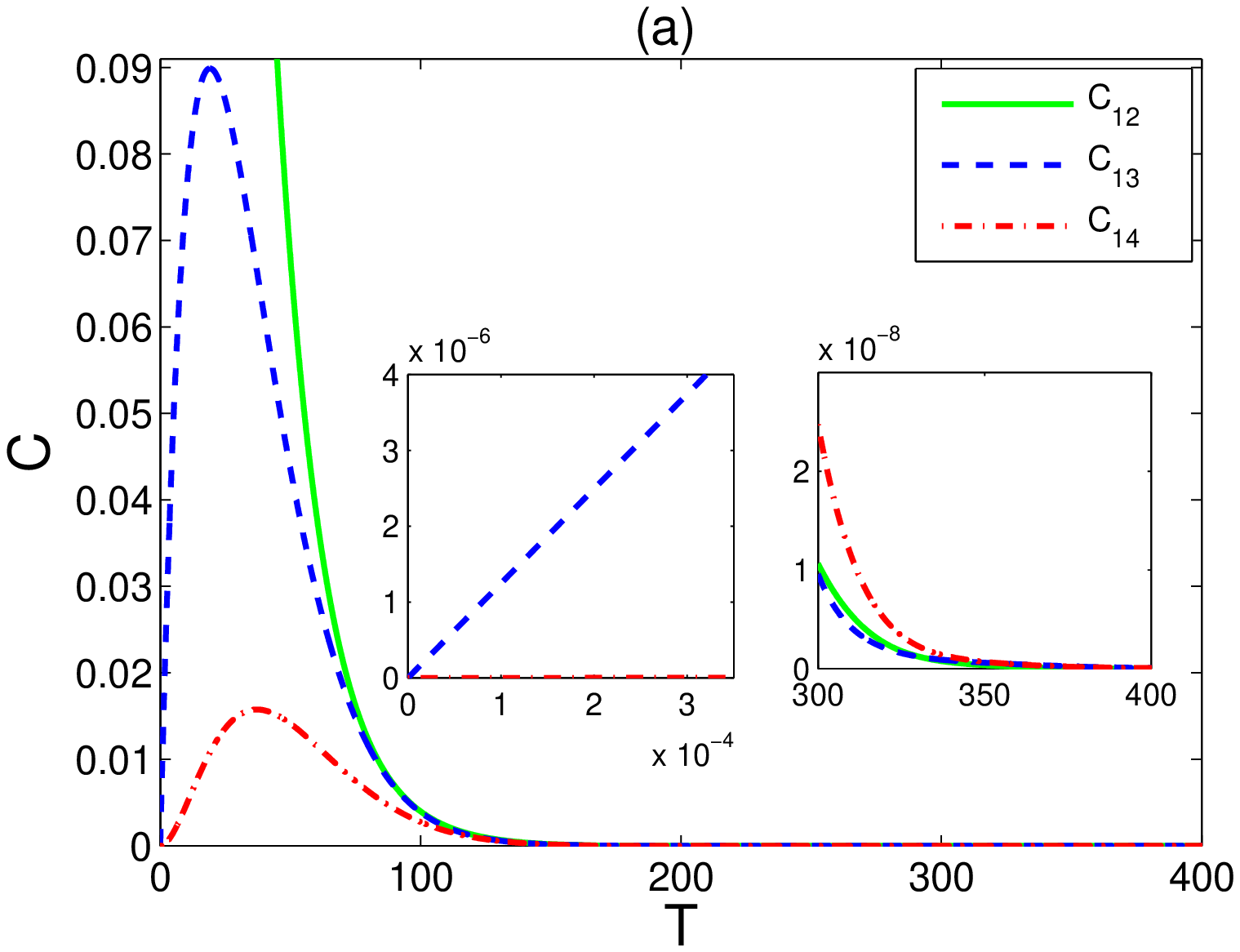}}\quad
  \subfigure{\includegraphics[width=8cm]{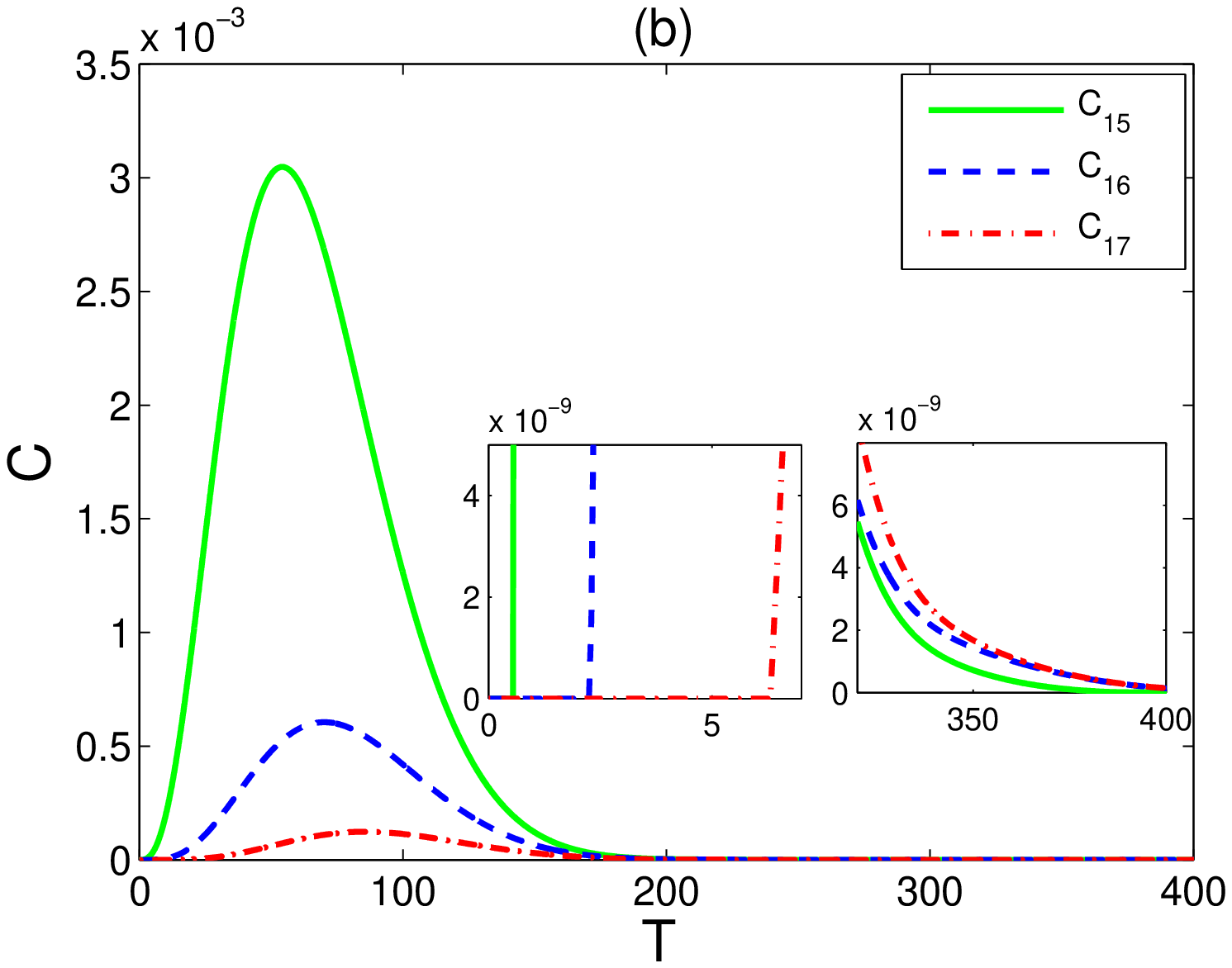}}\\
  \subfigure{\includegraphics[width=8cm]{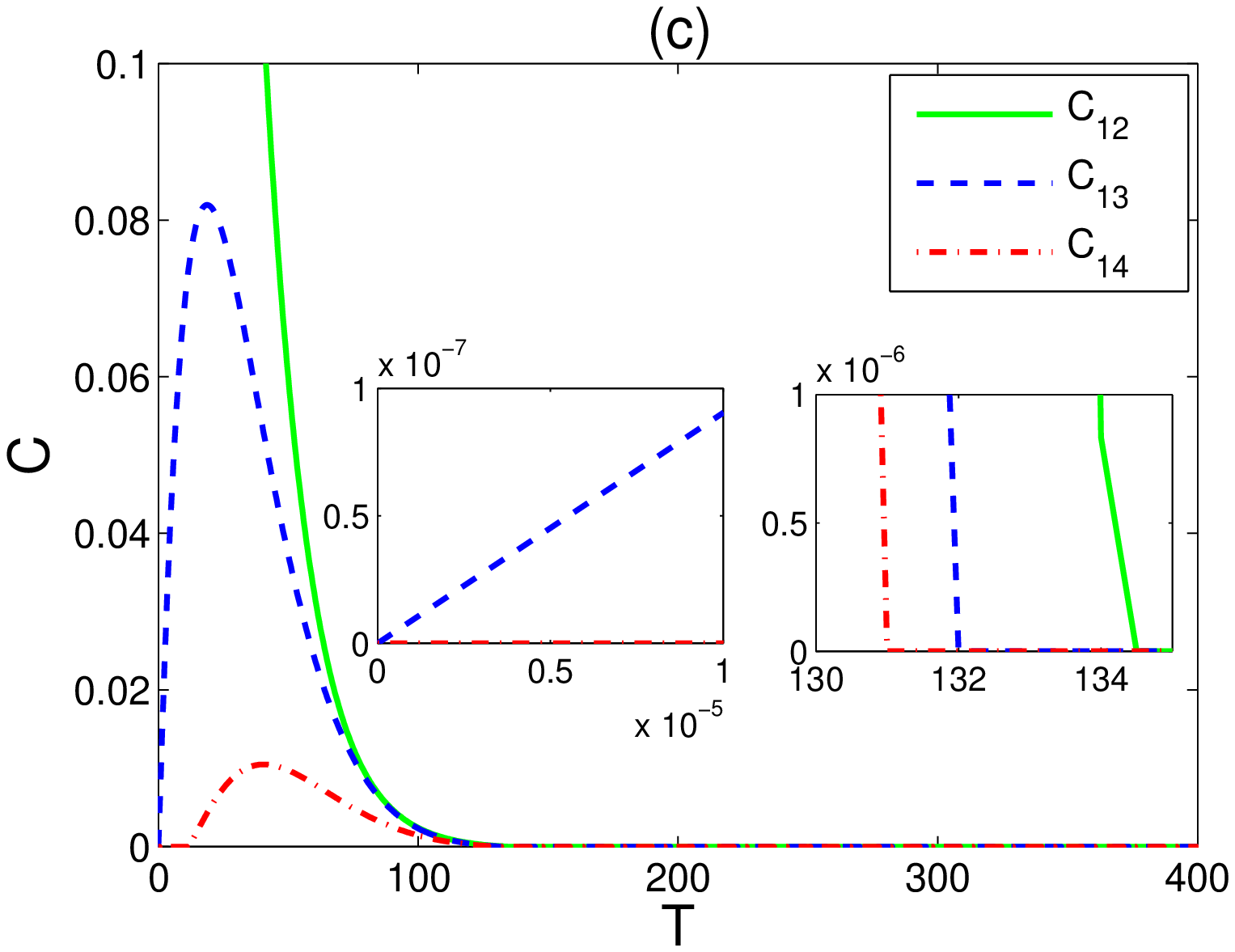}}\quad
  \subfigure{\includegraphics[width=8cm]{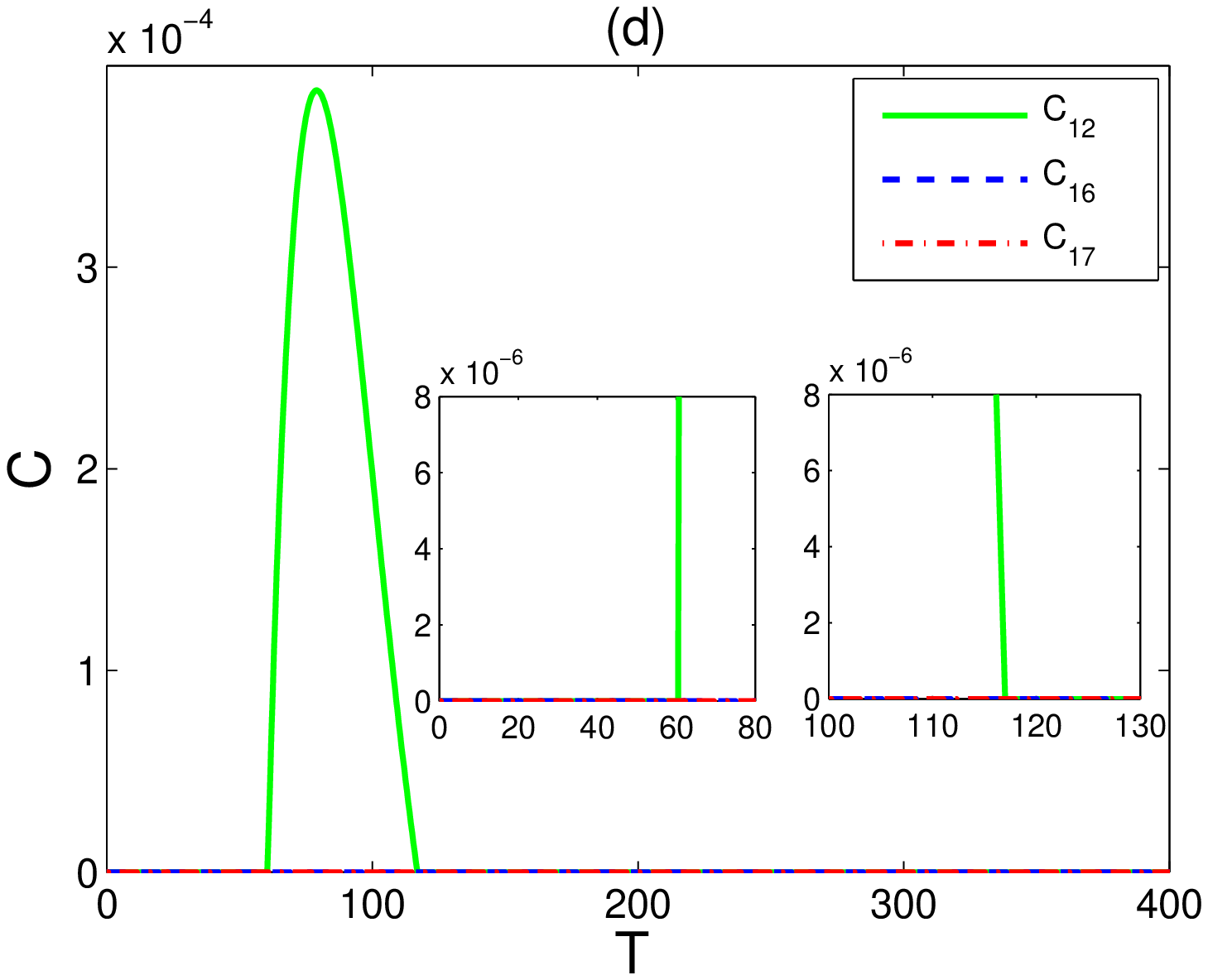}}\\
  \caption{{\label{N7_XX_open_G05_n_0_001} The dynamics of the entanglements $C_{12}$, $C_{13}$, $C_{14}$, $C_{15}$, $C_{16}$ and $C_{17}$ starting from a maximally entangled state in the XX system in presence of the environment ($\Gamma=0.05$) at $\bar{n}=0$ in (a) and (b) and $\bar{n}=0.01$ in (c) and (d), where N=7. The left inner panels illustrate the rise up of entanglement while the right ones illustrate its death.}}
 \end{minipage}
\end{figure}
The system is initially prepared in a state with two spins (1 and 2) at one end of the chain maximally entangled with each other and are completely disentangled from the rest of the spins in the chain, which are also disentangled from each other. We start with the Ising 7-spins system, which is explored in fig.~\ref{N7_Ising_XX_open_max_C_G0}(a) and (b). The (nn) entanglement $C_{12}$ starts at $t=0$ with a value of 1 but decays to zero before reviving and showing a non-uniform oscillatory behavior. The longer range entanglements $C_{13}$, $C_{14}$, $C_{15}$ $C_{16}$ and $C_{17}$, illustrated in fig.~\ref{N7_Ising_XX_open_max_C_G0}(a) and (b), start with a zero value at $t=0$ before rising up at latter times, the longer the range of the entanglement is the longer it takes to rise up as shown in the inner panels of fig.~\ref{N7_Ising_XX_open_max_C_G0}(a) and (b). In fig.~\ref{N7_Ising_XX_open_max_C_G0}(c) and (d), we turn to the entanglement dynamics in the $XX$ system, which shows one significant difference from what we have observed in the Ising system, the (nnn) entanglement $C_{13}$ starts to rise up immediately at $t=0$ but the other long range entanglements are delayed but not for as long as they were in the Ising system, as shown in the inner panels of the figure. Right after the different start all the entanglements, nearest neighbor and beyond, show very close profile of oscillation to that of the Ising case. So the entanglement dynamics in these cases are asymptotically very close.

To test the effect of the spin coupling in the $z-$direction on the entanglement transfer dynamics, we explore the $XXZ$ chain in fig.~\ref{N7_XXZ_XYZ_open_max_C_G0}(a) and (b) and the $XXZ$ chain in fig.~\ref{N7_XXZ_XYZ_open_max_C_G0}(c) and (d). Clearly, there is a good resemblance between the rise up of the beyond nearest neighbor entanglement in the $XX$ and $XXZ$ systems but asymptotically they have different oscillation profile. On the other hand, the $XYZ$ system has $C_{13}$ rising up immediately from zero but the other longer range entanglements rise up much latter compare with the previous cases with very strong oscillation and also show a different asymptotic oscillation profile.
\subsection{Entanglement transfer in presence of the environment}
Now we turn to examine the effect of the dissipative environment and thermal excitations on the entanglement transfer through the open boundary Heisenberg spin chains. We start with the Ising system at zero temperature in fig.~\ref{N7_Ising_open_G05_n_0_001}(a) and (b), where as can be noticed the (nn) entanglement $C_{12}$ starts from a maximum value of one and decays very rapidly to zero before reviving again to reach a steady state value of about 0.0123. The (nnn) entanglement $C_{13}$ is not created at $T=0$ but a very short time latter, $T\approx 1.5$, and rises up as shown in the left inner panel in fig.~\ref{N7_Ising_open_G05_n_0_001}(a), it reaches a maximum value, $C\approx 0.08$, before decaying again and vanishing at around $T=130$ as shown in the right inner panel of fig.~\ref{N7_Ising_open_G05_n_0_001}(a).
\begin{figure}[htbp]
\begin{minipage}[c]{\textwidth}
 \centering
   \subfigure{\includegraphics[width=8cm]{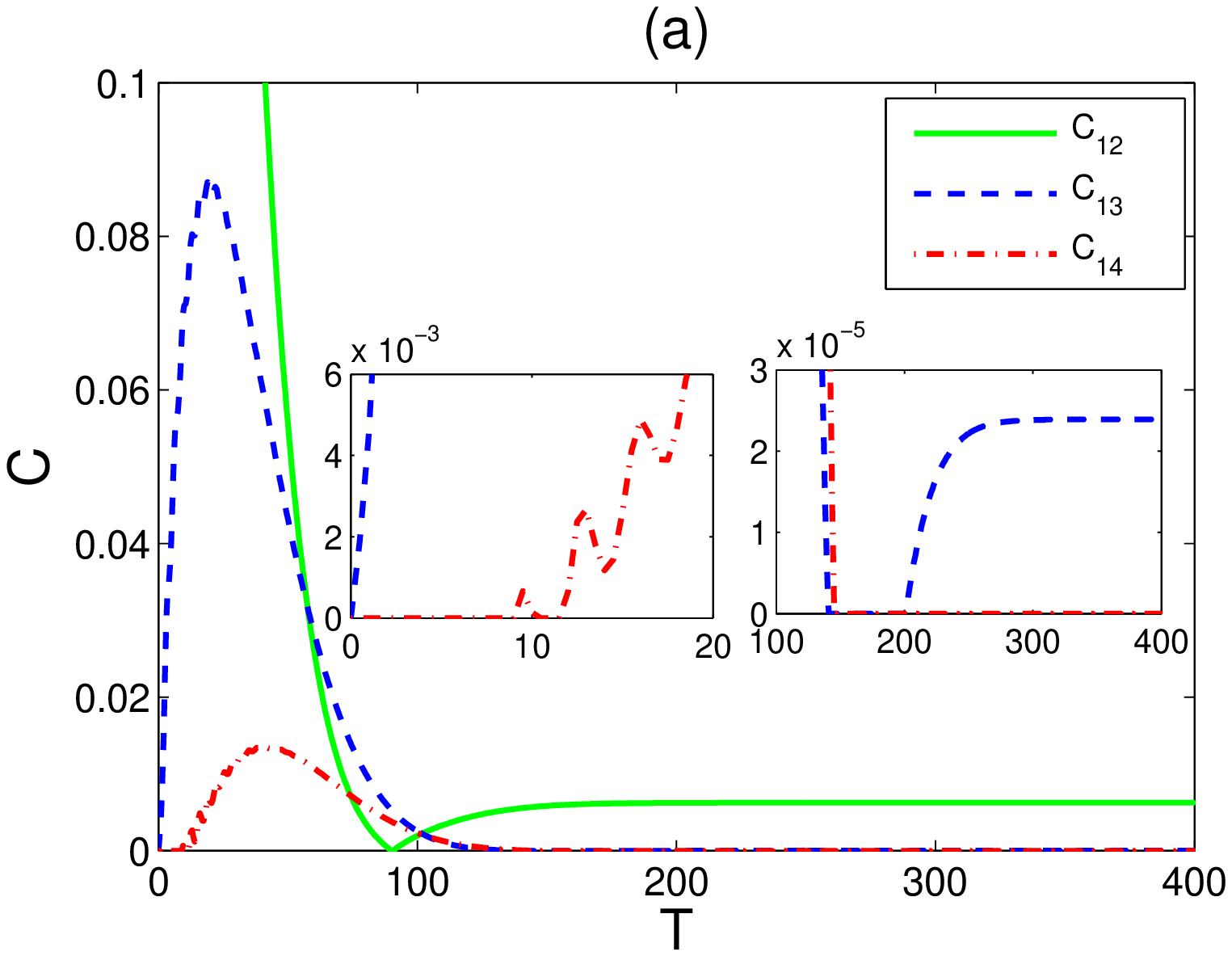}}\quad
   \subfigure{\includegraphics[width=8cm]{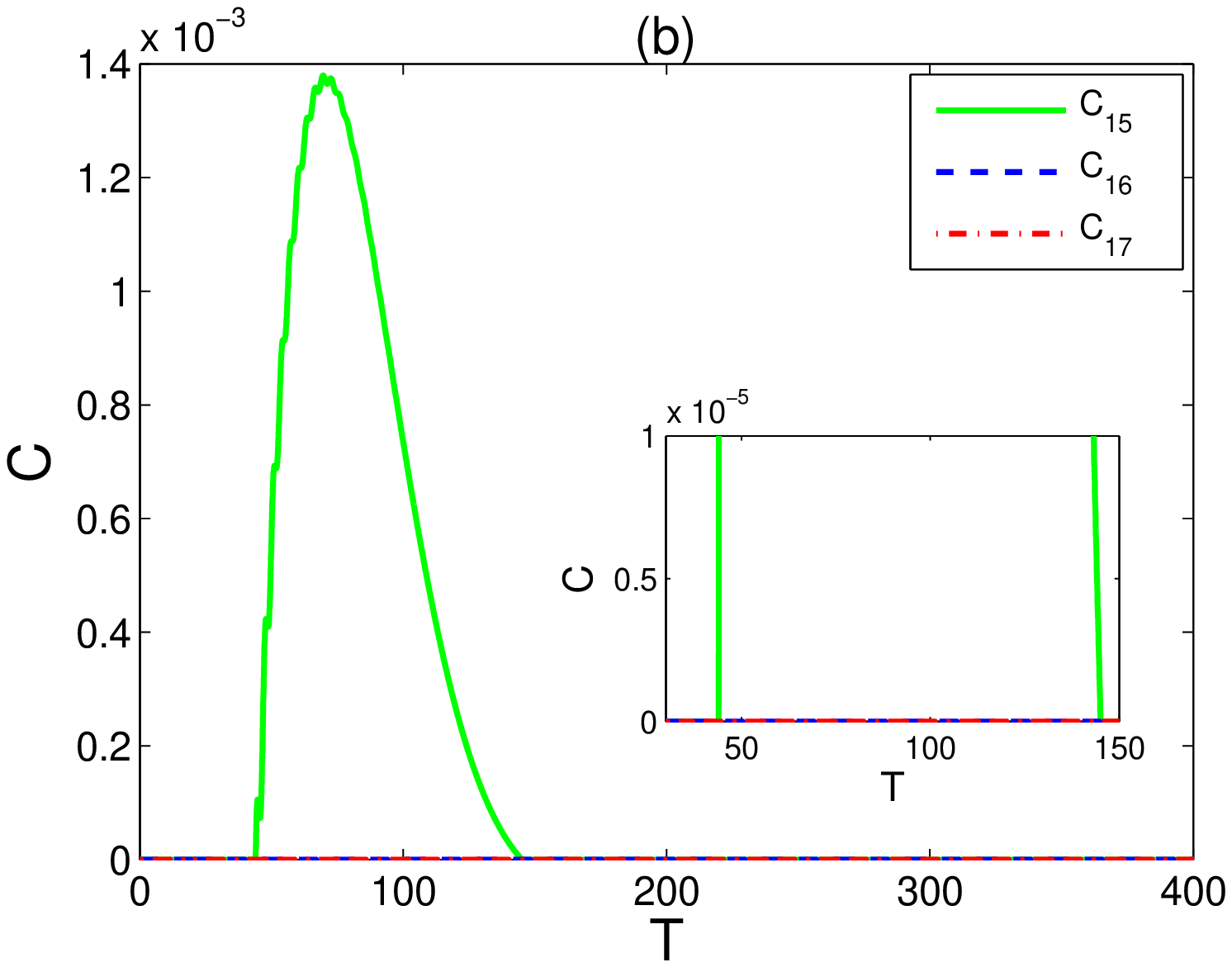}}\\
   \subfigure{\includegraphics[width=8cm]{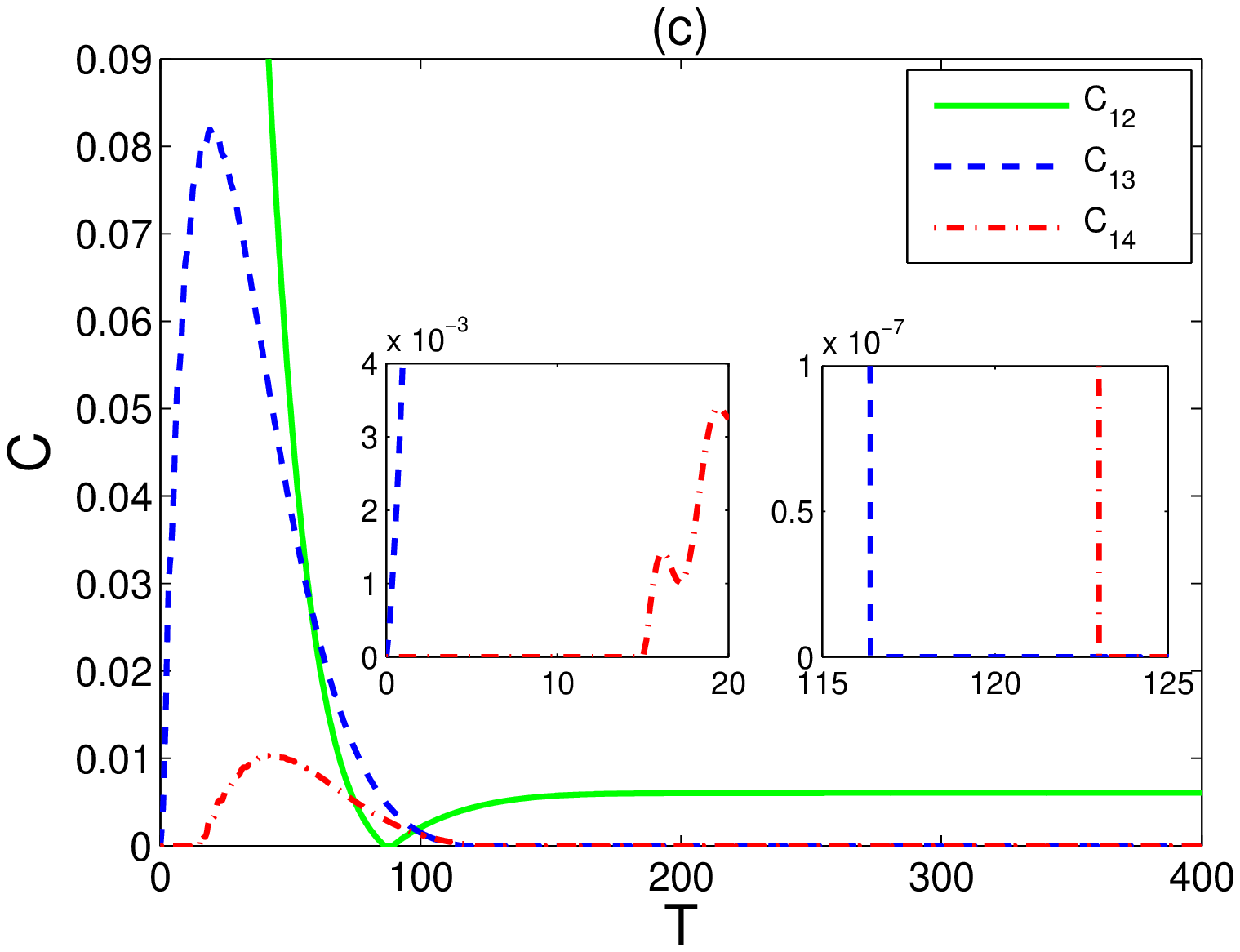}}\quad
   \subfigure{\includegraphics[width=8cm]{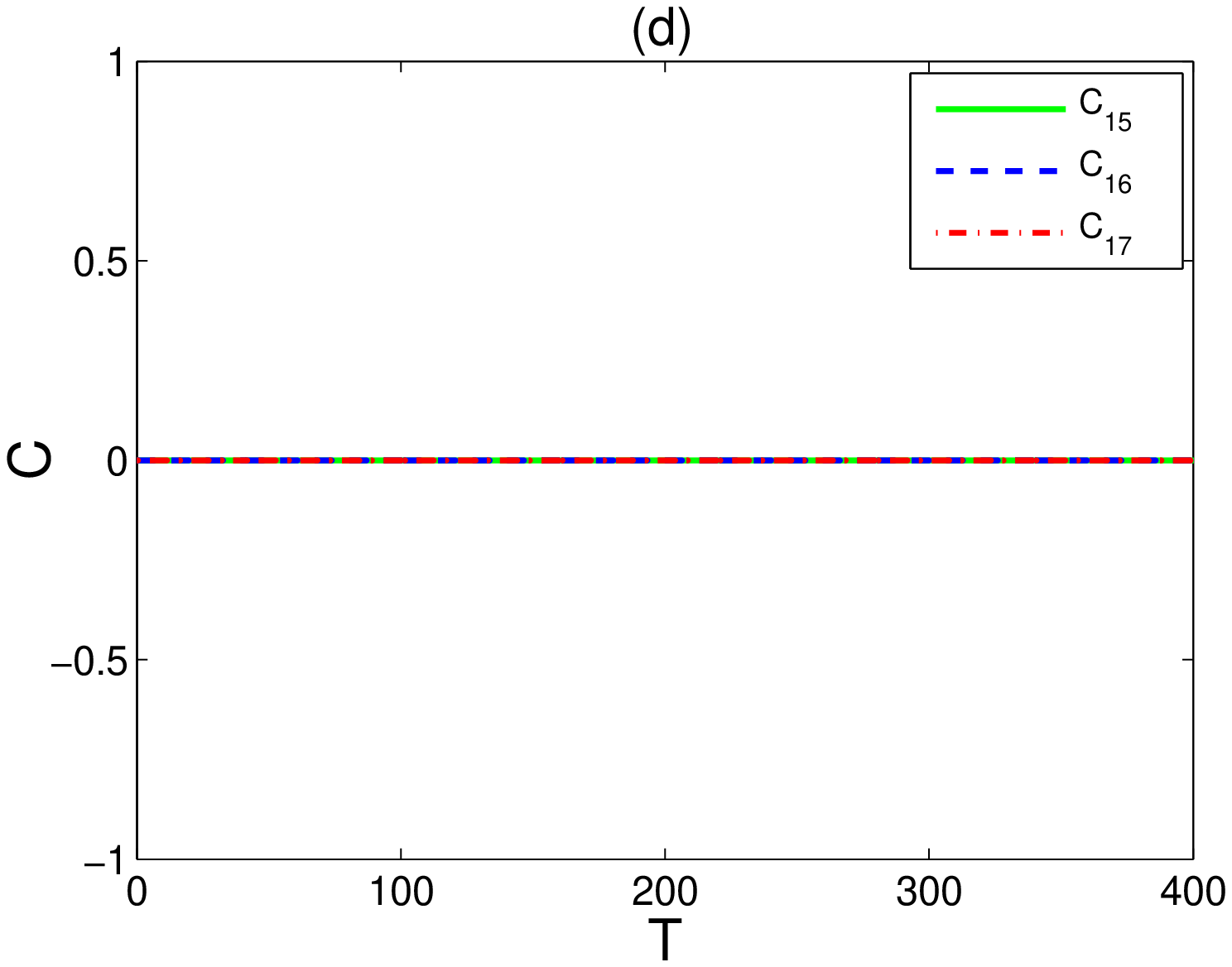}}\\
  \caption{{\label{N7_XYZ_open_G05_n_0_001} The dynamics of the entanglements $C_{12}$, $C_{13}$, $C_{14}$, $C_{15}$, $C_{16}$ and $C_{17}$ starting from a maximally entangled state in the XYZ system in presence of the environment at $\bar{n}=0$ in (a) and (b) and $\bar{n}=0.01$ in (c) and (d), where N=7. In (a) and (c), the left inner panels illustrate the rise up of entanglement while the right ones illustrate its death. In (b) the inner panel shows both of the rise up and death of entanglement.}}
 \end{minipage}
\end{figure}
The (nnnn) entanglement $C_{14}$ shows a very similar behavior to that of $C_{13}$, where it starts at a latter time $T\approx 16$ with a strong oscillation reaching a max value of about $0.0045$, then decaying and vanishing at about the same time as $C_{13}$, which is presented in the right inner panels of fig.~\ref{N7_Ising_open_G05_n_0_001}(a). The (nnnn) entanglement $C_{15}$ starts at even latter time $T \approx 78$, as illustrated in fig.~\ref{N7_Ising_open_G05_n_0_001}(b), and increases to reach a maximum value before decaying and vanishing at $T \approx 104$. The entanglements beyond $C_{15}$ are zero and never rise up. The effect of the finite temperature, $\bar{n}=0.01$, on the Ising system is tested in fig.~\ref{N7_Ising_open_G05_n_0_001}(c) and (d). The overall behavior of the entanglements $C_{12}$, $C_{13}$ and $C_{14}$ is very close to the zero temperature case, the main changes are the reduction in the maximum values of the entanglements and the vanishing times of $C_{13}$ $C_{14}$ become different from each other and earlier than before. The entanglement functions $C_{15}$, $C_{16}$, and $C_{17}$ never rise up from zero as the temperature is raised as can be noticed in fig.~\ref{N7_Ising_open_G05_n_0_001}(d). Of course, as the temperature is raised further all the entanglements vanish.
\begin{figure}[htbp]
\begin{minipage}[c]{\textwidth}
 \centering
   \subfigure{\includegraphics[width=8cm]{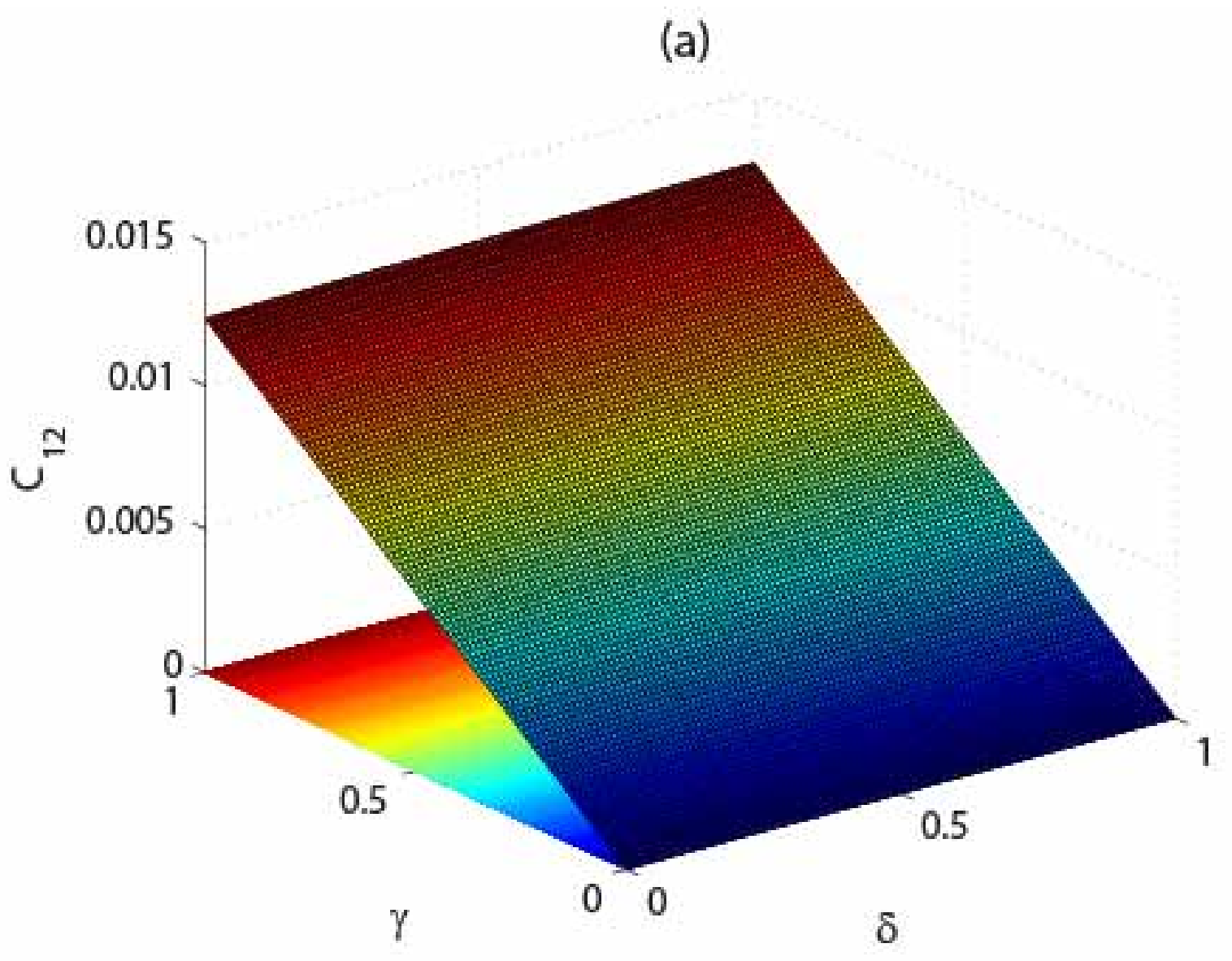}}\quad
   \subfigure{\includegraphics[width=8cm]{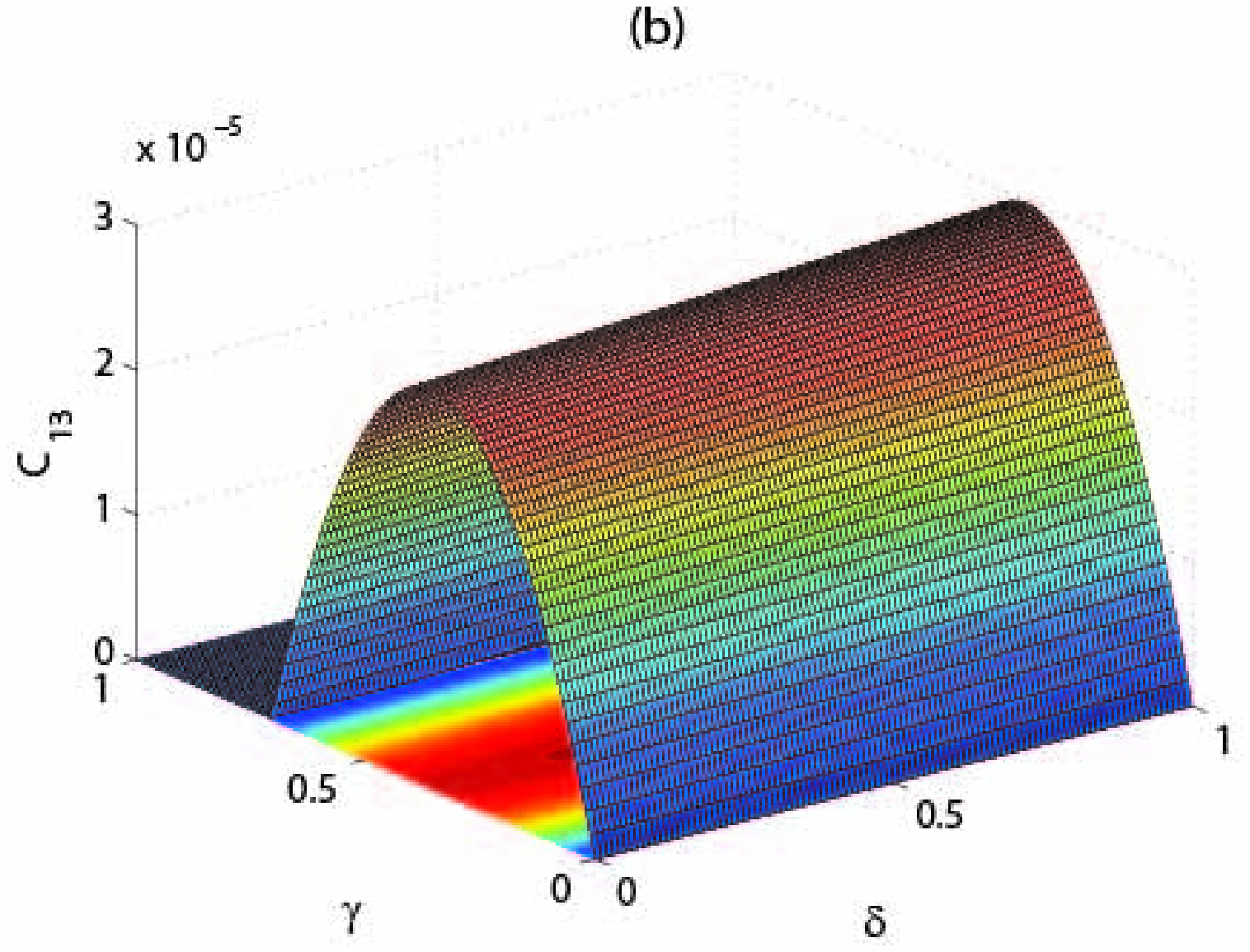}}\\
   \subfigure{\includegraphics[width=8cm]{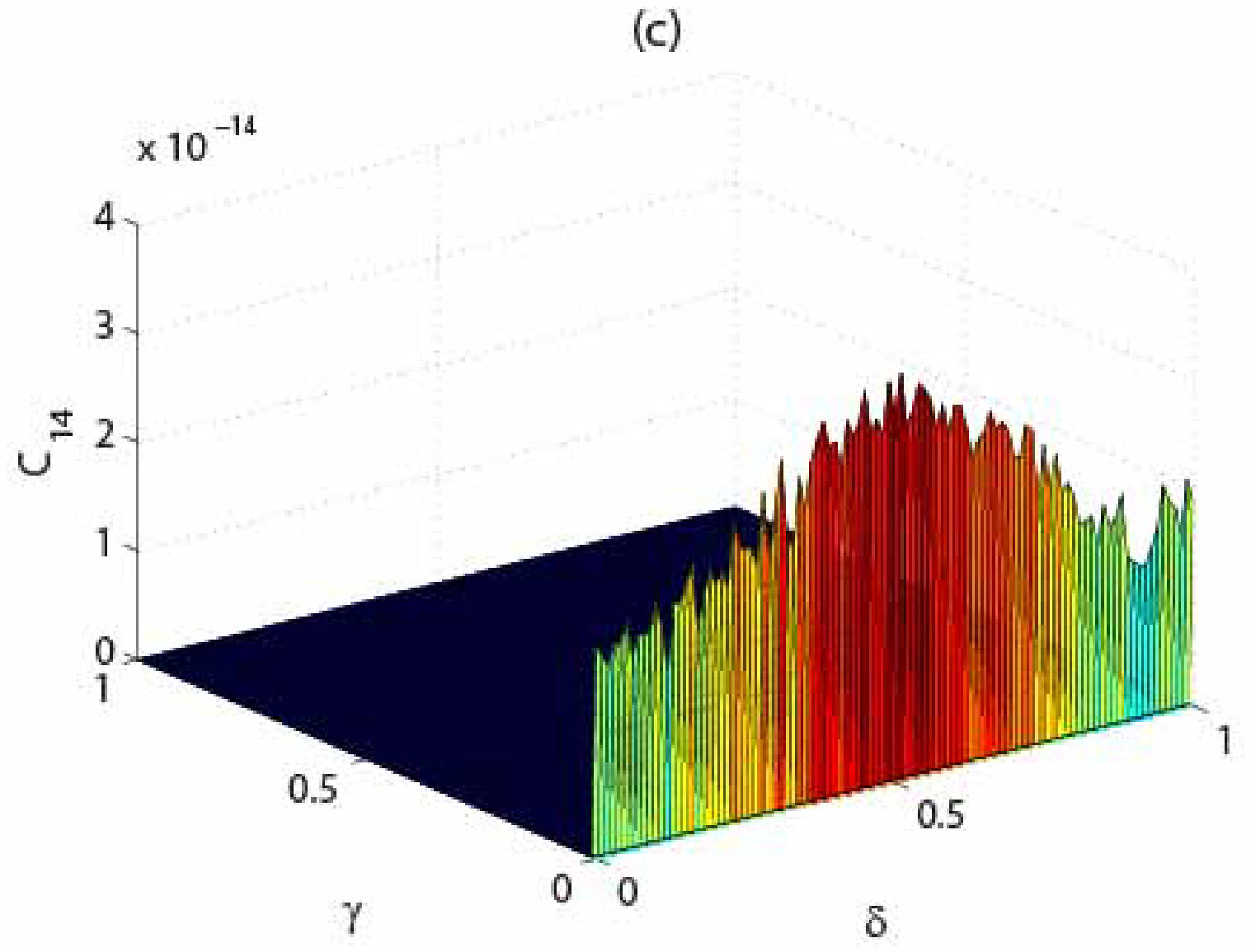}}\quad
   \subfigure{\includegraphics[width=8cm]{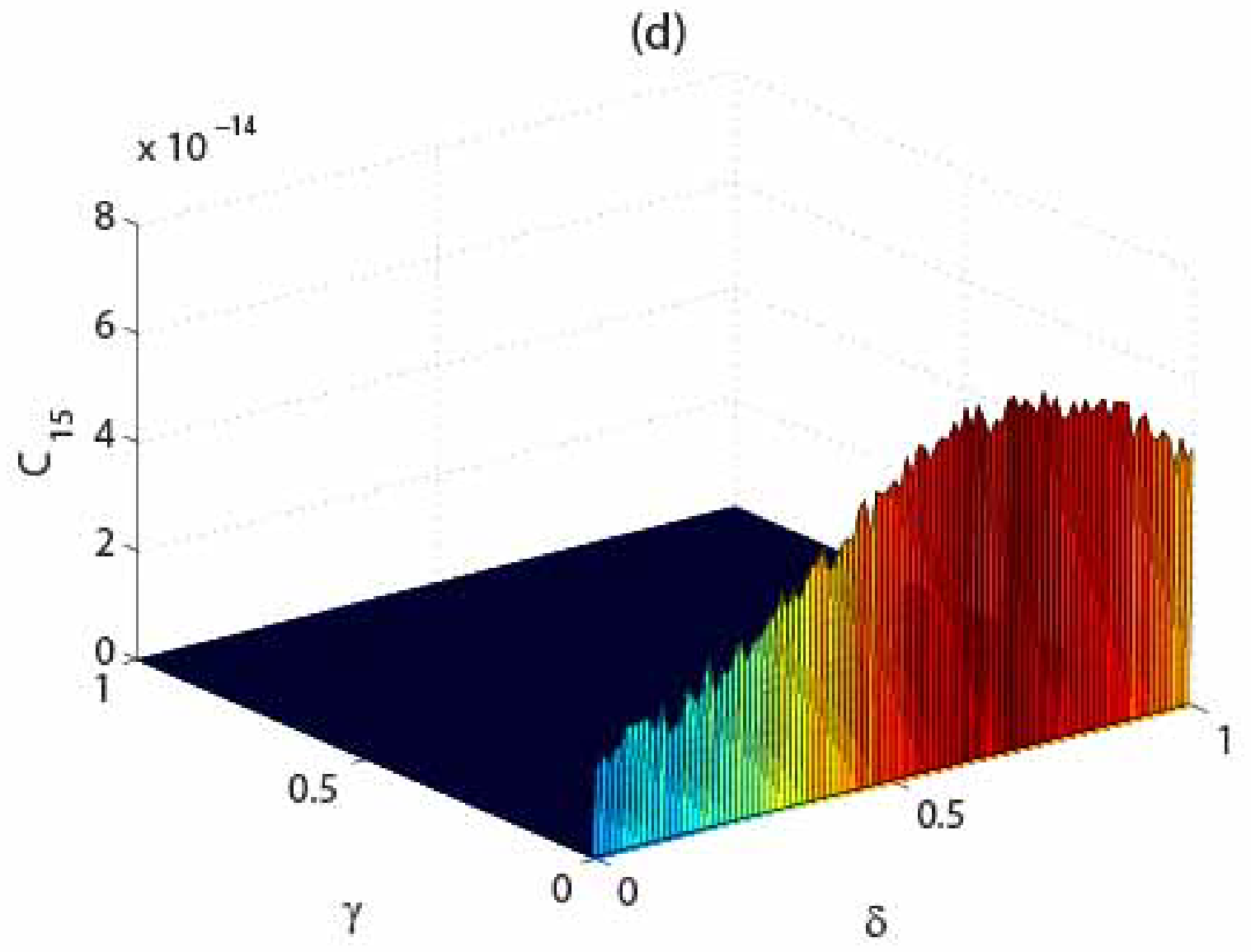}}\\
  \caption{{\label{N5_open_C_vs_gamma_delta_G05_n0} The asymptotic behavior of (a) $C_{12}$; (b) $C_{13}$ and the value of (c) $C_{14}$; (d) $C_{15}$ (at $T=300$) in $\gamma-\delta$ space of the Heisenberg $XYZ$ system in presence of the environment ($\Gamma=0.05$) starting from any initial state (disentangled, entangled or maximally entangled) at zero temperature, where N=5.}}
\end{minipage}
\end{figure}

The entanglement transfer in the $XX$ spin chain is explored in fig.~\ref{N7_XX_open_G05_n_0_001}. In contrary to the Ising case, at zero temperature, the (nnn) entanglement $C_{13}$ starts to rise up immediately at $t=0$ with no delay, as illustrated in fig.~\ref{N7_XX_open_G05_n_0_001}(a) and the left inner panel, but the other far entanglements, $C_{14}$, $C_{15}$, and even $C_{16}$ and $C_{17}$ start up latter on one after the other as shown in fig.~\ref{N7_XX_open_G05_n_0_001}(a) and (b). But all entanglements decay asymptotically and vanish. As the temperature is raised, $\bar{n}=0.01$, illustrated in fig.~\ref{N7_XX_open_G05_n_0_001}(c) and (d), the (nnn) entanglement $C_{13}$ still rises up at $t=0$ whereas $C_{14}$ and $C_{15}$ are created latter and $C_{16}$ and $C_{17}$ remain zero at all times.
There is a significant change in the behavior of entanglement transfer in the $XYZ$ system, depicted in fig.~\ref{N7_XYZ_open_G05_n_0_001}, the (nnn) entanglement $C_{13}$ at zero temperature reaches a steady state asymptotically, exactly like the (nn) entanglement $C_{12}$, as shown in fig.~\ref{N7_XYZ_open_G05_n_0_001}(a). The steady state of $C_{13}$ vanishes as the temperature is raised, $\bar{n}=0.01$, in contrary to that of $C_{12}$, which shows more robustness as illustrated in fig.~\ref{N7_XYZ_open_G05_n_0_001}(c). The time evolution of the longer range entanglements $C_{15}$, $C_{16}$ and $C_{17}$ is shown in fig.~\ref{N7_XYZ_open_G05_n_0_001}(d), where they never rise up from zero. Clearly, the degrees of anisotropy not only play a major role in controlling the entanglement transfer dynamics in the Heisenberg spin chains but also affect the different pairwise entanglements in different ways.
\begin{figure}[htbp]
\begin{minipage}[c]{\textwidth}
 \centering
 \subfigure{\includegraphics[width=8cm]{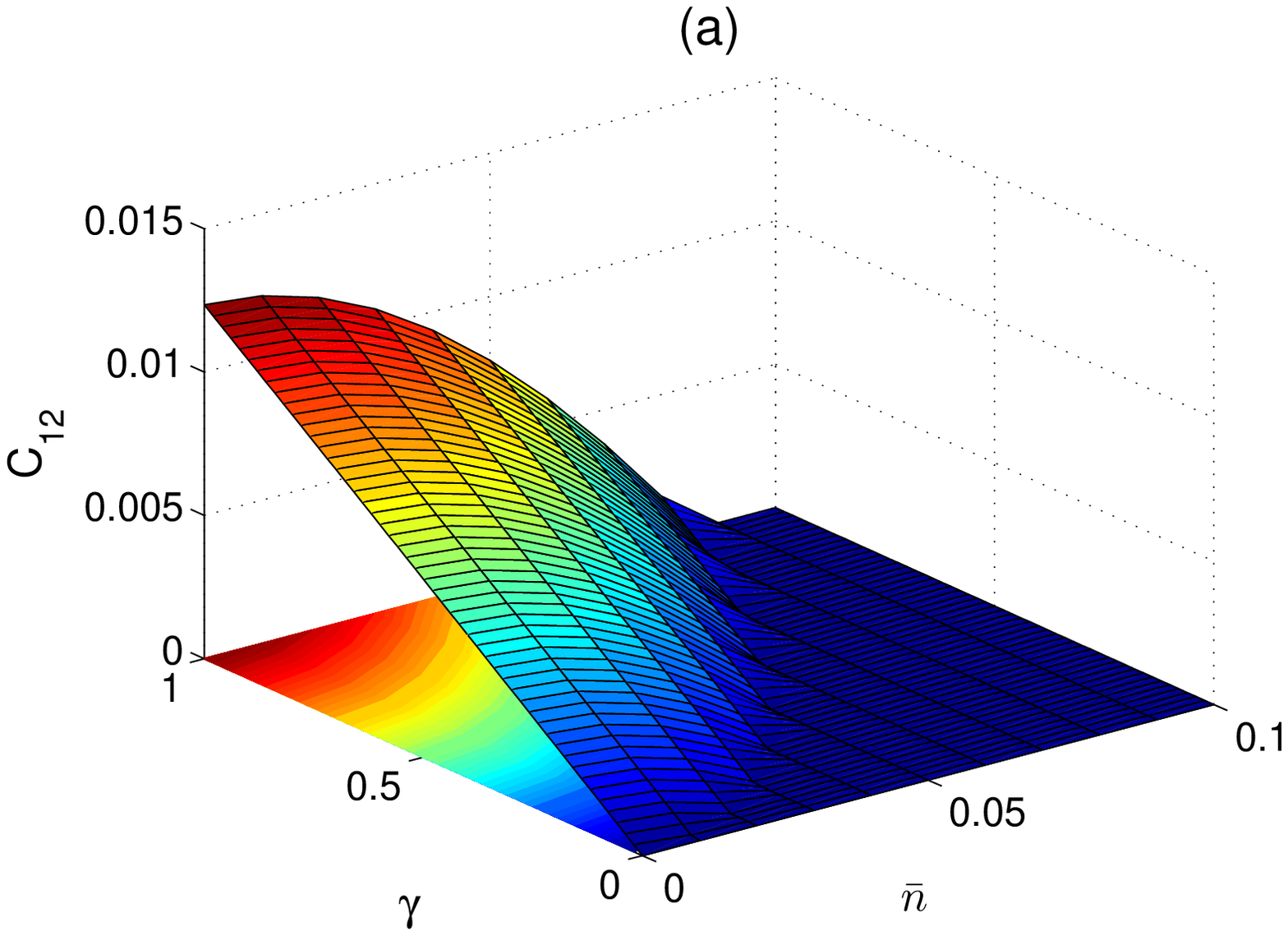}}\quad 
 \subfigure{\includegraphics[width=8cm]{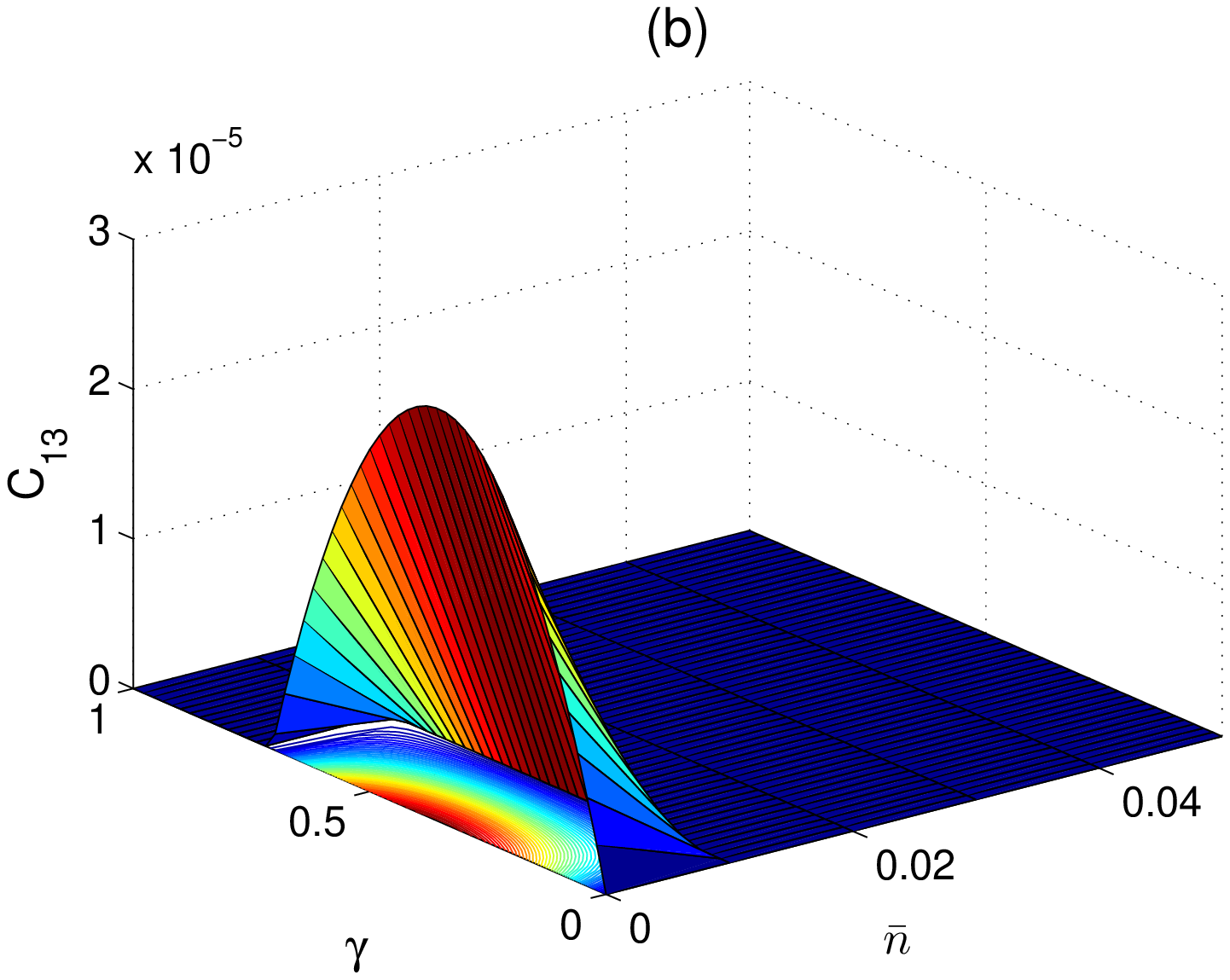}}\\
 \subfigure{\includegraphics[width=8cm]{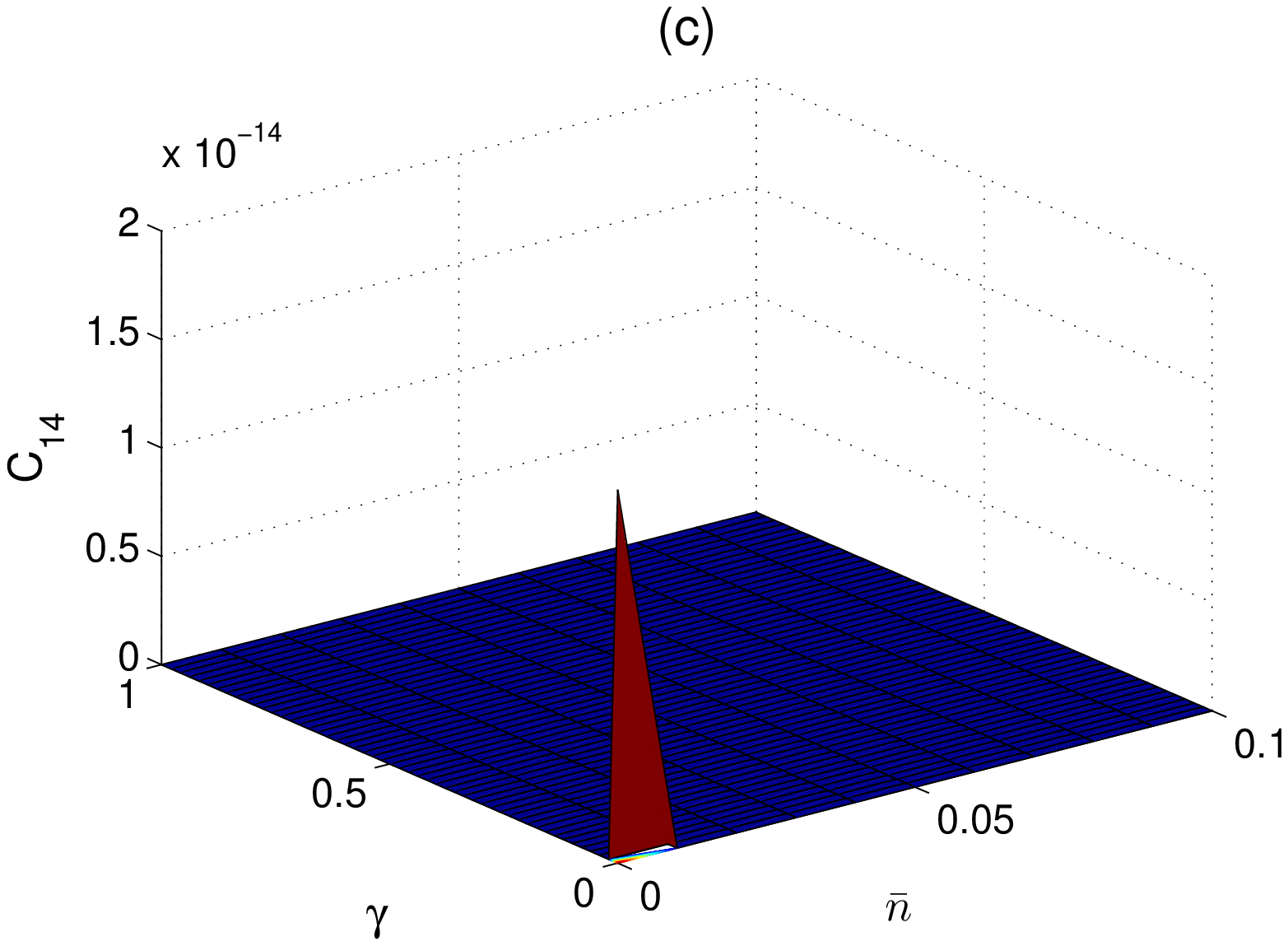}}\quad 
 \subfigure{\includegraphics[width=8cm]{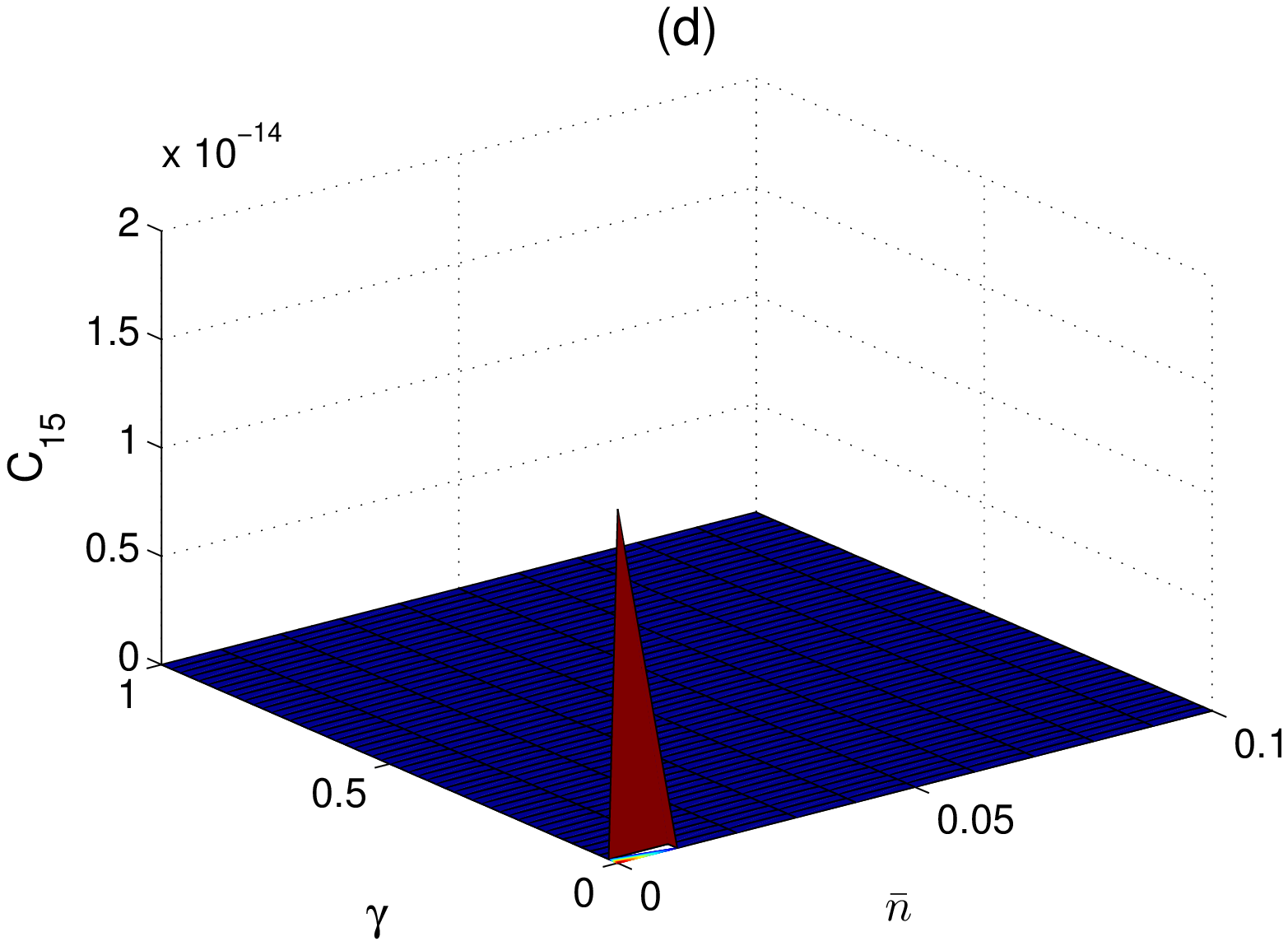}}\\
 \caption{{\label{N5_open_C_vs_n_gamma_G05} The asymptotic behavior of (a) $C_{12}$; (b) $C_{13}$ and the value of (c) $C_{14}$; (d) $C_{15}$ (at $T=300$) in the Heisenberg $XYZ$ system in presence of the environment ($\Gamma=0.05$) versus the anisotropy parameter $\gamma$ and the temperature parameter $\bar{n}$ starting from any initial state (disentangled, entangled or maximally entangled) and for $\delta = 1$.}}
 \end{minipage}
\end{figure}
In order to further examine the effect of the anisotropy of the system at zero temperature, we plot the value of the entanglements $C_{12}$ $C_{13}$ $C_{14}$ and $C_{15}$, at $T=300$ in the $\gamma-\delta$ space of Heisenberg spin system in fig.~\ref{N5_open_C_vs_gamma_delta_G05_n0}, for $N=5$. In general, the asymptotic behavior of the entanglements $C_{12}$ and $C_{13}$ looks very close to what has been observed in the closed boundary case except for few small changes. As can be noticed in fig.~\ref{N5_open_C_vs_gamma_delta_G05_n0}(a), the nn entanglement $C_{12}$ decreases monotonically as $\gamma$ decreases reaching zero at $\gamma=0$ whereas the $\delta$ parameter has no noticeable effect on $C_{12}$. The (nnn) entanglement $C_{13}$ shows a different behavior where it starts with a zero value, at $\gamma =1$, and sustains this value up to $\gamma \approx 0.72$ (not $\gamma \approx 0.5$ as in the closed boundary case) before increasing to reach a maximum value at $\gamma \approx 0.42$ then it decreases again to reach a zero value at $\gamma =0$, as illustrated in fig.~\ref{N5_open_C_vs_gamma_delta_G05_n0}(b). There is a quite small effect on $C_{13}$ due to the variation in the parameter $\delta$, where the entanglement value increases monotonically (but very slightly) as $\delta$ increases. In fig.~\ref{N5_open_C_vs_gamma_delta_G05_n0}(c) and (d), there are only non-zero values for $C_{14}$ and $C_{15}$ at $\gamma=0$ and varies as $\delta$ is varied with a maximum value around $\delta=0.5$ for $C_{14}$ and $\delta=0.75$ for $C_{15}$. In fact, the behavior of $C_{12}$ and $C_{13}$ don't change at latter times so what is shown in fig.~\ref{N5_open_C_vs_gamma_delta_G05_n0} (a) and (b) are their asymptotic steady sate values, which is not the case for $C_{14}$ and $C_{15}$ as they vanish at latter time and never revive again.
To test the entanglement robustness against thermal excitation at different degrees of anisotropy, we depict the values of entanglements $C_{12}$, $C_{13}$, $C_{14}$ and $C_{15}$ at $T=300$ versus both the anisotropic parameter $\gamma$ and the temperature parameter $\bar{n}$ in fig.~\ref{N5_open_C_vs_n_gamma_G05}. The resistance of the (nn) entanglement $C_{12}$ to the thermal effects decreases as the degree of anisotropy of the system decreases as shown in fig.~\ref{N5_open_C_vs_n_gamma_G05}(a) in a very similar fashion to the closed boundary case. On the other hand, the (nnn) entanglement $C_{13}$ shows no resistance at high anisotropy values but rises up at $\gamma \approx 0.72$ reaching a maximum value at $\gamma\approx 0.42$ before vanishing again at $\gamma=0$ as can be seen in fig.~\ref{N5_open_C_vs_n_gamma_G05}(b), where it survives within $\bar{n}<0.01$. The (nnnn) and (nnnnn) entanglements $C_{14}$ and $C_{15}$, plotted in fig.~\ref{N5_open_C_vs_n_gamma_G05}(c) and (d), exist only with a quite small value in the close vicinity of $\gamma=0$ and $\bar{n}=0$, which means these concurrences may survive only in the isotropic system very close to the zero temperature. At latter times, $T > 300$, the profiles of $C_{12}$ and $C_{13}$ don not change, whereas $C_{14}$ and $C_{15}$ vanish. Therefore, the quantum character and entanglement may persist in the Heisenberg spin chains even at non-zero temperature based mainly on the degree of spatial anisotropy in the system.
\section{Conclusions}
We have investigated the time evolution and transfer of short and long range quantum entanglement in a finite one-dimensional Heisenberg spin chains with nearest-neighbor spin interaction under the influence of thermal and dissipative Lindblad environments in presence of an external magnetic field. We have considered both cases of closed and open boundary spin chains with maximum number of 7 spins. We presented an exact numerical solution for the Lindblad master equation of the system in the Liouville space. In the closed boundary free Heisenberg spin chain (in absence of thermal or dissipative environments), the nearest neighbor and beyond nearest neighbor entanglement as well as the one-tangle $\tau_1$ and the overall bipartite entanglement $\tau_2$ were found to evolve in time in a non-uniform oscillatory form that changes significantly depending on the initial state, system size and the degree of spatial anisotropy. The oscillatory behavior of the entanglement in the spin chain is suppressed once the system is coupled to the dissipative environment. We showed how the asymptotic (long time) behavior of the entanglement in the system under the influence of the environment at zero temperature, particularly the nearest neighbor and the next to nearest neighbor, is very sensitive to the degree of the spatial anisotropy, which causes them to reach either a zero or a finite sustainable steady state value regardless of the initial state of the system. The steady state of the nearest neighbor and next to nearest neighbor entanglement shows robustness against temperature up to very small non-zero temperature, which varies significantly depending on the degree of anisotropy. The Robustness of the different ranges of entanglement against dissipative and thermal effects are enhanced at degrees of anisotropy that are different for each one of them. The open boundary spin chain was considered with a focus on the end to end entanglement transfer through the chain. We have studied the entanglement transfer starting from a maximally entangled pair of spins at one end, which is initially disentangled from the rest of the mutually disentangled spins. The entanglement transfer time and speed through the chain vary significantly depending on the degrees of anisotropy and the separation from the entangled pair for both of the free and environment-coupled systems. The transferred nearest neighbor and next to nearest neighbor entanglement through the chain under the influence of the dissipative environment may vanish or asymptotically reach a finite steady state value depending on the degrees of anisotropy of the system and the thermal effects in a close pattern to the closed chain case. The longer range transferred entanglement sustains very small values for a short period of time before completely vanishing. The thermal excitation has a devastating effect on the entanglement in both of the closed and open boundary chains.


\end{document}